\documentclass[prd,tightenlines,floatfix,showpacs,preprintnumbers,nofootinbib,eqsecnum,superscriptaddress,twocolumn]{revtex4-1}
\pdfoutput=1

\usepackage{color}      % Non-standard graphics support
\usepackage{amsbsy}     % Bolds (vectors, tensors, etc.)
\usepackage{amsfonts}       % AMS fonts
\usepackage{amsmath}        % AMS maths
\usepackage{amssymb}        % AMS symbols
\usepackage{wasysym}
\usepackage{multirow}
\usepackage{mciteplus}
\usepackage{psfrag}
\usepackage{slashed}
\usepackage{cancel}
\usepackage{graphicx}
\usepackage{array}
\usepackage{subfig}
\usepackage[percent]{overpic}
\usepackage{ulem}

\usepackage{slashed}  % Provides the Feynman Slash; used for the No-Pion fit notation

\usepackage{hyperref}

\usepackage{placeins}

\usepackage[english]{babel}

\newcommand{\ncteq}{{\tt nCTEQ}}
\newcommand{\cteq}{{\tt CTEQ }}

\newcommand{\ncteqfit}{{\tt nCTEQ15}}

\newcommand{\ncteqnp}{{\tt nCTEQ15-np}}
\newcommand{\GeV}{{\rm GeV}}
\newcommand{\gsim}{\gtrsim}
\newcommand{\lsim}{\lesssim}

\begin{document}

\preprint{
\vbox{
\null \vspace{0.3in}
\hbox{LPSC-15-153}
% \hbox{SMU-HEP-12-05}
\hbox{MS-TP-15-11}
\hbox{FERMILAB-PUB-15-375-ND-PPD-T}
}
}
\vspace*{1ex}

\title{nCTEQ15 -- Global analysis of nuclear parton distributions with uncertainties in the \cteq framework}

 \author{K.~Kova\v{r}\'{\i}k}
% \email{karol.kovarik@uni-muenster.de}
 \affiliation{
              Institut f{\"u}r Theoretische Physik, Westf{\"a}lische Wilhelms-Universit{\"a}t M{\"u}nster,
              Wilhelm-Klemm-Stra{\ss}e 9, D-48149 M{\"u}nster, Germany
}
 \author{A. Kusina}
% \email{kusina@lpsc.in2p3.fr}
 \affiliation{
              Laboratoire de Physique Subatomique et de Cosmologie,
              Universit\'e Grenoble-Alpes, CNRS/IN2P3,
              53 avenue des Martyrs, 38026 Grenoble, France
}

 \author{T.~Je\v{z}o}
% \email{tomas.jezo@mib.infn.it}
 \affiliation{
              Universit\`a di Milano-Bicocca and INFN, Sezione di Milano-Bicocca,\\
              Piazza della Scienza 3, 20126 Milano, Italy}

 \author{D.~B.~Clark}
% \email{dbclark@smu.edu}
 \affiliation{
  Southern Methodist University, Dallas, TX 75275, USA
}

 \author{C.~Keppel}
% \email{keppel@jlab.org}
 \affiliation{
  Thomas Jefferson National Accelerator Facility, Newport News, VA, 23606, USA
}

 \author{F.~Lyonnet}
% \email{flyonnet@smu.edu}
 \affiliation{
  Southern Methodist University, Dallas, TX 75275, USA
}

 \author{J.G.~Morf{\'i}n}
% \email{morfin@fnal.gov}
 \affiliation{
  Fermi National Accelerator Laboratory, Batavia, Illinois 60510, USA
}

 \author{F.~I.~Olness}
% \email{olness@smu.edu}
 \affiliation{
  Southern Methodist University, Dallas, TX 75275, USA
}

 \author{J.F.~Owens}
% \email{owens@hep.fsu.edu}
 \affiliation{
  Department of Physics, Florida State University, Tallahassee, Florida 32306-4350, USA
}

 \author{I.~Schienbein}
% \email{ingo.schienbein@lpsc.in2p3.fr}
 \affiliation{
              Laboratoire de Physique Subatomique et de Cosmologie,
              Universit\'e Grenoble-Alpes, CNRS/IN2P3,
              53 avenue des Martyrs, 38026 Grenoble, France
}

 \author{J.~Y.~Yu}
% \email{yu@physics.smu.edu}
 \affiliation{
  Southern Methodist University, Dallas, TX 75275, USA
}

\date{\today}

\begin{abstract}
We present the new  \ncteqfit\ set of nuclear parton distribution functions (nPDFs) 
with uncertainties.
This fit extends the CTEQ proton PDFs to include the nuclear dependence using data on nuclei all the way up to ${}^{208}$Pb.
The uncertainties are determined using the Hessian method with an optimal rescaling of the 
eigenvectors to accurately represent the uncertainties for the chosen tolerance criteria. 
In addition to the Deep Inelastic Scattering (DIS) and Drell-Yan (DY) processes, 
we also include inclusive pion production data from RHIC to help constrain the nuclear gluon PDF. 
Furthermore, we investigate the correlation of the data sets with specific nPDF flavor components,
and asses the impact of individual experiments. 
We also provide comparisons of the  \ncteqfit\ set with recent fits from other groups. 
\end{abstract}


\maketitle

\tableofcontents{}

%!TEX root=paper.tex
%+++++++++++++++++++++++++++++++++++++++++++++++++++++++
%\newpage
\section{Introduction}
\label{sec:intro}
%+++++++++++++++++++++++++++++++++++++++++++++++++++++++
%
% Proton PDFs
In the last thirty years, an impressive array of discoveries in particle physics has come from high energy hadron experiments.
These discoveries, along with many other key measurements, rely on our understanding of nucleon structure.
A nucleon can be described using the language of parton distribution
functions (PDFs) which is based on QCD factorization 
theorems~\cite{Collins:1985ue,Bodwin:1984hc,Collins:1998rz}.
PDFs are determined in global analyses of a variety of different hard scattering processes such as
deep inelastic scattering (DIS), Drell-Yan (DY) lepton pair production, vector boson production and the inclusive jet production.
The backbone of any global analysis are the very precise DIS structure function data from HERA which cover
a wide kinematic range in $(x,Q^2)$.
Several global analyses, based on an ever growing set of precise experimental data
and on next-to-next-to-leading order (NNLO) theoretical predictions, are regularly updated and 
maintained~\cite{Gao:2013xoa,Ball:2014uwa,Harland-Lang:2014zoa,Alekhin:2013nda,Owens:2012bv,Jimenez-Delgado:2014twa}.

Over the years, a series of global analysis studies have been performed within a single framework, or comparing different frameworks.
For example,  detailed studies of PDF uncertainties have been compared using Hessian, Lagrangian and Monte Carlo methods. Furthermore, the precision of experimental data and theoretical 
predictions in the proton case allows one to perform studies of smaller effects such as the difference between the treatment of heavy 
quarks in different analyses or the exact treatment of target-mass corrections and higher twist effects.
As a consequence, the nucleon structure is quite well known over a wide kinematic range.

% Nuclear PDFs
Similarly, the theoretical description of hard scattering processes in lepton--nucleus and proton--nucleus reactions
requires the knowledge of parton distribution functions inside nuclei characterized by the atomic number $A$ and the charge $Z$.
It has been known since the discovery of the EMC effect \cite{Aubert:1983xm}
more than 30 years ago 
that the nucleus cannot be considered as an ensemble of $Z$ free protons and $(A-Z)$ free neutrons. 
Consequently, the nuclear PDFs (nPDFs) will differ from the naive additive combination of free proton and neutron PDFs.
As in the proton case, nuclear PDFs have been determined in the literature
by  global fits to experimental data for hard scale processes
including deep inelastic scattering on nuclei and nuclear collision 
experiments~\cite{deFlorian:2011fp,Eskola:2009uj,Schienbein:2009kk,Hirai:2007sx}.
However, compared to the proton our knowledge of nuclear PDFs is much less advanced. There are several reasons.

On the theoretical side, the description of nuclear induced hard processes is more challenging
due to the  complex nuclear environment.
Still, all global nuclear PDF analyses rely on the assertion that the QCD factorization theorems remain valid for 
$\ell A$ and $p A$ hard scattering processes, see e.g.~\cite{Qiu:2003cg,Accardi:2004be}. 
In fact, it is only in this context that the universal parton distributions ($f_i^A(x,Q)$) are defined;
they are given as matrix elements of the same local twist-2 operators as in the proton case but on nuclear states.
The nuclear PDFs then account for nuclear effects (in particular EMC suppression, shadowing, anti-shadowing) at the twist-2 level
in a universal manner and the entire formalism becomes predictive.
However, higher twist contributions are expected to be enhanced in a nucleus ($\propto A^{1/3}$) \cite{Qiu:2003cg,Accardi:2004be}.
Here, final state re-scattering corrections due to the propagation of the outgoing partons through the nuclear medium,
which are higher twist, should be power suppressed but may be substantial and so must be either included in the analysis 
or eliminated by suitable kinematic cuts.\footnote{Needless to say that the final state interactions do not concern the fully inclusive DIS structure functions but may be relevant for less inclusive observables (single pion production, di-muon production in $\nu A$ DIS, \ldots).
On the other hand, power suppressed initial state interactions are expected to be numerically small.}
In addition, other effects like a different propagation of the hadronic fluctuations
of the exchange bosons in the nuclear medium,\footnote{There could be modifications of charged current neutrino scattering that are different than those for neutral current charged lepton scattering for instance due to the exchange of a charged massive vector boson \cite{Morfin:2012kn}.} gluon saturation, and deviations from
DGLAP evolution at small $x$ may play a more prominent role in the
nuclear case, see e.g. \cite{Armesto:2006ph,Kopeliovich:2012kw} and references therein.

Ultimately, the validity of the twist-2 factorization formalism will be tested phenomenologically by how well our approach 
based on the factorization assumption describes the data.
The existing global analyses generally lead to a good description of the data confirming this picture;
however, it  may  be challenged
by future precision data from the LHC and an Electron-Ion Collider (EIC) covering an extended kinematic plane.
It is notable that tensions between $\nu A$ DIS data 
and $\ell A$ DIS data have been reported 
\cite{Schienbein:2007fs,Kovarik:2010uv}
which might be due to higher twist contributions, or indicate a breaking of twist-2 factorization.
These  tensions largely disappear if the correlations between the NuTeV data points are discarded 
\cite{Paukkunen:2013grz}.
The other reason why nuclear PDFs lag behind the proton analyses can be traced back to the lack of precise experimental data.
For example the constraints on the nPDFs for any single nucleus are (so far) too scarce,
so that experimental data from scattering on multiple nuclei must be considered.
Since the nuclear effects are clearly dependent on the number of nucleons, this requires
modeling of the non-trivial nuclear $A$ dependence of the parton distributions.
Even after combining the data sets for different nuclei,
the precision of the nuclear PDFs is not yet comparable to the proton PDFs
where quark distributions for most flavors together with the gluon distribution are reliably determined 
over a broad kinematic range, due to the smaller number and hence smaller kinematic coverage of the current relevant nuclear data.
As a consequence, the nuclear PDFs in every analysis have large uncertainties as the parton distributions  
are not fully constrained by the available data. The nuclear PDFs still largely depend on assumptions inherent in every analysis. 
The dependence on assumptions, such as for example the parameterization form, leads to predictions where different analyses differ 
by more than the estimated uncertainties. It follows therefore that in order to assess the true uncertainty, all available
results and their uncertainties should be considered and combined.

In this paper we present a new analysis of nuclear PDFs in the CTEQ global PDF fitting framework. We use theoretical predictions at the 
next-to-leading order to fit all available data from charged lepton DIS and Drell-Yan di-lepton production as in our previous analysis 
\cite{Schienbein:2009kk}. In addition, we have added inclusive pion production data from RHIC and have performed a careful analysis 
of the uncertainties using the Hessian method.
Our framework differs considerably from other global analyses of nuclear PDFs which we compare our results with.

The remainder of this paper is organized as follows. 
In Section~\ref{sec:framework}, we introduce in detail the framework including the parametrization of 
the nPDFs at the input scale together with a review of the Hessian method which we use to estimate the uncertainties on the nPDFs.
In Section~\ref{sec:data}, we review the experimental data included in the fit. 
In Section~\ref{sec:res}, we present the results of our fit, compare with recent results from the literature, 
and examine the correlations between individual PDF flavors and the various experiments. 
Finally, in Section~\ref{sec:sum}, we summarize the obtained results.
Additionally we include two appendices.
In Appendix~\ref{app:Hess} we provide details on the Hessian rescaling method,
and in Appendix~\ref{app:usage} we comment on the usage and availability of our nPDFs.

%!TEX root=paper.tex
%+++++++++++++++++++++++++++++++++++++++++++++++++++++++
\section{The nPDF Framework}
\label{sec:framework}
%+++++++++++++++++++++++++++++++++++++++++++++++++++++++
In this section we describe in detail the framework of the \ncteq\ global analysis.
For the purpose of fitting nuclear parton distributions we will parameterize 
$f_{i}^{p/A}(x,Q_0)$, the PDFs of a proton bound in a nucleus $A$,
then construct the full distributions of partons in the nucleus using isospin
symmetry, and in the end perform a fit just like in the case of the free proton.
Indeed, isospin symmetry is used to construct the PDFs
of a bound neutron, $f_{i}^{n/A}(x,Q)$, from those of the proton by exchanging
up- and down-quark distributions. Afterwards the parton distributions of the nucleus are
constructed as:
\begin{equation}
f_{i}^{(A,Z)}(x,Q) = \frac{Z}{A}f_{i}^{p/A}(x,Q) + \frac{A-Z}{A}f_{i}^{n/A}(x,Q),
\label{eq:nucleusPDF}
\end{equation} 
where $Z$ is number of protons and $A$ number of protons and neutrons in the nucleus.\footnote{Note that 
the PDFs  of the nucleus, $f_{i}^{(A,Z)}(x,Q)$, are the objects of interest
which are constrained by the experimental data, whereas the $f_{i}^{p/A}(x,Q)$ and $f_{i}^{n/A}(x,Q)$
are just {\em effective} quantities used internally to decompose the nuclear PDFs.
They should not be interpreted literally as matrix elements of local operators where the free nucleon
states have been replaced by bound nucleon states in a nuclear medium since they also include effects from
multi-nucleon states. The notion of ``effective bound nucleon PDFs" is also used in the literature discussing the factorization
in the case of $pA$ interactions \protect\cite{Qiu:2003cg}.} 

The theoretical calculations in our global analysis make use of parton distributions of a 
particular nucleus $f_i^{(A,Z)}$ to determine the DIS structure functions, Drell-Yan cross sections or
the cross section for an inclusive pion production:
\begin{eqnarray}
	F_2^A(x,Q^2) &=& \sum_i f_i^{(A,Z)}(x,Q^2)\otimes C_{2,i}(x,Q^2)\,,\\
	d\sigma_{AB\rightarrow l\bar{l}X} &=& \sum_{ij} f_i^{(A_1,Z_1)}\otimes f_j^{(A_2,Z_2)}\otimes 
	d\hat{\sigma}^{ij\rightarrow l\bar{l}X},
	\nonumber\\[-0.5cm] \null \\
	d\sigma_{dA\rightarrow \pi X} &=& \sum_{ijk} f_i^{d}\otimes f_j^{(A,Z)}\otimes \label{eq:pion}
	d\hat{\sigma}^{ij\rightarrow kX}\otimes D_k^{\pi} ,
\end{eqnarray}
where $\otimes$ stands for a convolution integral over the momentum fraction.
The DIS structure functions calculations are carried out using the ACOT variable flavor number scheme 
\cite{Aivazis:1993kh,Aivazis:1993pi,Collins:1998rz,Kramer:2000hn} at next-to-leading order in QCD \cite{Kretzer:1998ju}.\footnote{For recent 
extensions of the ACOT scheme to higher orders, required for global analyses at next-to-next-to-leading order, see \protect\cite{Guzzi:2011ew,Stavreva:2012bs};
the massless limits have been validated with the help of QCDNUM.\cite{Botje:2010ay}
}
We take into account only the dominant target mass effects which are included in the structure 
function expressions in the ACOT scheme \cite{Aivazis:1993kh}. Full treatment of the target mass corrections \cite{Schienbein:2007gr} 
is not necessary in our analysis because they are relevant mostly at large $x$ 
and low $Q^2$, a region of phase-space which we exclude by kinematic cuts. Moreover, the target mass corrections are expected
to be of lesser importance in the ratios of structure functions.

In all theory calculations we identify the renormalization and factorization 
scales: $\mu=\mu_R=\mu_F$. The scale is set differently for different processes: in deep-inelastic scattering 
it is set to the virtuality of the exchanged vector boson $\mu^2=Q^2$; in Drell-Yan production processes it is set to 
the invariant mass of the produced lepton pair $\mu^2=M^2$; and in inclusive pion production the common scale is set 
equal to the final state fragmentation scale as $\mu=\mu_F^\prime=0.5 p_T$ where $p_T$ is the transverse momentum of the produced 
$\pi^0$.
To speed-up the evaluation of next-to-leading order cross sections in the fit, we have the ability to use K-factors; 
however for the final fitting the full NLO calculations are used.
In the case of inclusive pion production, we use the results of Ref.~\cite{Aurenche:1999nz,incNLL-website}
and speed up the calculation by using pre-computed grids already including
convolutions with one PDF and fragmentation function and leaving only one convolution (with the nuclear
PDFs) to be calculated during the fitting procedure.

%%%%%%%%%%%%%%%%%%%%%%%%%%%%%
\subsection{Parameterization}
\label{sec:parameterization}
%%%%%%%%%%%%%%%%%%%%%%%%%%%%%
%
The starting point of any determination of parton distribution functions is the parameterization of
individual distributions at the input scale $Q_0$.
The parameterization of the presented \ncteq\ nuclear PDFs is the same
as in our previous analyses~\cite{Schienbein:2009kk,Stavreva:2010mw,Kovarik:2010uv}.
It mimics the parameterization used in the free proton CTEQ
fits~\cite{Owens:2007kp,Pumplin:2002vw,Stump:2003yu},
and takes the following form:

\begin{equation}
\begin{split}
xf_{i}^{p/A}(x,Q_{0}) &= c_{0}\,x^{c_{1}}(1-x)^{c_{2}}e^{c_{3}x}(1+e^{c_{4}}x)^{c_{5}},
\\{\rm for}\quad & i=u_{v},d_{v},g,\bar{u}+\bar{d},s+\bar{s},s-\bar{s},\\[2mm]
\frac{\bar{d}(x,Q_{0})}{\bar{u}(x,Q_{0})}
&= c_{0}\,x^{c_{1}}(1-x)^{c_{2}}+(1+c_{3}x)(1-x)^{c_{4}}.
\end{split}\label{eq:param}
\end{equation}
The input scale is chosen to be the same as for the free proton fits~\cite{Owens:2007kp,Stump:2003yu}, namely $Q_0=1.3$ GeV.

However, this parameterization needs to be appropriately modified to accommodate the additional
nuclear degrees of freedom. As in other available nuclear PDFs~\cite{Eskola:2009uj,deFlorian:2011fp,Hirai:2007sx},
nuclear targets are characterized only by their atomic mass number $A$. However, in contrast to
those nPDFs where the nuclear effects are added on top of the free proton PDFs in form
of ratios, in our analysis we introduce the additional $A$ dependence directly to the $c$-coefficients of Eq.~(\ref{eq:param}):%
\begin{equation}
\begin{split}
c_{k}\to c_{k}(A) \equiv c_{k,0}+c_{k,1}\left(1-A^{-c_{k,2}}\right),
 \\
\quad k=\{1,\ldots,5\}.
\end{split}\label{eq:Adep}
\end{equation}
This parameterization is designed in such a way that for $A=1$ one recovers the underlying PDFs of a free proton.
The free proton PDFs are described by the coefficients $c_{k,0}$ which in our analysis are fixed to values of the fit of 
Ref.~\cite{Owens:2007kp} which is close to CTEQ6.1~\cite{Stump:2003yu} but has the advantage of having minimal
influence from nuclear data. 

Although in principle this framework can be used to determine the strange quark content of the bound nucleon,
there is not sufficient data available to reliably do that. Therefore we assume that at the initial scale $Q_0$
\begin{equation}\label{ssb}
	s^{p/A}(x,Q_0)=\bar{s}^{p/A}(x,Q_0) = \frac{\kappa(A)}{2}\Big(\bar{u}^{p/A}+\bar{d}^{p/A}\Big)\,,
\end{equation}
where $\kappa(A)$ is an $A$-dependent normalization factor parameterized as $\kappa(A) = \left(c^{s+\bar{s}}_{0,0}+ c^{s+\bar{s}}_{0,1}\big(1-A^{-c^{s+\bar{s}}_{0,2}}\big)\right)$.\footnote{This is a straightforward generalization of 
the approach employed in the underlying proton analysis which also
assumes that at the initial scale $Q_0$ the strange quark PDFs are constrained by $s=\bar{s}=\frac{\kappa}{2} (\bar{u}+\bar{d})$.}

The normalization coefficients $c_0$ in Eq.~\eqref{eq:param} are different than the other parameters. 
They are also dependent on the atomic number
but not all of them are free parameters that can be fitted. Most of them are constrained by sum rules.
The normalization coefficients for the valence quark PDFs are constrained for each atomic number $A$ by requiring
that they obey the number sum rules
\begin{equation}\label{eq:sum1}
	\int_0^1 dx\ f_{u_v}^{p/A}(x,Q_0) = 2\,,\quad\ \int_0^1 dx\ f_{d_v}^{p/A}(x,Q_0) = 1\,.
\end{equation}
The remaining normalization coefficients are constrained by the momentum sum rule 
\begin{equation}\label{eq:sum2}
	\int_0^1 dx\,\sum_i x f_{i}^{p/A}(x,Q_0) = 1\,,
\end{equation}
which however can only determine one of them. The rest of the normalization parameters are either considered as free parameters in the 
fit or are fixed using additional assumptions to simplify the analysis (e.g. like Eq.~\eqref{ssb}).
We choose to introduce free parameters for the momentum fraction of the gluon and for the momentum fraction of
$s+\bar{s}$ to be determined during the global fit together with the parameters from Eqs.~\eqref{eq:param} and \eqref{eq:Adep}.
The $A$-dependent momentum fraction of gluon is parametrized as
\begin{equation}
	\int_0^1 dx\,x g^{p/A}(x,Q_0) = M_g\exp\left[c^g_{0,0}+ c^g_{0,1}\left(1-A^{-c^g_{0,2}}\right)\right]\,,
\end{equation}
which modifies the momentum fraction of the gluon in a free proton (described by coefficients $M_g$ and $c^g_{0,0}$).

The momentum fraction of the $s+\bar{s}$ combination is then given by
\begin{eqnarray}
	&&\int_0^1 dx\,x \Big(s^{p/A}(x,Q_0) + \bar{s}^{p/A}(x,Q_0)\Big) = \\ \nonumber
&&
\frac{\kappa}{(2+\kappa)}
\Big(1-\int_0^1 dx\,\sum\limits_{i} x f_{i}^{p/A}\Big)\left[c^{s+\bar{s}}_{0,0}+ 
c^{s+\bar{s}}_{0,1}\left(1-A^{-c^{s+\bar{s}}_{0,2}}\right)\right],
\end{eqnarray}
where the sum runs through $i=u_v,d_v,g$. 
The remaining normalization parameters are taken care of by the 
momentum sum rule and do not introduce additional free parameters.

The parameterization of Eq.~\eqref{eq:param} together with the whole \ncteq\ nuclear PDF framework 
has been designed in analogy to the free proton PDFs where parton momentum $x$ is restricted to be in the range $(0,1)$. 
However, in the nuclear case, $x$ represents the parton fractional momentum with respect to the 
average momentum carried by a nucleon. Since a particular nucleon can have a momentum bigger than an average nucleon,
$x$ can extend up to $A$ in a nucleus with an atomic number $A$. If one were  to take this into account,
one would have to modify the sum rules in Eqs.~\eqref{eq:sum1} and \eqref{eq:sum2} together with the DGLAP evolution.
However, the structure functions at $x>1$ fall off rapidly and the contribution to the moments of the structure functions from
the region of $x>1$ is very small~\cite{Niculescu:2005rh,Fomin:2010ei}.
Therefore, all currently available nuclear PDFs have been obtained neglecting the $x>1$ region and we follow the same 
path.\footnote{In fact the first next-to-leading order nuclear PDF analysis~\cite{deFlorian:2003qf} used a framework 
which at least in principle allows to accommodate the case of $x>1$.}
%
%%%%%%%%%%%%%%%%%%%%%%%%%%%%%
\subsection{Finding the optimal PDFs}
\label{sec:fit}
%%%%%%%%%%%%%%%%%%%%%%%%%%%%%
%
The fitting procedure used to find PDFs that describe the considered data best
is based on minimizing the appropriate $\chi^2$ function, as described in~\cite{Pumplin:2002vw}.
The simplest definition of the $\chi^2$ function for $n$ experiments is
\begin{equation}\label{Eq:chi2}
\chi^2(\{a_j\}) = \sum_{i} \frac{[D_i-T_i(\{a_j\})]^2}{\sigma_i^2},
\end{equation} 
where $D_i$ are the measured experimental values, $T_i$ are the corresponding
theoretical predictions and $\sigma_i^2$ are the
systematic and statistical experimental errors added in quadrature. The parameters $\{a_j\}$ are a set of
free parameters which define the PDFs at the input scale (see Eq.~\eqref{eq:param}) and are varied in order to 
find the minimum of the $\chi^2$ function.

This simple $\chi^2$ definition, with slight modifications allowing
for the inclusion of overall changes to data normalization, is used by most of the
groups performing nuclear global analyses.
However, in the current analysis, as in the previous \ncteq\ 
fits~\cite{Schienbein:2007fs,Schienbein:2009kk,Stavreva:2010mw,Kovarik:2010uv}
this simple definition is modified to account for correlations in the experimental
uncertainties. 
We follow here the prescription suggested in Ref.~\cite{Pumplin:2002vw}.
The total $\chi^2$ for $n$ experiments with parameters $\{a_j\}$ is defined to be 
\begin{equation}
\chi^2(\{a_j\}) = \sum_{n} w_n \ \chi_n^2(\{a_j\}) \ ,
\end{equation}
where $w_n$ is the weight for experiment $n$; for our fits all weights are set to 1.
The  $\chi^2_n$ is a contribution from one individual experiment $n$, and this is given by
\begin{equation}
\chi_n^2(\{a_j\})  = \sum_{i} \frac{[D_i-T_i(\{a_j\})]^2}{\alpha_i^2}
                   - \sum_{k,k'} B_k \ A^{-1}_{kk'} \ B_{k'} \ ,
\end{equation}
where $i$ runs over data points and $k,k'$ run over sources of the correlated uncertainties.
For each experimental data point we sum the  statistical error $\sigma_i$ 
together with the uncorrelated systematic error $u_i$ in quadrature to obtain  $\alpha_i^2=\sigma_i^2+u_i^2$.
The components of the correlated uncertainties are given by \cite{Pumplin:2002vw}
\begin{equation}
\begin{split}
B_k(\{a_j\}) &= \sum_{i} \frac{\beta_{ik}\left[D_i-T_i(\{a_j\})\right]}{\alpha_i^2} \ ,
\\
A_{kk'} &= \delta_{kk'} + \sum_i \frac{\beta_{ik}\beta_{ik'}}{\alpha_i^2} \ ,
\end{split} 
\end{equation}
where $\beta_{ik}$ are the sources of correlated systematic errors.

We stress that in this procedure only the experimental uncertainties
are accounted for; all theoretical and model uncertainties (e.g. missing higher
order corrections, parameterization choice, etc.) are not taken into account.

Having defined the appropriate $\chi^2$ function it needs to be minimized with
respect to the fitting parameters $\{a_j\}$ that define the bound proton PDFs
at the initial scale $Q_0$. We perform the minimization using the {\tt pyMinuit}
package~\cite{pyminuit} which is a python interface to ``SEAL-Minuit''~\cite{Cminuit}
--- a C++ rewrite of the original Fortran {\tt Minuit} package~\cite{minuit}.

%
%%%%%%%%%%%%%%%%%%%%%%%%%%%%%
\subsection{Estimating uncertainties of PDFs}
\label{sec:hessian}
%%%%%%%%%%%%%%%%%%%%%%%%%%%%%
%
In section~\ref{sec:fit} we described how we obtain our best estimate (the central value) of the \ncteq\ nuclear PDFs 
as the minimum of the $\chi^2$ function defined in Eq.~(\ref{Eq:chi2}). 
Now we want to probe the vicinity of this minimum to be able to estimate uncertainties on our prediction.
This is done using the Hessian method~\cite{Pumplin:2000vx,Pumplin:2001ct}, which will be briefly described in the following. 
We follow the notation of Ref.~\cite{Pumplin:2000vx} and refer the reader to this publication for more
details on the Hessian formalism.

%++++++++++++++++++++++++++++
\subsubsection{Determination of the Hessian matrix}
\label{sec:Hessian_matrix}
%++++++++++++++++++++++++++++
%
The basic assumption of the Hessian method is that near
its minimum the $\chi^2$-function can be approximated by
a quadratic form of the fitting parameters $\{a_i\}$. Therefore, it can be written as
\begin{equation}
\chi^2 = \chi_0^2 + \sum_{i,j} H_{ij}\,y_i\, y_j ,
\label{eq:chi2_y}
\end{equation}
where $y_i=a_i-a_i^0$ are the parameter shifts from the minimum given by the $a_i^0$ parameters, 
$\chi_0^2\equiv\chi^2(\{a_i^0\})$ is the value of
the $\chi^2$-function in the minimum, and $H_{ij}$ is the Hessian matrix
defined as:%
\begin{equation}
H_{ij} = \frac{1}{2}\left(\frac{\partial^2\chi^2}{\partial y_i\partial y_j}\right)_{a_i=a_i^0}.
\end{equation}
Since the Hessian $H_{ij}$ is a symmetric $n\times n$ matrix
(where $n$ is the number of free parameters $a_i$)
it has $n$ orthogonal eigenvectors forming a basis in the
$\{y_i\}$-space.
The characteristic equation can be written as:
\begin{equation}
\sum_{j} H_{ij} V_j^{(k)} = \lambda_k V_i^{(k)}.
\end{equation}
The eigenvectors $V_i^{(k)}$ that we use can be normalized so that:
\begin{equation}
\sum_i V_i^{(j)} V_i^{(k)} = \delta_{jk}.
\end{equation}
For our later convenience we also introduce eigenvectors
normalized to the corresponding eigenvalues:
\begin{equation}
\tilde{V}_i^{(k)} = \frac{1}{\sqrt{\lambda_k}} V_i^{(k)}.
\end{equation}
The eigenvectors can be used to disentangle the original PDF parameters
and define a new basis ${\bf z}\equiv\{z_i\}$ where the Hessian is diagonal:%
\footnote{In the basis defined using the rescaled eigenvectors $\tilde{V}_i^{(k)}$, 
the Hessian is represented by a unit matrix.}
\begin{eqnarray}\nonumber
\sum_{i,j} H_{ij}\,y_i\, y_j &=& \sum_{i,j} H^D_{ij}\,z_i\, z_j = {\bf z}^T.D^{T}.H.D.{\bf z}
\\  &=& {\bf z}^T.\left(
\begin{array}{cccc}
\lambda_1 & 0         & \dots  & 0     \\
0         & \lambda_2 &        & \vdots\\
\vdots    &           & \ddots & 0     \\
0         & \dots     & 0      & \lambda_n
\end{array}
\right).{\bf z}\,.
\end{eqnarray}
The new coordinates are defined using a matrix $D$ as
\begin{equation}
{\bf z} = D^{-1} {\bf y}\, ,
\end{equation}
where $D$ is a matrix composed of eigenvectors:
\begin{equation}
\begin{split}
D = (V^{(1)},V^{(2)},...,V^{(n)}) \equiv
\left(
\begin{array}{cccc}
V^{(1)}_1 & V^{(2)}_1 & \dots  & V^{(n)}_1\\
V^{(1)}_2 & V^{(2)}_2 & \dots  & V^{(n)}_2\\
\vdots     & \vdots   & \ddots & \vdots    \\
V^{(1)}_n & V^{(2)}_n & \dots  & V^{(n)}_n
\end{array}
\right).
\\ \null 
\end{split}\label{eq:Dmatrix}
\end{equation}
Note that because the Hessian is symmetric
$D^{-1}=D^T$.
Using the index notation such as $D_{ij} = V_i^{(j)}$ we can write the relation between 
the original fitting parameters and the new parameters as:
\begin{equation}
y_i = \sum_{j} V_i^{(j)} z_j
\equiv \sum_{j} \tilde{V}_i^{(j)} \tilde{z}_j
= \sum_{j} \frac{1}{\sqrt{\lambda_j}} V_i^{(j)} \tilde{z}_j,
\label{eq:yTOz}
\end{equation}
where we introduced a new basis $\tilde{z}_i$
which corresponds to the rescaled eigenvectors $\tilde{V}_i^{(k)}$.
The inverse transformation is given by:
\begin{equation}
\begin{split}
z_i &= \sum_{j} y_j V_j^{(i)},\\
\tilde{z}_i &= \lambda_i \sum_{j} y_j \tilde{V}_j^{(i)}
= \sqrt{\lambda_i} \sum_{j} y_j V_j^{(i)}.
\end{split}
\label{eq:zTOy}
\end{equation}
In the new coordinates, $\Delta\chi^2 = \chi^2 - \chi_0^2$ has a particularly simple form:
\begin{equation}
\Delta\chi^2
 = \sum_{i} \lambda_i z_i^2 = \sum_{i} \tilde{z}_i^2\,.
\label{eq:Dchi2}
\end{equation}

Using the Hessian method to analyze the vicinity of the minimum of the $\chi^2$-function
seems straightforward in theory but in practice when applied to a global PDF analysis,
one encounters a few problems worth pointing out. As was already mentioned in the discussion 
of free proton PDFs \cite{Pumplin:2000vx} and as is the case in our 
analysis, the eigenvalues of the Hessian span several orders of magnitude. 
In order to correctly identify all eigenvalues, 
the precision with which the Hessian matrix is determined needs to be kept under control.

In practice, the Hessian matrix is calculated using finite differences to determine the second derivatives. 
A careful choice of the step in the finite difference definition of the second derivatives is crucial.
If the step is too large, one probes too large a neighborhood of the minimum where the $\chi^2$-function cannot be 
described by a quadratic approximation anymore. If the step is too small, numerical noise in the $\chi^2$-function
prevents a reliable determination of the second derivatives. Moreover, the step size has to be different for each of the 
parameters as the $\chi^2$-function depends differently on each of them. The relative step sizes $\Delta y_i$ to each of the
parameters are set as
\begin{equation}
 \Delta y_i=\sqrt{\frac{\Delta\chi^2}{H_{ii}}},
\end{equation}
where $\Delta\chi^2 = \chi^2 - \chi^2_0$ defines the small neighborhood from which the
derivatives of the $\chi^2$-function are calculated.

It turns out that the numerical noise in the $\chi^2$-function is larger than expected for the case of a global PDF ana\-ly\-sis. 
Contrary to what one would expect, the $\chi^2$-function is not smooth which influences the determination of 
the second derivatives for all step sizes. 
It all comes down to the fact that one evaluation of the $\chi^2$-function requires several hundred evaluations of 
different next-to-leading order theory calculations which, in their numerical implementations, are not smooth functions of the 
fit parameters.

To reduce the influence of the noise on the derivatives of the $\chi^2$-function, the standard definition 
of the derivative using the central differences
\begin{equation}\label{eq:dfcent}
	\frac{df}{dx} = \frac{f_{+1}-f_{-1}}{2h}
\end{equation}
in which $f_{k} = f(x_0 + kh)$, is replaced by noise reducing derivatives (see \cite{lanczosdif}).
The central differences approach to derivatives is based on interpolating the $\chi^2$-function by a polynomial which coincides
with the $\chi^2$-function in several chosen points e.g. a quadratic polynomial interpolating the $\chi^2$-function in 3 points leads 
to the derivative in Eq.~(\ref{eq:dfcent}). If the $\chi^2$-function suffers from numerical noise,
the interpolated polynomial suffers as much if not more.

We adopt a different approach and instead of interpolating $N$ points by a polynomial of the order $N-1$, we allow
a polynomial to assume different values in these $N$ points and approximate the $\chi^2$-function by the method of least squares.
This approach assumes that the order of the polynomial $M$ has to be strictly less than $N-1$ where $N$ is the number of points.
If we use a quadratic polynomial to fit 7 symmetrically chosen, equidistant points of the $\chi^2$-function, we obtain
the following prescriptions for the 7-point low-noise derivative
\begin{equation}
	\frac{df}{dx} = \frac{f_1-f_{-1} + 2(f_2-f_{-2}) + 3(f_3-f_{-3})}{28h}\,.
\end{equation}
Using these derivatives instead of the standard derivative from Eq.~(\ref{eq:dfcent}) and extending this approach to the 
second derivatives allows us to determine the Hessian with sufficient precision and to eliminate the influence of the numerical noise.
%
%++++++++++++++++++++++++++++
\subsubsection{Error PDFs}
\label{subsec:errorPDFs}
%++++++++++++++++++++++++++++
To translate the uncertainties contained in the data to the underlying PDF parameters, we use the fact that the $\chi^2$-function
in the diagonalized Hessian approximation is a simple function of the parameters $\tilde{z}_k$. 
Varying data within their errors corresponds to a change in $\chi^2$ (denoted by $\Delta\chi^2$) which can then in turn be interpreted 
as a shift in the parameters $\tilde{z}_k$ 
\begin{equation}
\begin{split}
\tilde{z}_k &= \pm\sqrt{\Delta\chi^2},\\
z_k &= \pm\sqrt{\frac{\Delta\chi^2}{\lambda_k}}, \qquad k=1,2,\dots,n\,.
\end{split}
\label{eq:z-pm}
\end{equation}
A specific change in $\chi^2$ can be obtained by varying the parameters using $n$ independent
directions in the parameter space.%
    \footnote{If one allows only positive changes of parameters, there are $2n$ directions.}
In the $\tilde{z}_k$ space
all directions are equivalent so we can choose the $n$ independent directions to coincide 
with the directions where one single parameter is varied. A change in one direction along one single parameter $\tilde{z}_k$ leads to a simultaneous change in all original parameters $a_i$
\begin{equation}
y_i \equiv \Delta a_i = \pm
\sqrt{\frac{\Delta\chi^2}{\lambda_k}} V_i^{(k)}.
\label{eq:y_error}
\end{equation}
The parameter shifts along the direction of the $\tilde{z}_k$ parameter
are used to generate $2n$ error PDFs for a specified $\Delta\chi^2$
\begin{equation}
\begin{split}
&f_k^\pm\equiv f\left(a_i^0 \pm
\sqrt{\frac{\Delta\chi^2}{\lambda_k}} V_i^{(k)}\right),
 \quad \text{for}\; k=1,2,\dots,n\,.
\label{eq:errorPDFdef}
\end{split}
\end{equation}
The error PDFs can be used to determine the PDF uncertainty of any observable $X$ which depends on PDFs.
% The uncertainty is denoted by $\Delta X$ and can be determined as
This uncertainty, which we denote as $\Delta X$, can be determined in different ways
and in this work we define it by adding errors in quadrature
\begin{equation}
\Delta X = \frac{1}{2}\sqrt{\sum_k \left( X(f_k^{+}) - X(f_k^{-}) \right)^2 }\quad .
\label{eq:errorsQuad}
\end{equation}

The PDF uncertainty $\Delta X$ clearly depends on the exact value chosen for $\Delta\chi^2$. 
In an ideal case, an increase of $\chi^2$
corresponding to one standard deviation from the central value is $\Delta\chi^2 = 1$.
However, 
in our fit we combine results from 
different experiments which are not necessarily uncorrelated or compatible, 
so the standard argument does not apply and $\Delta\chi^2$ may be
different from one. 
To estimate what is the appropriate value for the $\Delta\chi^2$ (often referred to as the tolerance)
we use a criterion similar to the one advocated in~\cite{Stump:2001gu,Martin:2009iq,Eskola:2009uj},
which results in the value $\Delta\chi^2 = 35$.
Additionally, since the value of our tolerance is far from 1, the quadratic approximation
of the Hessian method becomes less precise. We account for it by introducing an additional
procedure of rescaling of the Hessian matrix. Both the rescaling procedure and the criterion
for choosing the $\Delta\chi^2$ tolerance are described in detail in Appendix~\ref{app:Hess}.

%!TEX root=nCTEQ15.tex
%+++++++++++++++++++++++++++++++++++++++++++++++++++++++
\section{Experimental data}
\label{sec:data}
%+++++++++++++++++++++++++++++++++++++++++++++++++++++++

% F2AD table
%--------------------
\begin{table}[th]
\begin{center}
\begin{tabular}{|l|l|c|c|c|c|c|}
\hline
{\scriptsize $\mathbf{F_{2}^{A}/F_{2}^{D}:}$ }  &  &  &  &  & {\scriptsize $\#$ data }  &\tabularnewline
{\scriptsize Observable }  & {\scriptsize Experiment }  & {\scriptsize ID} & {\scriptsize Ref. }  & {\scriptsize $\#$ data }  & {\scriptsize after cuts }  & {\scriptsize $\chi^2$}\tabularnewline
\hline
\hline
{\scriptsize D}   & {\scriptsize NMC-97}  &  {\scriptsize 5160}  &  {\scriptsize \cite{Arneodo:1996qe}}     & {\scriptsize 292}   &  {\scriptsize 201}   &  {\scriptsize 247.73}\tabularnewline
{\scriptsize He/D}   & {\scriptsize Hermes}  &  {\scriptsize 5156}  &  {\scriptsize \cite{Airapetian:2002fx}}     & {\scriptsize 182}   &  {\scriptsize 17}   &  {\scriptsize 13.45}\tabularnewline
                     & {\scriptsize NMC-95,re}  &  {\scriptsize 5124}  &  {\scriptsize \cite{Amaudruz:1995tq}}     & {\scriptsize 18}   &  {\scriptsize 12}   &  {\scriptsize 9.78}\tabularnewline
                     & {\scriptsize SLAC-E139}  &  {\scriptsize 5141}  &  {\scriptsize \cite{Gomez:1993ri}}     & {\scriptsize 18}   &  {\scriptsize 3}   &  {\scriptsize 1.42}\tabularnewline
{\scriptsize Li/D}   & {\scriptsize NMC-95}  &  {\scriptsize 5115}  &  {\scriptsize \cite{Arneodo:1995cs}}     & {\scriptsize 24}   &  {\scriptsize 11}   &  {\scriptsize 6.10}\tabularnewline
{\scriptsize Be/D}   & {\scriptsize SLAC-E139}  &  {\scriptsize 5138}  &  {\scriptsize \cite{Gomez:1993ri}}     & {\scriptsize 17}   &  {\scriptsize 3}   &  {\scriptsize 1.37}\tabularnewline
{\scriptsize C/D}   & {\scriptsize FNAL-E665-95}  &  {\scriptsize 5125}  &  {\scriptsize \cite{Adams:1995is}}     & {\scriptsize 11}   &  {\scriptsize 3}   &  {\scriptsize 1.44}\tabularnewline
                     & {\scriptsize SLAC-E139}  &  {\scriptsize 5139}  &  {\scriptsize \cite{Gomez:1993ri}}     & {\scriptsize 7}   &  {\scriptsize 2}   &  {\scriptsize 1.36}\tabularnewline
                     & {\scriptsize EMC-88}  &  {\scriptsize 5107}  &  {\scriptsize \cite{Ashman:1988bf}}     & {\scriptsize 9}   &  {\scriptsize 9}   &  {\scriptsize 7.41}\tabularnewline
                     & {\scriptsize EMC-90}  &  {\scriptsize 5110}  &  {\scriptsize \cite{Arneodo:1989sy}}     & {\scriptsize 9}   &  {\scriptsize 0}   &  {\scriptsize 0.00}\tabularnewline
                     & {\scriptsize NMC-95}  &  {\scriptsize 5113}  &  {\scriptsize \cite{Arneodo:1995cs}}     & {\scriptsize 24}   &  {\scriptsize 12}   &  {\scriptsize 8.40}\tabularnewline
                     & {\scriptsize NMC-95,re}  &  {\scriptsize 5114}  &  {\scriptsize \cite{Amaudruz:1995tq}}     & {\scriptsize 18}   &  {\scriptsize 12}   &  {\scriptsize 13.29}\tabularnewline
{\scriptsize N/D}   & {\scriptsize Hermes}  &  {\scriptsize 5157}  &  {\scriptsize \cite{Airapetian:2002fx}}     & {\scriptsize 175}   &  {\scriptsize 19}   &  {\scriptsize 9.92}\tabularnewline
                     & {\scriptsize BCDMS-85}  &  {\scriptsize 5103}  &  {\scriptsize \cite{Bari:1985ga}}     & {\scriptsize 9}   &  {\scriptsize 9}   &  {\scriptsize 4.65}\tabularnewline
{\scriptsize Al/D}   & {\scriptsize SLAC-E049}  &  {\scriptsize 5134}  &  {\scriptsize \cite{Bodek:1983ec}}     & {\scriptsize 18}   &  {\scriptsize 0}   &  {\scriptsize 0.00}\tabularnewline
                     & {\scriptsize SLAC-E139}  &  {\scriptsize 5136}  &  {\scriptsize \cite{Gomez:1993ri}}     & {\scriptsize 17}   &  {\scriptsize 3}   &  {\scriptsize 1.14}\tabularnewline
{\scriptsize Ca/D}   & {\scriptsize NMC-95,re}  &  {\scriptsize 5121}  &  {\scriptsize \cite{Amaudruz:1995tq}}     & {\scriptsize 18}   &  {\scriptsize 12}   &  {\scriptsize 11.54}\tabularnewline
                     & {\scriptsize FNAL-E665-95}  &  {\scriptsize 5126}  &  {\scriptsize \cite{Adams:1995is}}     & {\scriptsize 11}   &  {\scriptsize 3}   &  {\scriptsize 0.94}\tabularnewline
                     & {\scriptsize SLAC-E139}  &  {\scriptsize 5140}  &  {\scriptsize \cite{Gomez:1993ri}}     & {\scriptsize 7}   &  {\scriptsize 2}   &  {\scriptsize 1.63}\tabularnewline
                     & {\scriptsize EMC-90}  &  {\scriptsize 5109}  &  {\scriptsize \cite{Arneodo:1989sy}}     & {\scriptsize 9}   &  {\scriptsize 0}   &  {\scriptsize 0.00}\tabularnewline
{\scriptsize Fe/D}   & {\scriptsize SLAC-E049}  &  {\scriptsize 5131}  &  {\scriptsize \cite{Bodek:1983qn}}     & {\scriptsize 14}   &  {\scriptsize 2}   &  {\scriptsize 0.78}\tabularnewline
                     & {\scriptsize SLAC-E139}  &  {\scriptsize 5132}  &  {\scriptsize \cite{Gomez:1993ri}}     & {\scriptsize 23}   &  {\scriptsize 6}   &  {\scriptsize 7.76}\tabularnewline
                     & {\scriptsize SLAC-E140}  &  {\scriptsize 5133}  &  {\scriptsize \cite{Dasu:1993vk}}     & {\scriptsize 10}   &  {\scriptsize 0}   &  {\scriptsize 0.00}\tabularnewline
                     & {\scriptsize BCDMS-87}  &  {\scriptsize 5101}  &  {\scriptsize \cite{Benvenuti:1987az}}     & {\scriptsize 10}   &  {\scriptsize 10}   &  {\scriptsize 5.77}\tabularnewline
                     & {\scriptsize BCDMS-85}  &  {\scriptsize 5102}  &  {\scriptsize \cite{Bari:1985ga}}     & {\scriptsize 6}   &  {\scriptsize 6}   &  {\scriptsize 2.56}\tabularnewline
{\scriptsize Cu/D}   & {\scriptsize EMC-93}  &  {\scriptsize 5104}  &  {\scriptsize \cite{Ashman:1992kv}}     & {\scriptsize 10}   &  {\scriptsize 9}   &  {\scriptsize 4.71}\tabularnewline
                     & {\scriptsize EMC-93(chariot)}  &  {\scriptsize 5105}  &  {\scriptsize \cite{Ashman:1992kv}}     & {\scriptsize 9}   &  {\scriptsize 9}   &  {\scriptsize 4.88}\tabularnewline
                     & {\scriptsize EMC-88}  &  {\scriptsize 5106}  &  {\scriptsize \cite{Ashman:1988bf}}     & {\scriptsize 9}   &  {\scriptsize 9}   &  {\scriptsize 3.39}\tabularnewline
{\scriptsize Kr/D}   & {\scriptsize Hermes}  &  {\scriptsize 5158}  &  {\scriptsize \cite{Airapetian:2002fx}}     & {\scriptsize 167}   &  {\scriptsize 12}   &  {\scriptsize 9.79}\tabularnewline
{\scriptsize Ag/D}   & {\scriptsize SLAC-E139}  &  {\scriptsize 5135}  &  {\scriptsize \cite{Gomez:1993ri}}     & {\scriptsize 7}   &  {\scriptsize 2}   &  {\scriptsize 1.60}\tabularnewline
{\scriptsize Sn/D}   & {\scriptsize EMC-88}  &  {\scriptsize 5108}  &  {\scriptsize \cite{Ashman:1988bf}}     & {\scriptsize 8}   &  {\scriptsize 8}   &  {\scriptsize 17.20}\tabularnewline
{\scriptsize Xe/D}   & {\scriptsize FNAL-E665-92}  &  {\scriptsize 5127}  &  {\scriptsize \cite{Adams:1992nf}}     & {\scriptsize 10}   &  {\scriptsize 2}   &  {\scriptsize 0.72}\tabularnewline
{\scriptsize Au/D}   & {\scriptsize SLAC-E139}  &  {\scriptsize 5137}  &  {\scriptsize \cite{Gomez:1993ri}}     & {\scriptsize 18}   &  {\scriptsize 3}   &  {\scriptsize 1.74}\tabularnewline
{\scriptsize Pb/D}   & {\scriptsize FNAL-E665-95}  &  {\scriptsize 5129}  &  {\scriptsize \cite{Adams:1995is}}     & {\scriptsize 11}   &  {\scriptsize 3}   &  {\scriptsize 1.20}\tabularnewline
\hline
\hline
\textbf{\scriptsize Total:}{\scriptsize {} }  &  &  &  & {\scriptsize 1205}  & {\scriptsize 414}  & {\scriptsize 403.70}\tabularnewline
\hline
\end{tabular}
\caption{{ The DIS $F_{2}^{A}/F_{2}^{D}$ data sets used in the \texttt{nCTEQ15} fit.
The table details values of $\chi^2$ for each experiment, the specific nuclear targets, references,
and the number of data points with and without kinematic cuts.
}}
\label{tab:exp1}
\end{center}
\end{table}
%--------------------

% F2AAp table
%--------------------
\begin{table}[h!]
\begin{center}
\begin{tabular}{|l|l|c|c|c|c|c|}
\hline
{\scriptsize $\mathbf{F_{2}^{A}/F_{2}^{A'}:}$ }  &  &  &  &  & {\scriptsize $\#$ data }  &\tabularnewline
{\scriptsize Observable }  & {\scriptsize Experiment }  & {\scriptsize ID} & {\scriptsize Ref. }  & {\scriptsize $\#$ data }  & {\scriptsize after cuts }  & {\scriptsize $\chi^2$}\tabularnewline
\hline
\hline
{\scriptsize C/Li}   & {\scriptsize NMC-95,re}  &  {\scriptsize 5123}  &  {\scriptsize \cite{Amaudruz:1995tq}}     & {\scriptsize 25}   &  {\scriptsize 7}   &  {\scriptsize 5.56}\tabularnewline
{\scriptsize Ca/Li}   & {\scriptsize NMC-95,re}  &  {\scriptsize 5122}  &  {\scriptsize \cite{Amaudruz:1995tq}}     & {\scriptsize 25}   &  {\scriptsize 7}   &  {\scriptsize 1.11}\tabularnewline
{\scriptsize Be/C}   & {\scriptsize NMC-96}  &  {\scriptsize 5112}  &  {\scriptsize \cite{Arneodo:1996rv}}     & {\scriptsize 15}   &  {\scriptsize 14}   &  {\scriptsize 4.08}\tabularnewline
{\scriptsize Al/C}   & {\scriptsize NMC-96}  &  {\scriptsize 5111}  &  {\scriptsize \cite{Arneodo:1996rv}}     & {\scriptsize 15}   &  {\scriptsize 14}   &  {\scriptsize 5.39}\tabularnewline
{\scriptsize Ca/C}   & {\scriptsize NMC-95,re}  &  {\scriptsize 5120}  &  {\scriptsize \cite{Amaudruz:1995tq}}     & {\scriptsize 25}   &  {\scriptsize 7}   &  {\scriptsize 4.32}\tabularnewline
                     & {\scriptsize NMC-96}  &  {\scriptsize 5119}  &  {\scriptsize \cite{Arneodo:1996rv}}     & {\scriptsize 15}   &  {\scriptsize 14}   &  {\scriptsize 5.43}\tabularnewline
{\scriptsize Fe/C}   & {\scriptsize NMC-96}  &  {\scriptsize 5143}  &  {\scriptsize \cite{Arneodo:1996rv}}     & {\scriptsize 15}   &  {\scriptsize 14}   &  {\scriptsize 9.78}\tabularnewline
{\scriptsize Sn/C}   & {\scriptsize NMC-96}  &  {\scriptsize 5159}  &  {\scriptsize \cite{Arneodo:1996ru}}     & {\scriptsize 146}   &  {\scriptsize 111}   &  {\scriptsize 64.44}\tabularnewline
{\scriptsize Pb/C}   & {\scriptsize NMC-96}  &  {\scriptsize 5116}  &  {\scriptsize \cite{Arneodo:1996rv}}     & {\scriptsize 15}   &  {\scriptsize 14}   &  {\scriptsize 7.74}\tabularnewline
\hline
\hline
\textbf{\scriptsize Total:}{\scriptsize {} }  &  &  &  & {\scriptsize 296}  & {\scriptsize 202}  & {\scriptsize 107.85}\tabularnewline
\hline
\end{tabular}
\caption{The DIS $F_{2}^{A}/F_{2}^{A'}$ data sets used in the \texttt{nCTEQ15} fit.
We list the same details for each data set as in Tab.~\ref{tab:exp1}.
}
\label{tab:exp2}
\end{center}
\end{table}
%--------------------

% Drell-Yan table
%--------------------
\begin{table}[h!]
\begin{center}
\begin{tabular}{|l|l|c|c|c|c|c|}
\hline
{\scriptsize $\mathbf{\sigma_{DY}^{pA}/\sigma_{DY}^{pA'}:}$ }  &  &  &  &  & {\scriptsize $\#$ data }  &\tabularnewline
{\scriptsize Observable }  & {\scriptsize Experiment }  & {\scriptsize ID} & {\scriptsize Ref. }  & {\scriptsize $\#$ data }  & {\scriptsize after cuts }  & {\scriptsize $\chi^2$}\tabularnewline
\hline
\hline
{\scriptsize C/H2}   & {\scriptsize FNAL-E772-90}  &  {\scriptsize 5203}  &  {\scriptsize \cite{Alde:1990im}}     & {\scriptsize 9}   &  {\scriptsize 9}   &  {\scriptsize 7.92}\tabularnewline
{\scriptsize Ca/H2}   & {\scriptsize FNAL-E772-90}  &  {\scriptsize 5204}  &  {\scriptsize \cite{Alde:1990im}}     & {\scriptsize 9}   &  {\scriptsize 9}   &  {\scriptsize 2.73}\tabularnewline
{\scriptsize Fe/H2}   & {\scriptsize FNAL-E772-90}  &  {\scriptsize 5205}  &  {\scriptsize \cite{Alde:1990im}}     & {\scriptsize 9}   &  {\scriptsize 9}   &  {\scriptsize 3.17}\tabularnewline
{\scriptsize W/H2}   & {\scriptsize FNAL-E772-90}  &  {\scriptsize 5206}  &  {\scriptsize \cite{Alde:1990im}}     & {\scriptsize 9}   &  {\scriptsize 9}   &  {\scriptsize 7.28}\tabularnewline
{\scriptsize Fe/Be}   & {\scriptsize FNAL-E886-99}  &  {\scriptsize 5201}  &  {\scriptsize \cite{Vasilev:1999fa}}     & {\scriptsize 28}   &  {\scriptsize 28}   &  {\scriptsize 23.09}\tabularnewline
{\scriptsize W/Be}   & {\scriptsize FNAL-E886-99}  &  {\scriptsize 5202}  &  {\scriptsize \cite{Vasilev:1999fa}}     & {\scriptsize 28}   &  {\scriptsize 28}   &  {\scriptsize 23.62}\tabularnewline
\hline
\hline
\textbf{\scriptsize Total:}{\scriptsize {} }  &  &  &  & {\scriptsize 92}  & {\scriptsize 92}  & {\scriptsize 67.81}\tabularnewline
\hline
\end{tabular}
\caption{{ The Drell-Yan process data sets used in the \texttt{nCTEQ15} fit.
We list the same details for each data set as in Tab.~\ref{tab:exp1}.
}}
\label{tab:exp3}
\end{center}
\end{table}
%--------------------

% Pion (RdAu) table
%--------------------
\begin{table}[h!]
\begin{center}
\begin{tabular}{|l|l|c|c|c|c|c|}
\hline
{\scriptsize $\mathbf{R_{dAu}^{\pi}/R_{pp}^{\pi}:}$ }  &  &  &  &  & {\scriptsize $\#$ data }  &\tabularnewline
{\scriptsize Observable }  & {\scriptsize Experiment }  & {\scriptsize ID} & {\scriptsize Ref. }  & {\scriptsize $\#$ data }  & {\scriptsize after cuts }  & {\scriptsize $\chi^2$}\tabularnewline
\hline
\hline
{\scriptsize dAu/pp}   & {\scriptsize PHENIX}  &  {\scriptsize PHENIX}  &  {\scriptsize \cite{Adler:2006wg}}     & {\scriptsize 21}   &  {\scriptsize 20}   &  {\scriptsize 6.63}\tabularnewline
                     & {\scriptsize STAR-2010}  &  {\scriptsize STAR}  &  {\scriptsize \cite{Abelev:2009hx}}     & {\scriptsize 13}   &  {\scriptsize 12}   &  {\scriptsize 1.41}\tabularnewline
\hline
\hline
\textbf{\scriptsize Total:}{\scriptsize {} }  &  &  &  & {\scriptsize 34}  & {\scriptsize 32}  & {\scriptsize 8.04}\tabularnewline
\hline
\end{tabular}
\caption{{ The pion production data sets used in the \texttt{nCTEQ15} fit.
We list the same details for each data set as in Tab.~\ref{tab:exp1}.
}}
\label{tab:exp4}
\end{center}
\end{table}
%--------------------
%
In the current analysis we use  deep inelastic scattering data (DIS), Drell-Yan lepton pair production data (DY)
and inclusive pion production data from RHIC (for nuclei with $A>2$).
The details of particular experiments such as the number of data points, measured observables, etc. are summarized in 
Tables~\ref{tab:exp1}-\ref{tab:exp4}.

The reason to include data from different processes is that each process helps constrain different 
combinations of parton distributions. The bulk of our data are from DIS which help
pin down the valence and sea distributions, however they are not very sensitive to different quark
flavors and gluons. The DY data can be used to differentiate between $u$ and $d$ quark flavors,
and the inclusive pion data have a potential to better constrain the gluon distribution.%
\footnote{Note that the inclusive pion production observable is different in the sense that 
it has an additional dependence on a fragmentation function.}

We introduce kinematic cuts on the included data which limit possible effects of higher 
twist contributions and target mass corrections and at the same time are compatible with the kinematic
cuts used in the underlying free proton analysis. The cuts used in this analysis are:
\begin{itemize}
\item DIS: $Q>2$ GeV and $W>3.5$ GeV,
\item DY: $2<M<300$ GeV,\\
(where $M$ is the invariant mass of the produced lepton pair)  
\item  $\pi^0$ production: $p_T>1.7$ GeV.
\end{itemize}
%
%----------------
\begin{figure}[t]
\centering{}
\includegraphics[width=0.47\textwidth,bb=7 11 521 408]{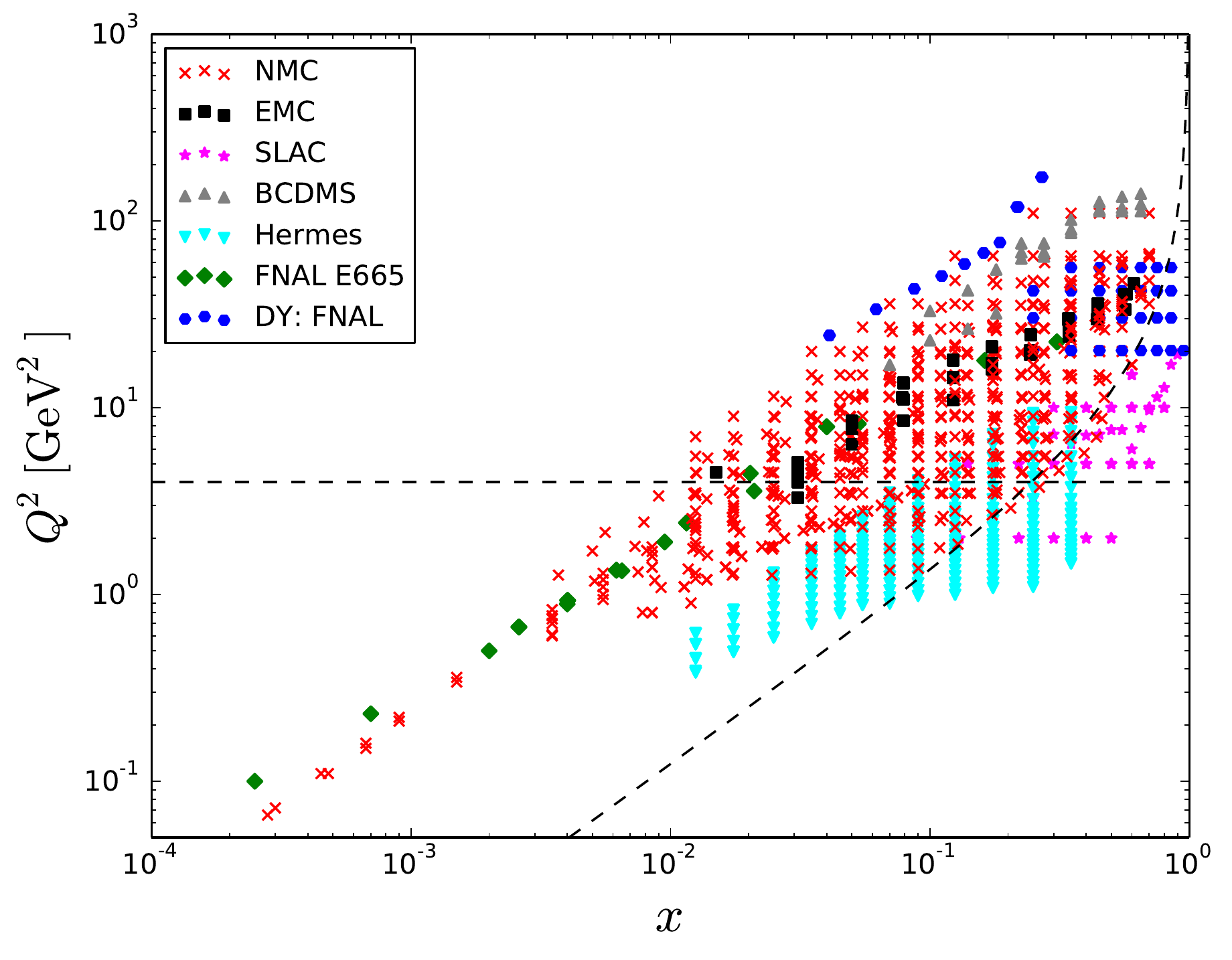}
\caption{Kinematic reach of DIS and DY data used in the presented \ncteq\ fits.
The dashed lines represent the kinematic cuts employed in this analysis ($Q>2$ GeV, $W>3.5$ GeV).
Only the data points lying above both of these lines are included in the fits.}
\label{fig:kinem}
\end{figure}
%----------------
%----------------
\begin{figure}[t]
\centering{}
\includegraphics[width=0.47\textwidth,bb=22 2 569 416]{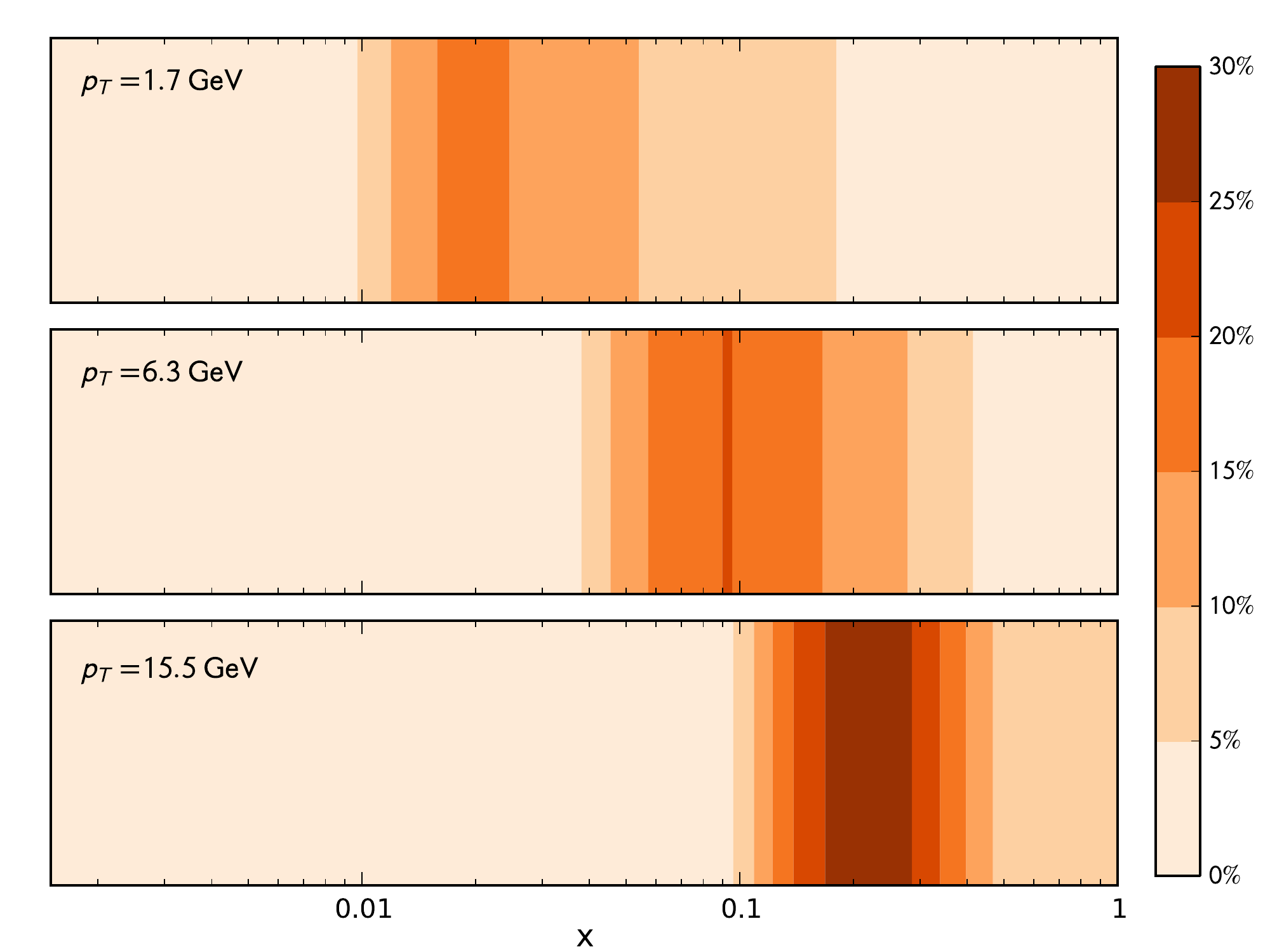}
\caption{Approximate $x$-range for the pion data with the Binnewies-Kniehl-Kramer fragmentation function.}
\label{fig:pionXval}
\end{figure}
%----------------

After the cuts are applied, 740 data points remain, including 616 DIS, 92 DY and
32 pion production data points. 

Note that the overall number of data points we use is considerably smaller compared to the number of data fitted
by other groups (e.g., EPS~\cite{Eskola:2009uj} has 929 data points).
One reason is that the other analyses employ less stringent kinematic cuts on $Q^2$:
\begin{itemize}
\item EPS~\cite{Eskola:2009uj}: $Q>1.3$ GeV,
\item HKN~\cite{Hirai:2007sx}: $Q>1$ GeV,
\item DSSZ~\cite{deFlorian:2011fp}:  $Q>1$ GeV.
\end{itemize}
In addition, none of the analyses mentioned above employ a cut on $W$. 
Whereas the looser cuts allow one to use more
data in the fit, there are possible disadvantages connected to this choice. 
In particular, if one adopts loose cuts, one runs into the danger that the contributions from the
target mass effects or higher twist effects can get enhanced. Especially the latter effects
may be more important in the nuclear case due to 
the higher density of spectator partons in the nucleus~\cite{Qiu:2003cg,Accardi:2004be} and so their effect can be
easily underestimated.
However, the effect of higher twist and target mass corrections 
have been shown to be weakened in ratios of observables \cite{Arrington:2003nt,Malace:2014uea}.

The kinematic reach of the DIS and DY data sets used in our fit is summarized in Fig.~\ref{fig:kinem}, where individual 
experimental points are shown in the $(x,Q^2)$ plane. Note that the two dashed lines indicate the kinematic cuts; 
points lying below these lines are excluded from our analysis.

In Fig.~\ref{fig:pionXval} we estimate the kinematic impact of the pion data by plotting
the cross section for inclusive pion production before convoluting it with gold PDFs, see Eq.~\eqref{eq:pion}. 
Fig.~\ref{fig:pionXval} shows the normalized cross-section as a function of the Bjorken-$x$ of a parton inside a nucleon of 
a gold atom. This is only an estimate which uses the leading order (LO) prediction and it also depends on the
fragmentation function (FF) that is used. Nevertheless, it is useful and allows us to
see that the $x$-values probed by the pion data depend quite substantially on the $p_T$.
In particular, for higher $p_T$, higher $x$ values are probed, e.g. for
$p_T\sim 15$ GeV we are mostly sensitive to $x\in(0.2,0.3)$, whereas for lower $p_T$
the probed $x$ values are more diffused, e.g. for $p_T\sim2$ GeV $x\in(0.01,0.04)$.

One should mention that there are still experimental data that could have been included in our analysis 
but we have decided for different reasons to exclude them from the current work.
We comment briefly on the two most important examples.

First, there are neutrino DIS data from CDHSW \cite{Berge:1989hr}, CHORUS \cite{Onengut:2005kv},
and in particular from the NuTeV collaboration~\cite{Tzanov:2005kr}.
Since they include a considerable number of data points and probe more
flavor combinations than the charged lepton data, they can be used to differentiate individual flavors.
However, tensions between the inclusive charged current $\nu A$ DIS data from NuTeV and the neutral current $\ell^\pm A$ data 
found in \cite{Schienbein:2007fs,Schienbein:2009kk,Kovarik:2010uv} indicate that some additional effort is required to understand
how these discrepancies can be resolved so that all data could be used in one fit simultaneously.
Since these discrepancies appear only if one takes into account the full information contained in the correlated error matrix, 
neglecting these correlations makes it possible to combine $\nu A$ and $\ell^\pm A$ DIS in one fit 
\cite{Paukkunen:2010hb,Paukkunen:2013grz,deFlorian:2011fp}. 
We plan to revisit the neutrino data in a future publication but decided not to include them
in our present PDF release.

Another important set of data which could be included are the already available LHC data. 
In particular the cleanest probe of nuclear effects at the LHC comes from the vector boson, $W^{\pm}$, $Z$,
production~\cite{Aad:2014bha,ATLAS-CONF-2014-020,Chatrchyan:2014csa,Khachatryan:2015hha}.
Results on asymmetries in pPb collisions~\cite{Khachatryan:2015hha} in particular
have a potential to provide valuable input for  nuclear PDF analyses.
These data are not included in the current release as we first want to provide
a baseline analysis without any LHC data. 

%!TEX root=paper.tex
%
%%%%%%%%%%%%%%%%%%%%%%%%%%%%%%%%%%%%%%%%%%%%%%%%%%%%%%%%%%%%%%%%%%%%%%%%%%%%%%%%%%%%%%%%%%%%%%%%%%%%%%%
\begin{table*}[t!]
\begin{center}
\begin{tabular}{|c|c|c|c|c|c|c|c|c|c|c|c|}
\hline
Par.   & Value & Par. & Value & Par. & Value & Par. & Value & Par. & Value & Par. & Value\\
\hline
\hline
$M^g$       &  (0.382)        & $M^{u_v}$       & (0.327)      & $M^{d_v}$       & (0.136)      & $M^{\bar{d}+\bar{u}}$       & (0.129)       & $M^{s+\bar{s}}$       & (0.026)  & $M^{\bar{d}/\bar{u}}$       & (0.000)\\
\hline                                                                                                                                                                    
$c_{0,0}^g$ &  (0.000)       &  --             &   --          &   --            &   --          &   --                        &  --           & $c_{0,0}^{s+\bar{s}}$ & (0.500) &   --                        &  --   \\
$c_{1,0}^g$ &  (0.523)       & $c_{1,0}^{u_v}$ &  (0.630)       & $c_{1,0}^{d_v}$ &  (0.513)       & $c_{1,0}^{\bar{d}+\bar{u}}$ & (-0.324)        & $c_{1,0}^{s+\bar{s}}$ & (-0.324) & $c_{1,0}^{\bar{d}/\bar{u}}$ & (10.075)\\
$c_{2,0}^g$ &  (3.034)       & $c_{2,0}^{u_v}$ &  (2.934)       & $c_{2,0}^{d_v}$ &  (4.211)       & $c_{2,0}^{\bar{d}+\bar{u}}$ & (8.116)        & $c_{2,0}^{s+\bar{s}}$ & (8.116) & $c_{2,0}^{\bar{d}/\bar{u}}$ & (4.957)\\
$c_{3,0}^g$ &  (4.394)       & $c_{3,0}^{u_v}$ &  (-2.369)       & $c_{3,0}^{d_v}$ &  (-2.375)       & $c_{3,0}^{\bar{d}+\bar{u}}$ & (0.413)        & $c_{3,0}^{s+\bar{s}}$ & (0.413) & $c_{3,0}^{\bar{d}/\bar{u}}$ & (15.167)\\
$c_{4,0}^g$ &  (2.359)       & $c_{4,0}^{u_v}$ &  (1.266)       & $c_{4,0}^{d_v}$ &  (0.965)       & $c_{4,0}^{\bar{d}+\bar{u}}$ & (4.754)        & $c_{4,0}^{s+\bar{s}}$ & (4.754) & $c_{4,0}^{\bar{d}/\bar{u}}$ & (17.000)\\
$c_{5,0}^g$ &  (-3.000)       & $c_{5,0}^{u_v}$ &  (1.718)       & $c_{5,0}^{d_v}$ &  (3.000)       & $c_{5,0}^{\bar{d}+\bar{u}}$ & (0.614)        & $c_{5,0}^{s+\bar{s}}$ & (0.614) & $c_{5,0}^{\bar{d}/\bar{u}}$ & (9.948)\\
\hline
\hline
Par.   & Value & Par. & Value & Par. & Value & Par. & Value & Par. & Value & Par. & Value\\
\hline
\hline
$c_{0,1}^g$ &        (-0.256) &  --             &   --          &    --           &   --          &      --                     &   --          & $c_{0,1}^{s+\bar{s}}$ & (0.167) &     --                      &  --   \\
$c_{1,1}^g$ & \textbf{-0.001} & $c_{1,1}^{u_v}$ & \textbf{-2.729} & $c_{1,1}^{d_v}$ & \textbf{0.272} & $c_{1,1}^{\bar{d}+\bar{u}}$ & \textbf{0.411} & $c_{1,1}^{s+\bar{s}}$ & (0.411) & $c_{1,1}^{\bar{d}/\bar{u}}$ & (0.000)\\
$c_{2,1}^g$ &        (0.000) & $c_{2,1}^{u_v}$ & \textbf{-0.162} & $c_{2,1}^{d_v}$ & \textbf{-0.198} & $c_{2,1}^{\bar{d}+\bar{u}}$ &        (0.415) & $c_{2,1}^{s+\bar{s}}$ & (0.415) & $c_{2,1}^{\bar{d}/\bar{u}}$ & (0.000)\\
$c_{3,1}^g$ &        (0.383) & $c_{3,1}^{u_v}$ &        (0.018) & $c_{3,1}^{d_v}$ &        (0.085) & $c_{3,1}^{\bar{d}+\bar{u}}$ &        (-0.759) & $c_{3,1}^{s+\bar{s}}$ & (0.000) & $c_{3,1}^{\bar{d}/\bar{u}}$ & (0.000)\\
$c_{4,1}^g$ & \textbf{0.055} & $c_{4,1}^{u_v}$ & \textbf{12.176} & $c_{4,1}^{d_v}$ &        (3.874) & $c_{4,1}^{\bar{d}+\bar{u}}$ &        (-0.203) & $c_{4,1}^{s+\bar{s}}$ & (0.000) & $c_{4,1}^{\bar{d}/\bar{u}}$ & (0.000)\\
$c_{5,1}^g$ & \textbf{0.002} & $c_{5,1}^{u_v}$ & \textbf{-1.141} & $c_{5,1}^{d_v}$ & \textbf{-0.072} & $c_{5,1}^{\bar{d}+\bar{u}}$ & \textbf{-0.087} & $c_{5,1}^{s+\bar{s}}$ & (0.000) & $c_{5,1}^{\bar{d}/\bar{u}}$ & (0.000)\\
\hline
\hline
Par.   & Value & Par. & Value & Par. & Value & Par. & Value & Par. & Value & Par. & Value\\
\hline
\hline
$c_{0,2}^g$ & \textbf{-0.037} &     --          &         --   &     --          &         --     &      --                     &        --    & $c_{0,2}^{s+\bar{s}}$  & (0.104) &      --                     &    --  \\
$c_{1,2}^g$ & \textbf{-1.337} & $c_{1,2}^{u_v}$ &       (0.006) & $c_{1,2}^{d_v}$ &       (0.466)   & $c_{1,2}^{\bar{d}+\bar{u}}$ &       (0.172) & $c_{1,2}^{s+\bar{s}}$  & (0.172) & $c_{1,2}^{\bar{d}/\bar{u}}$ &  (0.000)\\
$c_{2,2}^g$ &        (0.000) & $c_{2,2}^{u_v}$ &       (0.524) & $c_{2,2}^{d_v}$ &       (0.440)   & $c_{2,2}^{\bar{d}+\bar{u}}$ &       (0.290) & $c_{2,2}^{s+\bar{s}}$  & (0.290) & $c_{2,2}^{\bar{d}/\bar{u}}$ &  (0.000)\\
$c_{3,2}^g$ &        (0.520) & $c_{3,2}^{u_v}$ &       (0.073) & $c_{3,2}^{d_v}$ &       (0.107)   & $c_{3,2}^{\bar{d}+\bar{u}}$ &       (0.298) & $c_{3,2}^{s+\bar{s}}$  & (0.000) & $c_{3,2}^{\bar{d}/\bar{u}}$ &  (0.000)\\
$c_{4,2}^g$ & \textbf{-0.514} & $c_{4,2}^{u_v}$ &       (0.038) & $c_{4,2}^{d_v}$ &       (-0.018)   & $c_{4,2}^{\bar{d}+\bar{u}}$ &       (0.888) & $c_{4,2}^{s+\bar{s}}$  & (0.000) & $c_{4,2}^{\bar{d}/\bar{u}}$ &  (0.000)\\
$c_{5,2}^g$ & \textbf{-1.417} & $c_{5,2}^{u_v}$ &       (0.615) & $c_{5,2}^{d_v}$ &       (-0.236)   & $c_{5,2}^{\bar{d}+\bar{u}}$ &       (1.353) & $c_{5,2}^{s+\bar{s}}$  & (0.000) & $c_{5,2}^{\bar{d}/\bar{u}}$ &  (0.000)\\
\hline
\end{tabular}
\end{center}
\caption{Values of the parameters of the \ncteqfit\ fit
at the initial scale $Q_0=1.3$~GeV. 
Values in bold represent the free parameters and values in parentheses are fixed in the fit. 
The not listed normalization parameters
are determined by the momentum and number sum rules as discussed in the text.
For completeness, we provide the full set of the free proton parameters $c_{k,0}$ (first set of rows).
The $M^i$ parameters (first row) show the (fixed) momentum fraction carried by different flavors
in the case of a free proton.
}
\label{tab:params_nCTEQ15g}
\end{table*}
%%%%%%%%%%%%%%%%%%%%%%%%%%%%%%%%%%%%%%%%%%%%%%%%%%%%%%%%%%%%%%%%%%%%%%%%%%%%%%%%%%%%%%%%%%%%%%
%
%+++++++++++++++++++++++++++++++++++++++++++++++++++++++
\section{Results}
\label{sec:res}
%+++++++++++++++++++++++++++++++++++++++++++++++++++++++

A key result of the current \ncteqfit\  fit compared to the previous
\ncteq\ releases~\cite{Schienbein:2007fs,Schienbein:2009kk} is the inclusion
of PDF uncertainties using  the Hessian method, cf.\ Sec.~\ref{sec:hessian}.

The second significant addition is the inclusion of a new type of experimental data,
namely the pion production data from the PHENIX and STAR collaborations.
Since these data have the potential to provide information on the gluon distribution
(which otherwise is weakly constrained) it is important to precisely estimate
their impact on the resulting PDFs. For this purpose the \ncteqfit\ fit will be compared with
a reference fit \ncteqnp\ which is identical except it does {\em not} include the pion data.

The full set of data we consider is listed in 
\hbox{Tables~\ref{tab:exp1}---\ref{tab:exp4}.}
Note that we have included QED radiative corrections for the DIS FNAL-E665-95 (Pb/D, Ca/D, C/D)
data sets and this significantly improves the description of these data.%
    \footnote{For example, the $\chi^2$ for the FNAL-E665-95 Pb/D data (ID~5129) 
    is reduced from $5.91$ to $1.20$ (for 3 data points) when the QED radiative corrections
    are included.}
In the following,  we discuss the results of this \ncteqfit\ analysis
and compare it with other available sets of nuclear PDFs.

%%%%%%%%%%%%%%%%%%%%%%%%%%%%%
\subsection{The \ncteqfit\ Fit}
%%%%%%%%%%%%%%%%%%%%%%%%%%%%%
%

\subsubsection{PDF Parameterization}

The PDFs in our fit are parameterized at the input scale $Q_0=1.3$ GeV according to 
Eqs.~\eqref{eq:param} and~\eqref{eq:Adep}. 
%
%
%--------------
\begin{figure*}[th]
\begin{center}
\subfloat[Gluon]
{
\includegraphics[width=0.36\textwidth]{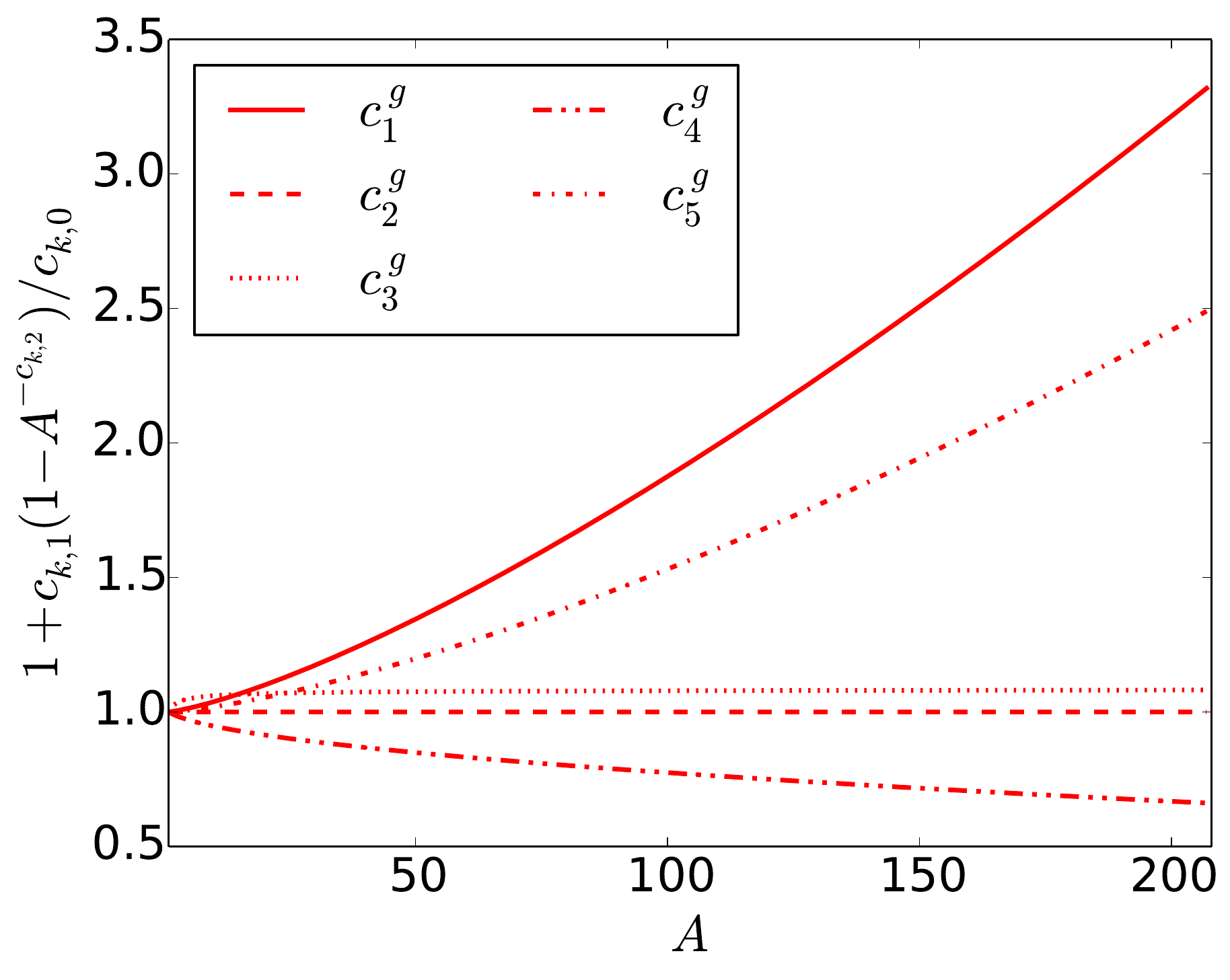}
}
\qquad\qquad
\subfloat[$\bar{d}+\bar{u}$]
{
\includegraphics[width=0.36\textwidth]{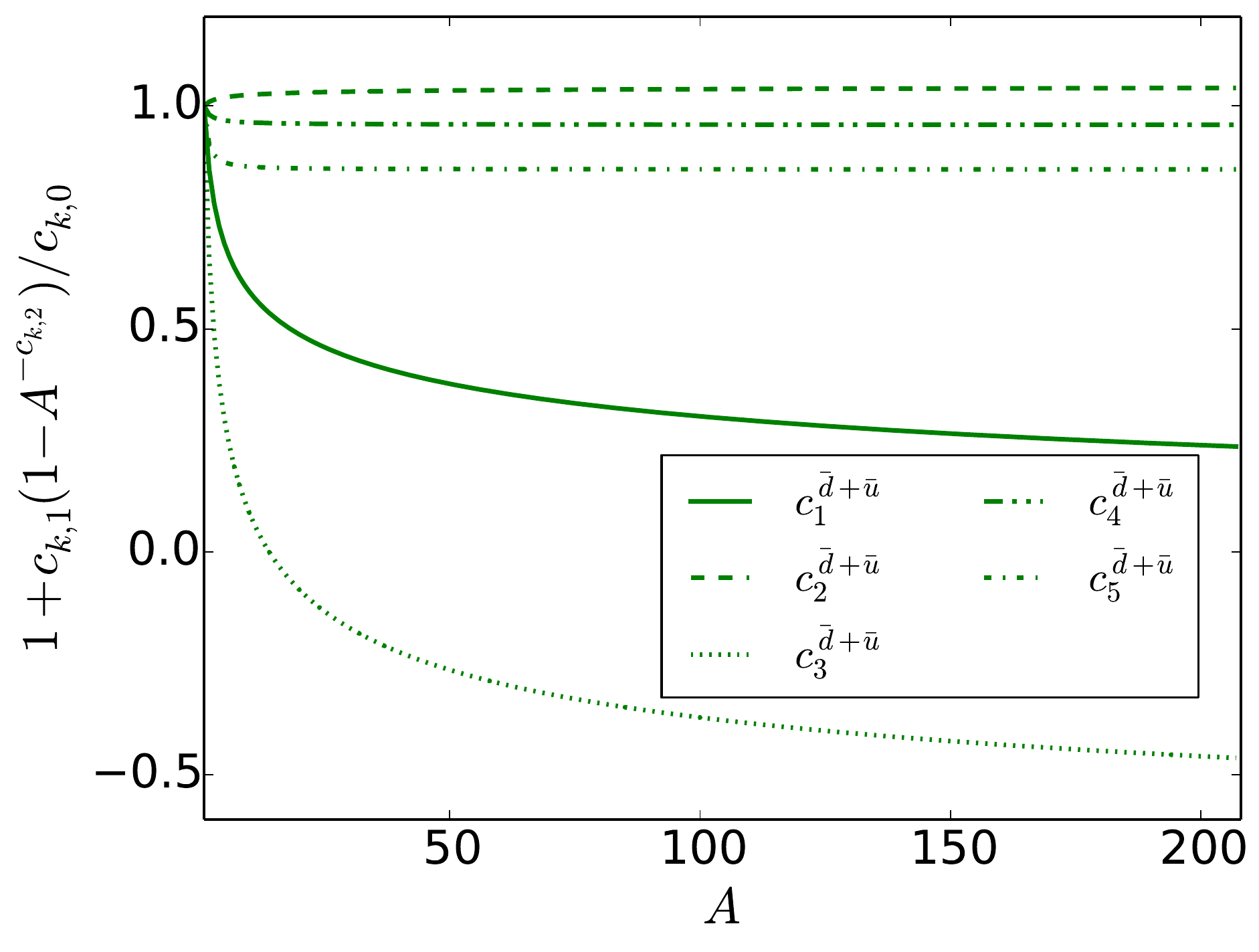}
}
\\
\subfloat[$u$-valence]
{
\includegraphics[width=0.36\textwidth]{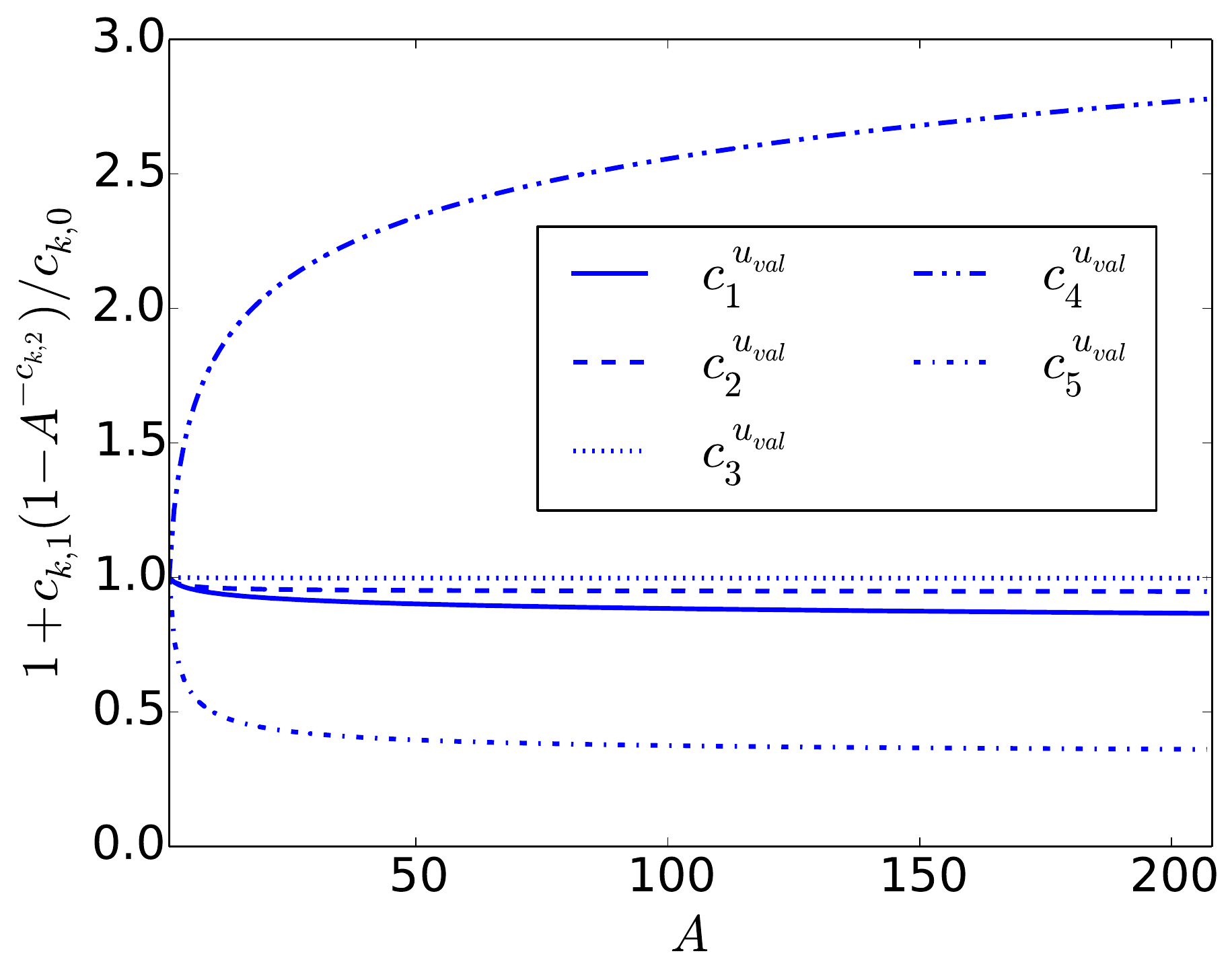}
}
\qquad\qquad
\subfloat[$d$-valence]
{
\includegraphics[width=0.36\textwidth]{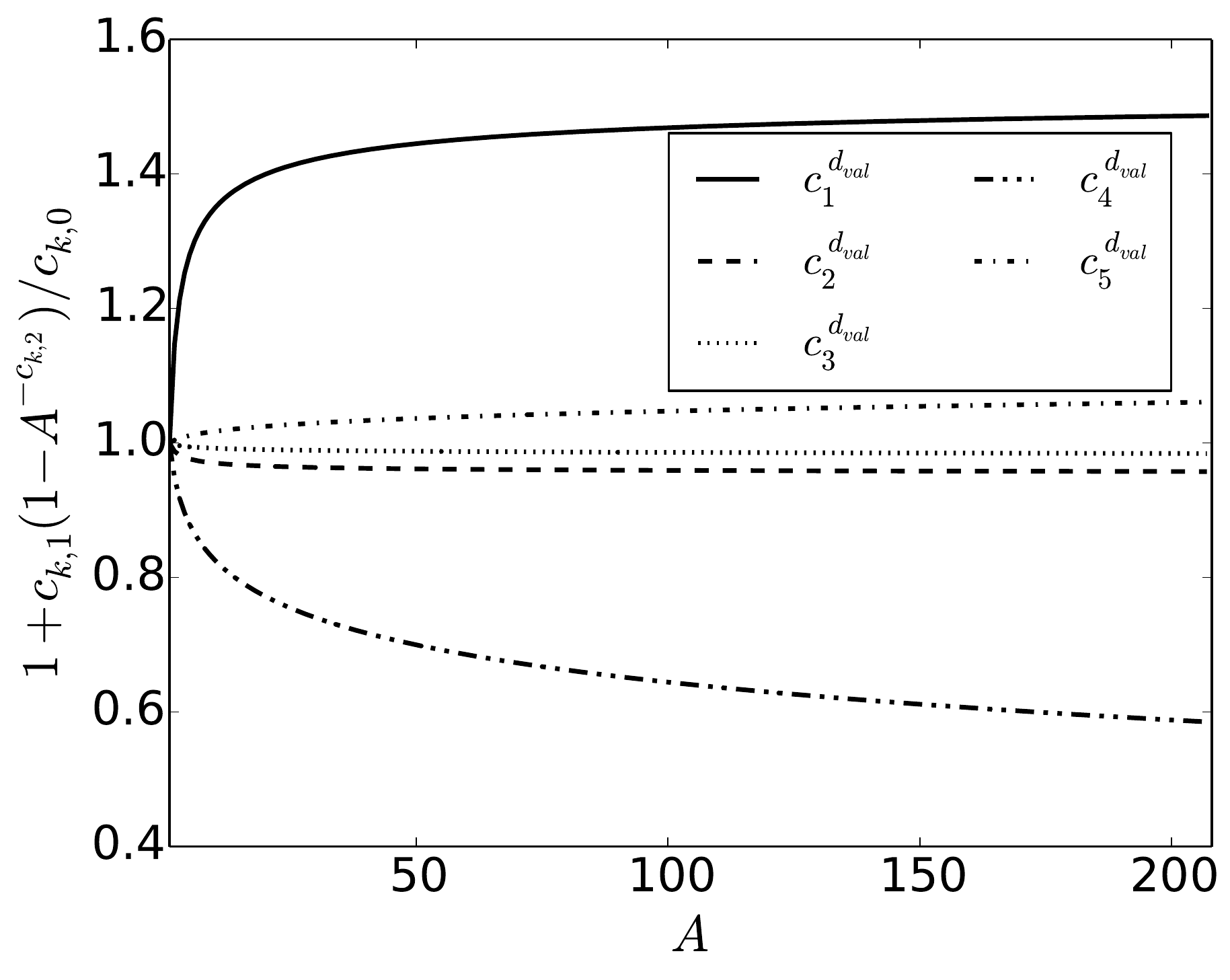}
}
\caption{
$A$-dependence of the fit parameters as given in Eq.~\eqref{eq:Adep}. 
Specifically, we plot 
$c_k(A)=c_{k,0} + c_{k,1} (1-A^{-c_{k,2}})$
for each flavor normalized to the corresponding free proton parameter $c_{k,0}$.
The superscripts \{1, 2, ...\} in the legend correspond to the parameters \{$c_1$, $c_2$, ... \} in
Eq.~\eqref{eq:param}. 
}
\label{fig:param-Adep}
\end{center}
\end{figure*}
%--------------
%
This provides considerable flexibility as
each of the seven flavor combinations can have $\sim$10  free parameters
to describe the $x$ and $A$ 
dependence.\footnote{
For each of the 5 flavor combinations $\{u_v, d_v, g, \bar{u}+\bar{d},s=\bar{s}\}$
of  Eq.~\protect\eqref{eq:param}
we have 10 parameters  $\{c_{k,1}, c_{k,2}\}$ for $k=\{1...5\}$ 
in addition to the  normalization parameters $c_{0}$
that are partly fixed by the number and momentum sum rules.
For $\{\bar{d}/\bar{u}\}$, we have 8 parameters at our disposal.}
However, the  available experimental data are not sufficient to 
constrain such a flexible parameterization.
Therefore, 
we limit our  actual fit to 16 parameters; 
specifically, we include  7 gluon, 4 $u$-valence,
3 $d$-valence and 2 $\bar{d}+\bar{u}$ free parameters. 
The details of the fit are summarized  in Table~\ref{tab:params_nCTEQ15g}
which shows the best fit values of the free  parameters, as well as the values of the fixed parameters. 

For the pion data, we allow for the normalization to vary and we obtain 1.031 for
the  PHENIX data~\cite{Adler:2006wg} and 0.962 for the STAR data~\cite{Abelev:2009hx}.\footnote{
Note that the data normalization parameters do
not enter the Hessian analysis of uncertainties.
}
Our obtained normalization shifts of $\sim$4\% lie well within the experimental
normalization uncertainty.\footnote{
See Table~1 and Fig.~2 in Ref.~\cite{Adler:2006wg} for PHENIX, and 
Fig.~25 in Ref.~\cite{Abelev:2009hx} and Table~5 in Ref.~\cite{STARdata} for STAR.
}

Our parameterization smoothly interpolates between different nuclei 
as a function of the nuclear mass number $A$;
the number of protons $Z$ and neutrons $(A-Z)$ 
enters only through the isospin composition of a nucleus, 
cf.\ Eq.~\eqref{eq:nucleusPDF}.
Fig.~\ref{fig:param-Adep} shows the $A$ dependence of the fitting parameters normalized by the corresponding values of the free proton baseline parameters $c_{k,0}$.
(Note, some of these parameters are fixed, cf.\ Table~\ref{tab:params_nCTEQ15g}.)
Many of the para\-meters change rapidly 
in the region of light nuclei $A\lesssim 25$ and are relatively 
stable for heavy nuclei $A\gtrsim 50$.
Also, we observe that the parameters responsible for the small $x$ behavior $\{c_1\}$,
typically exhibit a strong $A$ dependence,  whereas the large $x$ parameters $\{c_2\}$ 
are comparably insensitive to the type of nucleus.
In particular, the biggest effect occurs for the gluon where the $c_1^g$ parameter
describing the low $x$ gluon PDF and $c_5^g$ parameter (responsible for mid-$x$)
are changing linearly throughout the whole range of $A$.

%%%%%%%%%%%%%%%%%%%%%%%%%%%%%%%%%%%%%%%%%%%%%%%%%%%%%%%%%%%%%%%%%%%%%%%
% chi2 per experiment/target
\begin{figure*}[th]
\begin{center}
\subfloat[Value of $\chi^2/$dof for the individual experiments which are
identified by the IDs that are listed in \hbox{Tables~\ref{tab:exp1}---\ref{tab:exp4}.}
The {\tt 51xx} IDs correspond to the DIS experiments, 
the {\tt 52xx} IDs are the DY data, and 
the pion  data are labeled by the collaboration name.
The experiments are sorted left-to-right: \{DIS, DY, $\pi^0$\} and sub-sorted by the nuclear mass number $A$.
]
{
\includegraphics[width=0.8\textwidth]{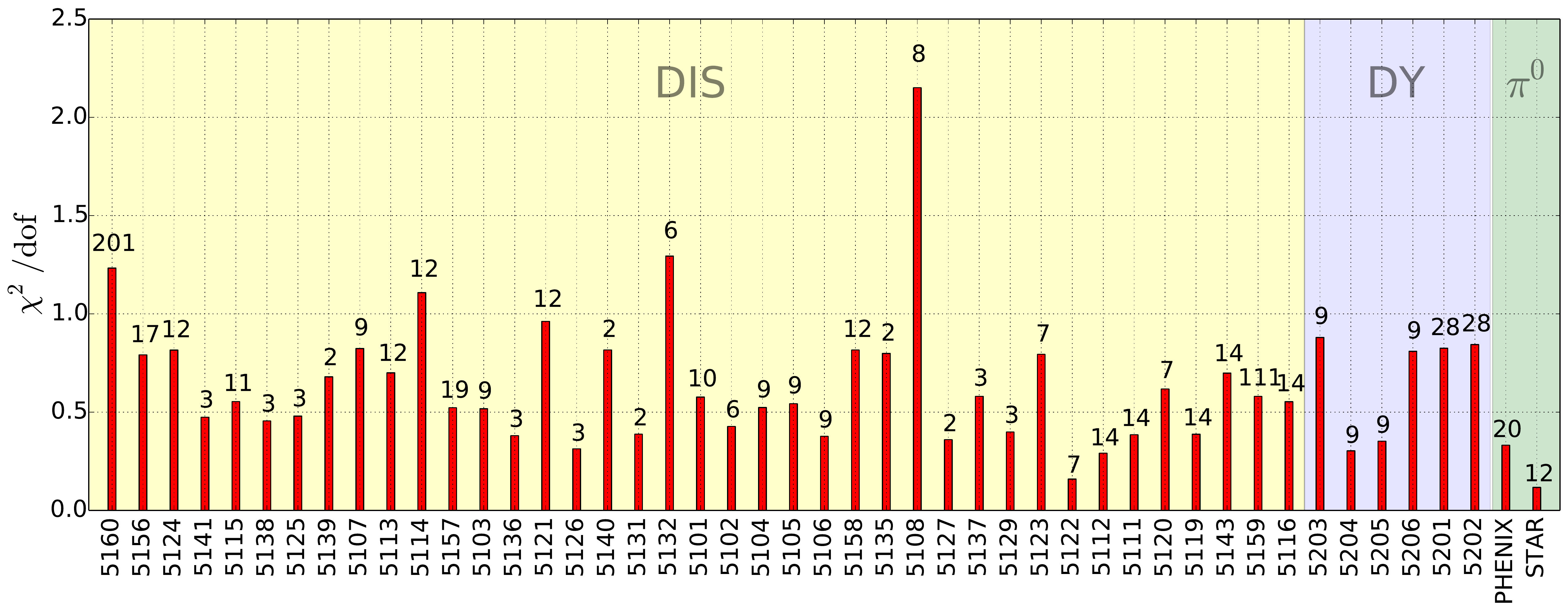}
\label{fig:chi2Dof_PE}
}
\\
\subfloat[Value of $\chi^2/$dof per nuclear target used in the \ncteqfit\ fit
sorted left-to-right by the nuclear mass number $A$.
]
{
\includegraphics[width=0.8\textwidth]{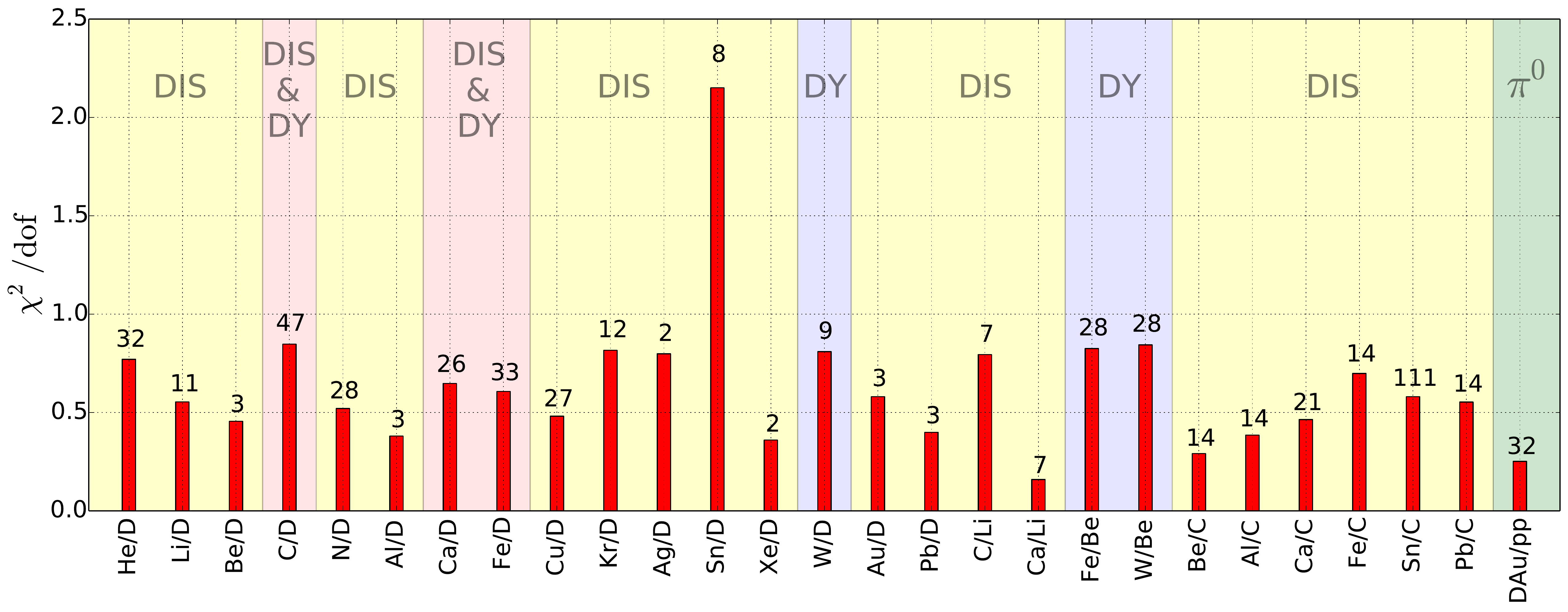}
\label{fig:chi2Dof_PT}
}
\caption{
Value of $\chi^2/{\rm dof}$ for (a) individual experiments and
(b) per nuclear target used in the \ncteqfit\ fit.
The numbers on top of the bars represent the number of data points (after
kinematic cuts).
}
\label{fig:chi2oDof}
\end{center}
\end{figure*}
%%%%%%%%%%%%%%%%%%%%%%%%%%%%%%%%%%%%%%%%%%%%%%%%%%%%%%%%%%%%%%%%%%%%%%%

%%%%%%%%%%%%%%%%%%%%%%%%%%%%%%%%%%%%%%%%%%%%%%%%%%%%%%%%%%%%%%%%%%%%%%%%%%%%%%%%%%%%%%%%%%%%%%
\subsubsection{$\chi^2$ of the fit}
\label{sec:reschi2}

We now examine the overall statistical quality of the fit as measured by the  $\chi^2$.
For the \ncteqfit\ fit we obtain a total $\chi^2$ of 587.4 with 740 data points (after kinematic cuts). 
With 18 free parameters (including 2 data normalization parameters) this leads to a $\chi^2/dof=0.81$
which indicates a good fit. Furthermore, this  $\chi^2/dof$ is not too small which
could indicate deficiencies of the fit such as over-fitting.

To better evaluate the fit quality, 
in Fig.~\ref{fig:chi2Dof_PE} we plot the $\chi^2/dof$ for the individual experiments 
and check that the majority of experiments has a  $(\chi^2/dof)\simeq 1$.
While  most experiments satisfy this ``goodness of fit'' criterion, 
there is one  experiment that stands out as having a poor fit:  
the DIS EMC-88 data for Sn/D (ID~5108).
Several previous global analyses have also found it  challenging to 
accommodate the Sn/D data  \cite{Hirai:2007sx,deFlorian:2011fp}. 

In Fig.~\ref{fig:chi2Dof_PT}, we show again the $\chi^2/dof$, but this time the experiments are grouped by nuclear target
and are sorted by increasing nuclear mass number $A$.
This allows us to see that there are no systematic effects 
associated with our choice of the $A$ parameterization.
With the noted exception of Sn/D, 
all other nuclear targets from helium up to lead
are described very well
with a $\chi^2/dof \simeq 1$.

%%%%%%%%%%%%%%%%%%%%%%%%%%%%%%%%%%%%%%%%%%%%%%%%%%%%%%%%%%%%%%%%%%%%%%%%%%%%%%%%%%%%%%%%%%%%%%
% 1D-scans (rescaled-Hessian)
%----------------
\begin{figure}[th]
\centering{}
\includegraphics[width=0.48\textwidth]{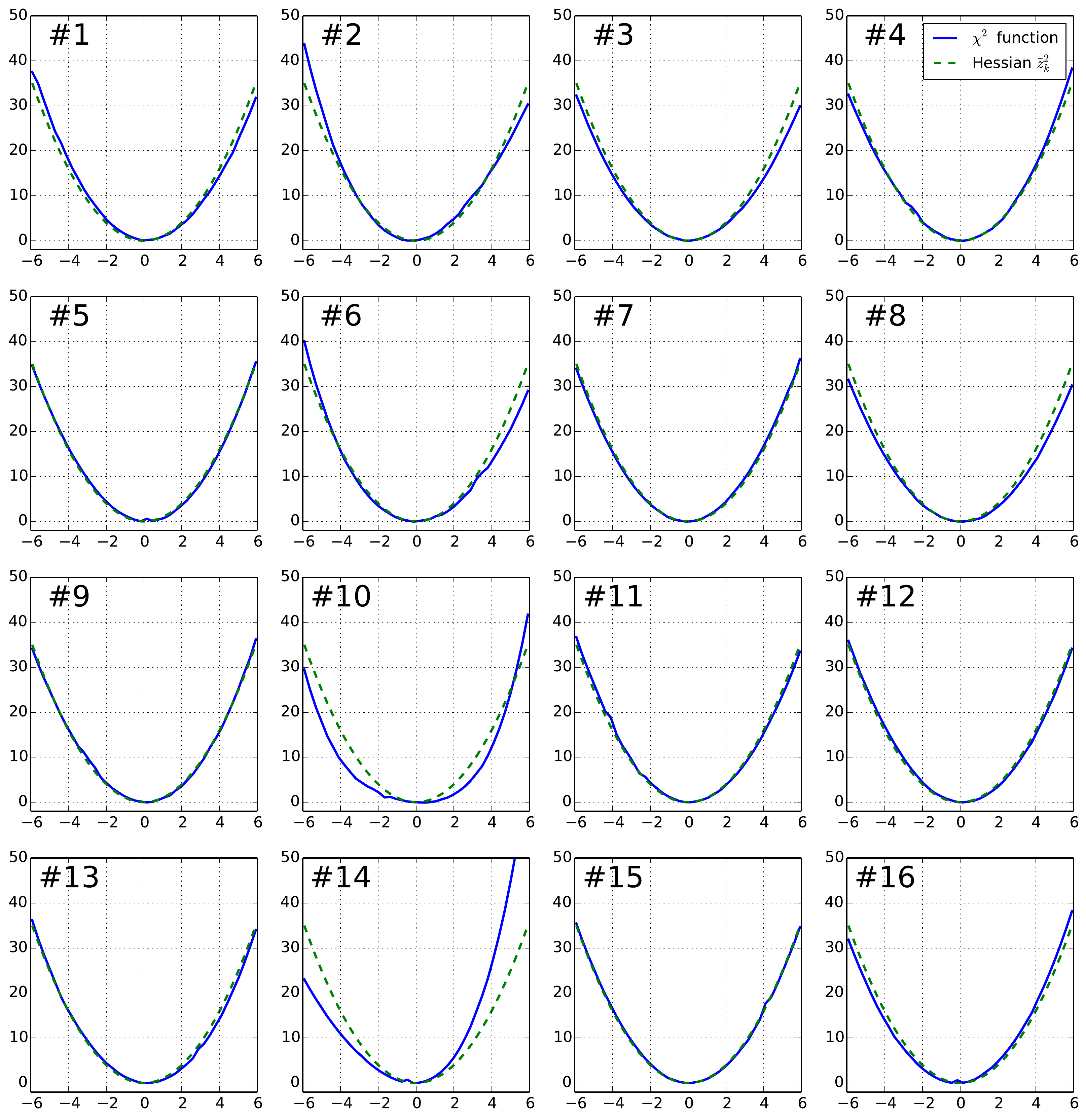}
\caption{
$\chi^2$ function relative to its value at the minimum, $\Delta\chi^2=\chi^2-\chi^2_0$,
plotted along the 16 error directions in the eigenvector space, $\tilde z_i^2$.
We display the true $\chi^2$ function (solid lines) 
and the quadratic approximation given by Hessian method $\Delta\chi^2=\tilde z_i^2$ (dashed lines). 
The eigenvector directions are ordered from the largest to the smallest eigenvalue.
}
\label{fig:1Dscans_compact}
\end{figure}
%----------------
%%%%%%%%%%%%%%%%%%%%%%%%%%%%%%%%%%%%%%%%%%%%%%%%%%%%%%%%%%%%%%%%%%%%%%%%%%%%%%%%%%%%%%%%%%%%%%
%%%%%%%%%%%%%%%%%%%%%%%%%%%%%%%%%%
% Nuclear dependence
%
%----------------
\begin{figure}[th]
\centering{}
\includegraphics[width=0.48\textwidth]{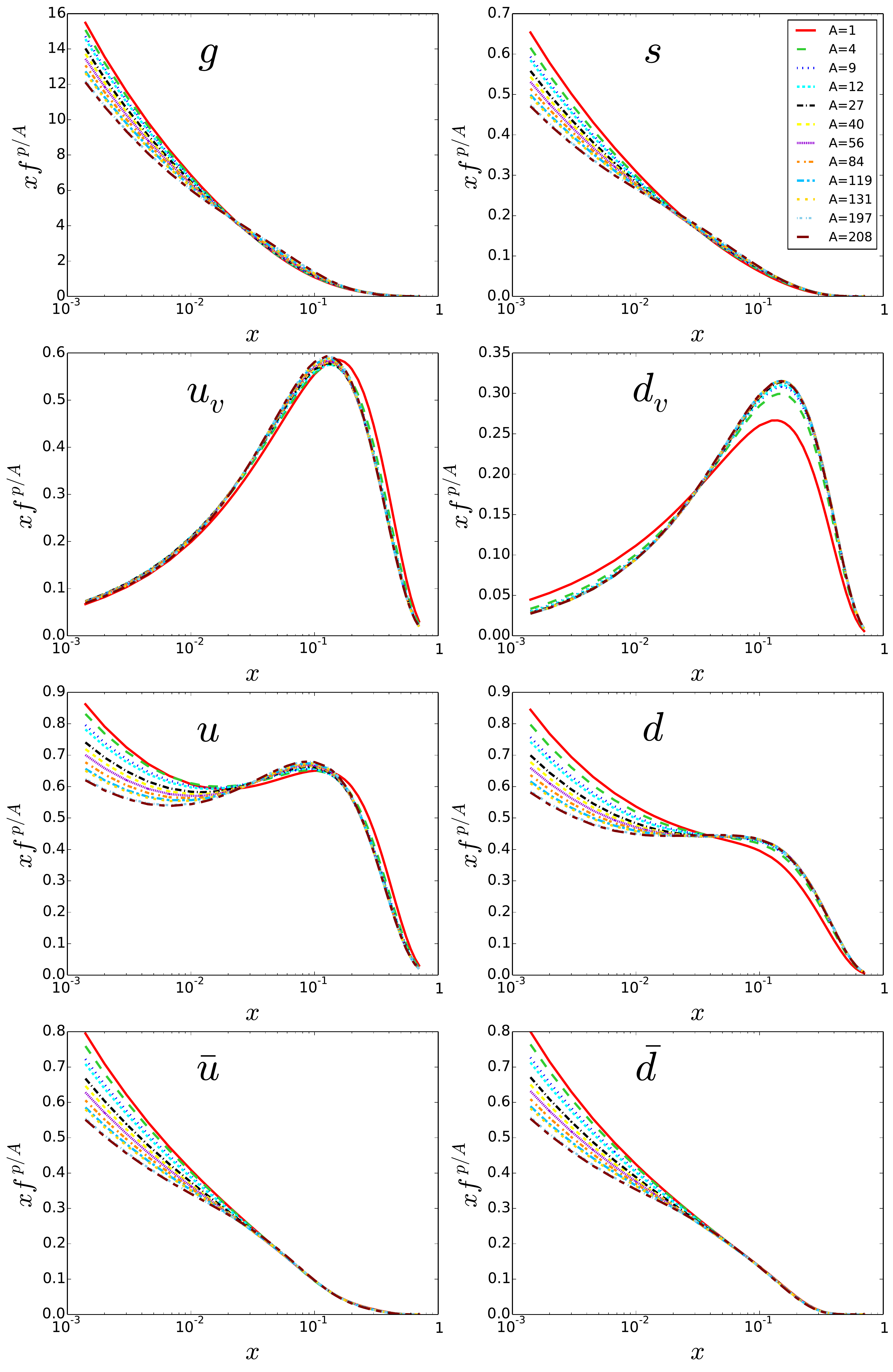}
\caption{\ncteqfit\ bound proton PDFs at the scale $Q=10$ GeV 
for a range of nuclei from the free proton ($A=1$) to lead ($A=208$).
}
\label{fig:compar_PDFs_diff-nuc}
\end{figure}
%----------------

%%%%%%%%%%%%%%%%%%%%%%%%%%%%%%%%%%%%%%%%%%%%%%%%%%%%%%%%%%%%%%%%%%%%%%%%%%%%%%%%%%%%%%%%%%%%%%
\subsubsection{Error PDF reliability.}
Before we examine the actual \ncteqfit\ predictions, 
we first investigate the quality of the Hessian error analysis. 
This will allow us to judge the reliability of our error estimates and,
in turn, the  quality of our predictions.\footnote{
 Note that by construction, the Hessian method can only  probe the local
 minimum connected to the ``best fit'' (central prediction), and  is not sensitive
 to a landscape with multiple minima. Unfortunately, in case of nPDFs fits multiple minima are 
 possible as there is not sufficient data to fully constrain the nPDFs.}
There are two factors that need to be assessed:
\begin{itemize}
	\item[(i)] the quality of the quadratic approximation,
	\item[(ii)] how well the Hessian approximation describes the actual $\chi^2$ function
	in a region around the minimum given by our tolerance criterion, $\Delta\chi^2=35$.
\end{itemize}
To estimate these factors, we plot the $\chi^2$ function
relative to its value at the minimum ($\Delta\chi^2=\chi^2-\chi^2_0$) along the 16 
error directions in the eigenvector space (see Fig.~\ref{fig:1Dscans_compact}).
For comparison, we also display the Hessian approximation given by the quadratic form $\Delta\chi^2=\tilde z_i^2$.
The plots are ordered according to the decreasing values of the eigenvalues corresponding to the $\tilde z_i$ directions; 
the largest eigenvalue is of order $10^9$, and the smallest of order $10$.
For the largest few eigenvalues of Fig.~\ref{fig:1Dscans_compact} 
the quadratic approximation works extremely well; however,  
for the smaller eigenvalues $\{{\it e.g.},  \#10, \#14\}$
it can deviate from the $\chi^2$ function.
Nevertheless, in all the cases we are able to obtain a
good description of the actual $\chi^2$ function for  $\tilde z_i \sim [-6,6]$
which corresponds to our tolerance criterion $\sqrt{\Delta\chi^2}=\sqrt{35}\sim6$.
This analysis verifies that the error PDFs defined using the modified Hessian formalism will
closely reflect the actual $\chi^2$ function determined by the experimental data, and will
not be severely affected by the imperfections of the quadratic approximation that
occurs for directions corresponding to lower 
eigenvalues.\footnote{
In the modified Hessian approach that we use,  the discrepancies at $\Delta\chi^2=35$ originate mostly
    from the non-symmetric behavior of the $\chi^2$ function; see Sec.~\ref{sec:hessian}
    and Appendix~\ref{app:Hess} for details.}
%
%--------------
% nCTEQ15g
%----------------
\begin{figure*}[th]
\centering{}
\includegraphics[clip,width=0.48\textwidth]{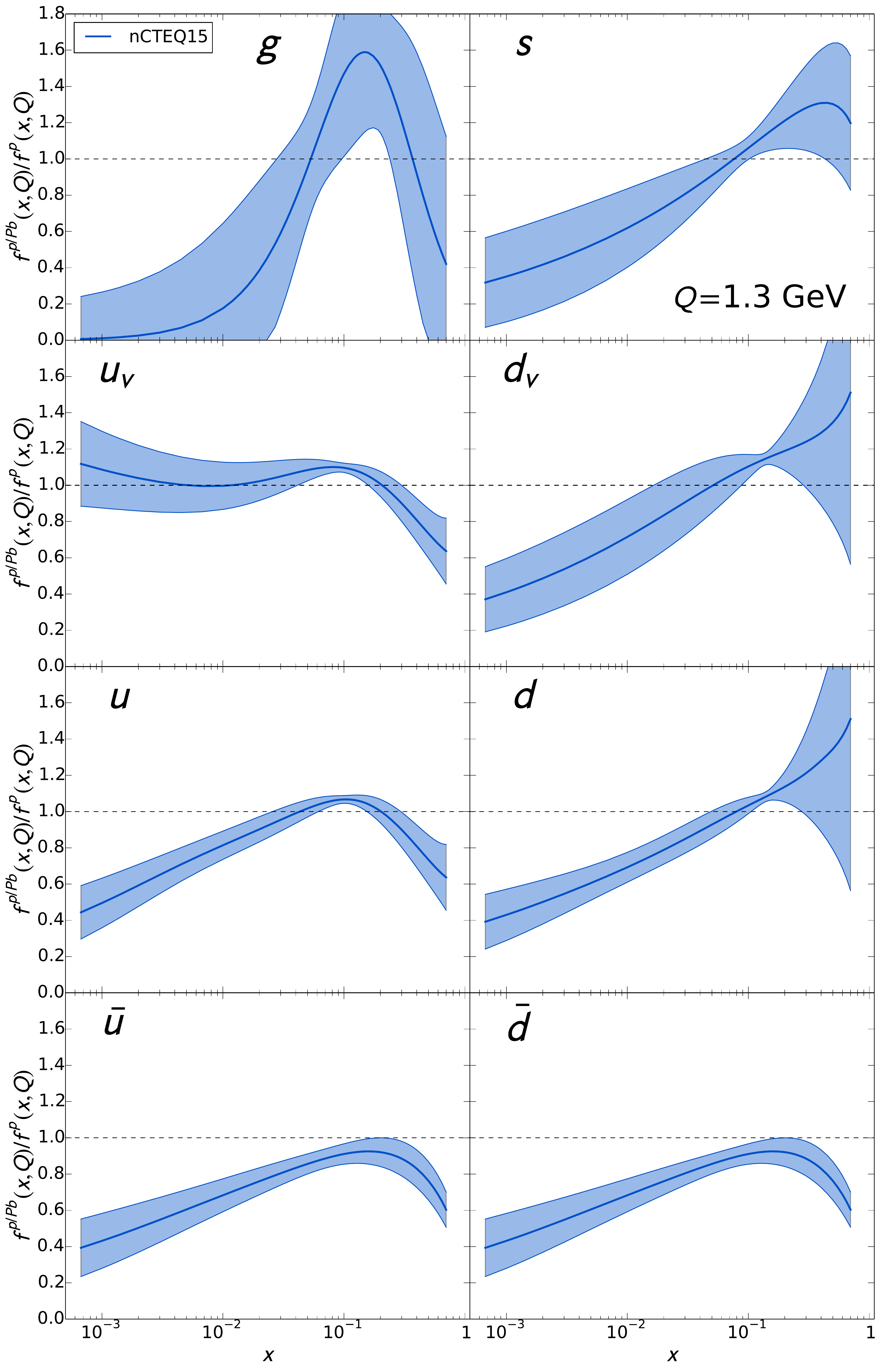}
\quad{}
\includegraphics[width=0.49\textwidth]{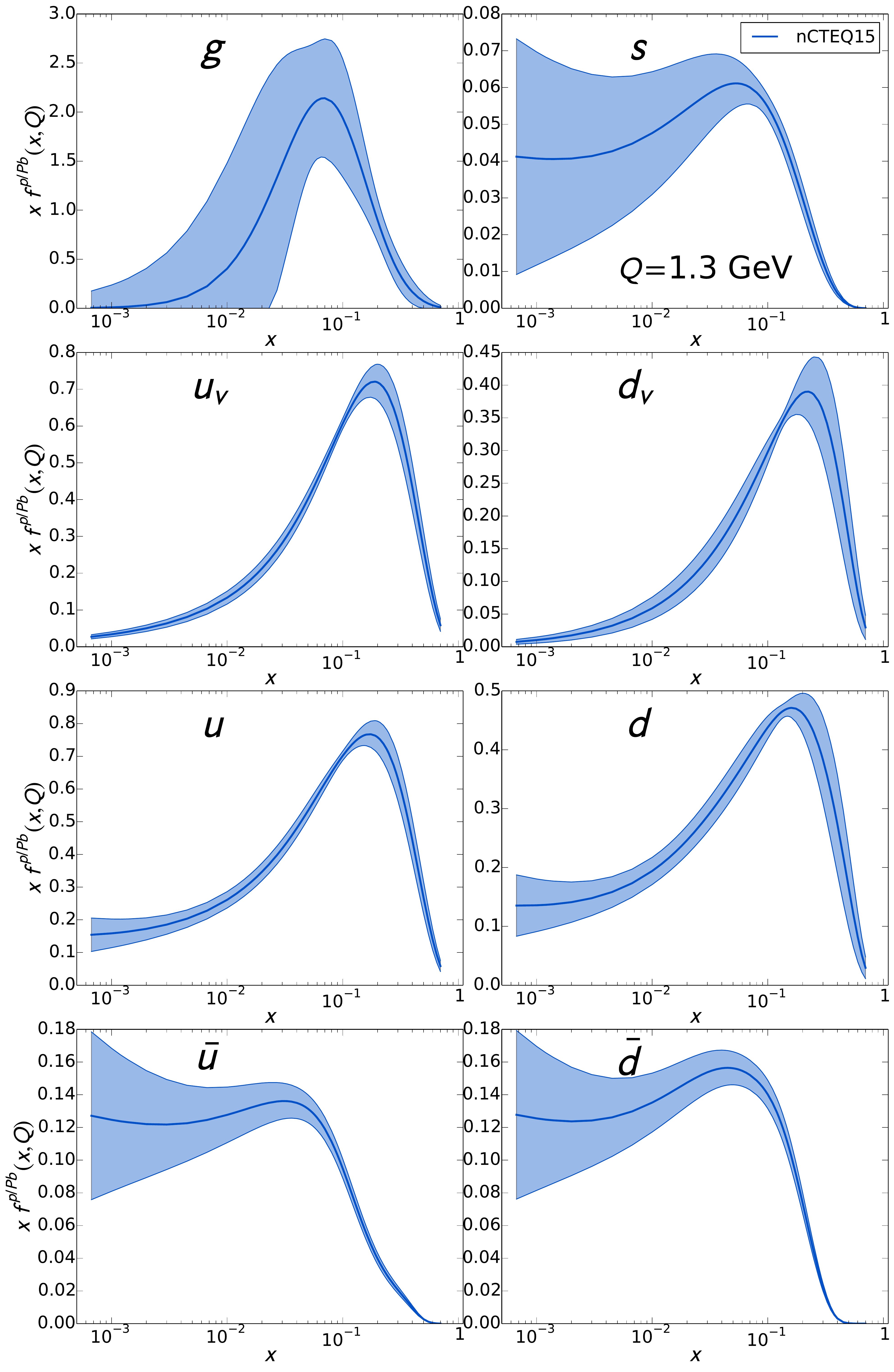}
\caption{Results of the \ncteqfit\ fit.
On the left we show nuclear modification factors defined as ratios of proton PDFs
bound in lead to the corresponding free proton PDFs, and on the right we show the
actual bound proton PDFs for lead. In both cases the scale is equal to $Q=1.3$ GeV.}
\label{fig:nCTEQ15g_1}
\end{figure*}
%----------------
%----------------
\begin{figure*}[th]
\centering{}
\includegraphics[clip,width=0.48\textwidth]{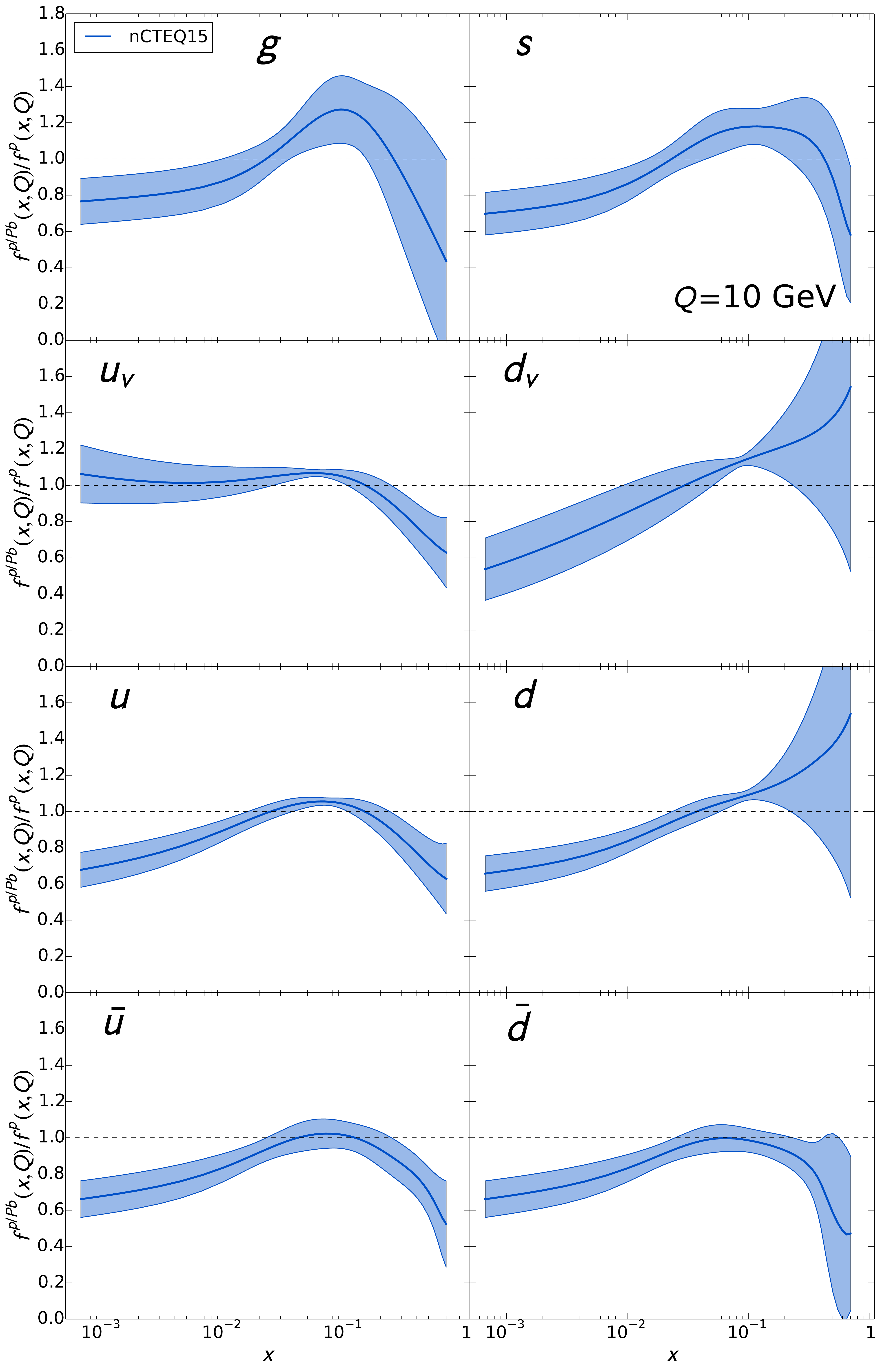}
\quad{}
\includegraphics[width=0.48\textwidth]{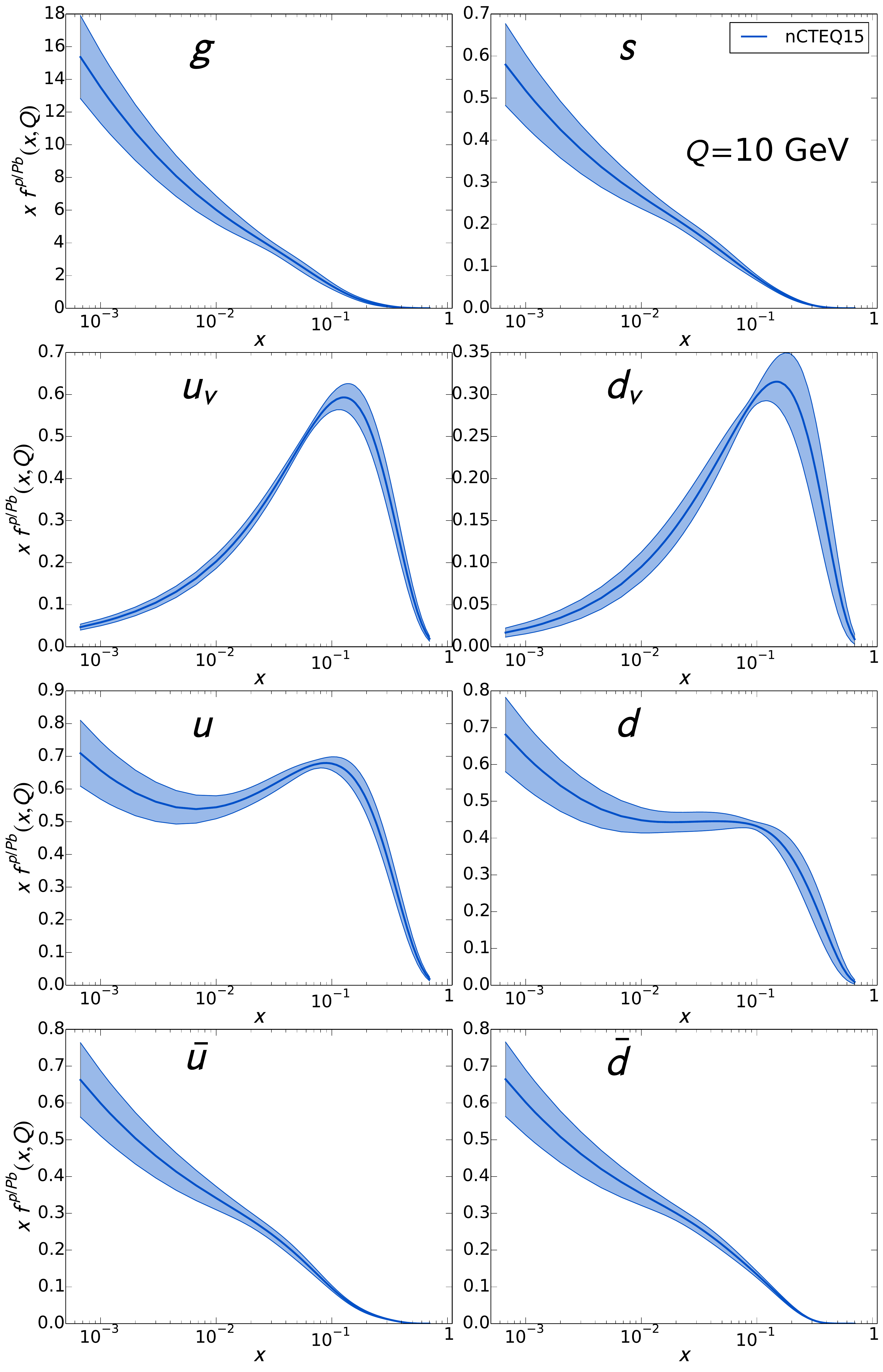}
\caption{Results of the \ncteqfit\ fit.
On the left we show nuclear modification factors defined as ratios of proton PDFs
bound in lead to the corresponding free proton PDFs, and on the right we show the
actual bound proton PDFs for lead. In both cases the scale is equal to $Q=10$ GeV.}
\label{fig:nCTEQ15g_10}
\end{figure*}
%----------------

%----------------
\begin{figure*}[th]
\centering{}
\includegraphics[width=0.96\textwidth]{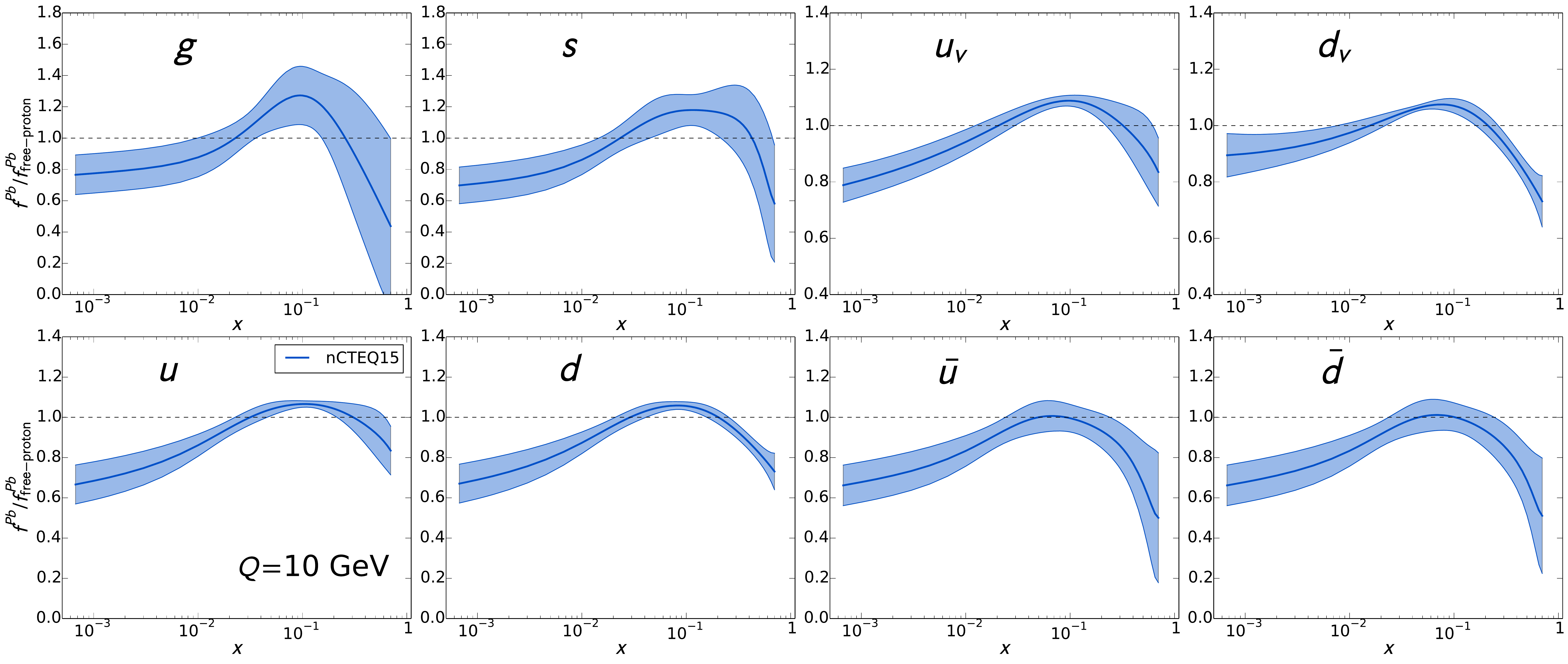}
\caption{
We show nuclear modification factors defined as ratios of lead PDFs
compared to a lead PDF constructed from free-proton PDFs.  
The PDFs are constructed using  $f^{A} = \frac{Z}{A}f^{p/A} + \frac{A-Z}{A}f^{n/A}$
for ${}^{207}\text{Pb}$ and the free proton, with a scale of  $Q=10$~GeV.}
\label{fig:nCTEQ15g_lead}
\end{figure*}
%----------------

%%%%%%%%%%%%%%%%%%%%%%%%%%%%%%%%%%%%%%%%%%%%%%%%%%%%%%%%%%%%%%%%%%%%%%%%%%%%%%%%%%%%%%%%%%%%%%
\subsubsection{nPDFs vs. nuclear $A$}

We  now examine the results of the \ncteqfit\ fit starting with the $A$-dependence of the
various nPDF flavors.
In Fig.~\ref{fig:compar_PDFs_diff-nuc} we display the  central fit predictions 
for a range of nuclear $A$ values from $A=1$ (proton) to $A=208$ (lead). 
When examining the $A$-dependence we observe 
that as we move to larger $A$ the gluon and sea-distributions
$\{g, \bar{u}, \bar{d}, s \}$ decrease at small $x$ values. 
This trend is also present for the $\{u, d\}$ PDFs.
On the other hand, the $A$-dependence of $\{u_v, d_v\}$ distributions
is reduced relative to the other flavor  components.

Finally, Figs.~\ref{fig:nCTEQ15g_1} and~\ref{fig:nCTEQ15g_10}, show our nPDFs ($f^{p/Pb}$) for a lead nucleus
together with the nuclear correction factors at the input scale $Q=Q_0=1.3$ GeV and at $Q=10$ GeV
to show the evolution effects when the PDFs are probed at a typical hard scale. We have chosen
to present results for the rather heavy lead nucleus because of its relevance for the heavy ion
program at the LHC. In all cases, we display the uncertainty band arising from 
the error PDF sets based upon our eigenvectors and the tolerance criterion.
It should be noted that the uncertainty bands for $x\lsim10^{-2}$ and $x\gsim0.7$ are
not directly constrained by data but only by the momentum and number sum rules.
The uncertainty bands are the result of extrapolating the functional form of our parametrization into
these unconstrained regions.

Some comments are in order:
\begin{itemize}
 \item As can be seen from Fig.~\ref{fig:nCTEQ15g_1} (a), our input gluon is strongly suppressed/shadowed
 with respect to the free proton in the $x\lsim0.04$ region. In fact, it has a valence-like
 structure (see Fig.~\ref{fig:nCTEQ15g_1} (b)) which vanishes at small $x$.
 Consequently, the steep small $x$ rise of the gluon distribution at $Q=10$ GeV (see Fig.~\ref{fig:nCTEQ15g_10})
 is entirely due to the QCD evolution.
 However, we should note that there is no data constrints below $x\sim0.01$ and the gluon
 uncertainty in this region is underestimated.
 In addition, our gluon has an anti-shadowing peak around $x\sim0.1$ and then exhibits
 suppression in the EMC region $x\sim0.5$. However, the large $x$ gluon features wide uncertainty band
 reflecting the fact that there are no data constraints.
 
\item In our analysis we determine the $\bar{u}+\bar{d}$ combination and assume that there is
no nuclear modification to the $\bar{d}/\bar{u}$ combination (see Sec.~\ref{sec:framework}
and Table~\ref{tab:params_nCTEQ15g}). As a result the $\bar{u}$ and $\bar{d}$ PDFs are very similar,
the small difference between the two comes from the underlying free proton PDFs.

\item In this analysis we do not fit the strange distribution but relate it to the light quarks
sea distribution, see Eq.~\eqref{ssb}. As a result the strange quark distribution is very similar to the
$\bar{u}$ and $\bar{d}$ distributions.

\item Contrary to the other existing nPDFs where the nuclear correction factors for the
valence distributions are assumed to be the same, we treat $u_v$ and $d_v$ as independent.
This leads to an interesting feature of our result where $u_v$ is suppressed and $d_v$
is enhanced in the EMC region. This behavior is not entirely unexpected, there are nuclear models 
predicting a flavor dependence for the EMC effect~\cite{Cloet:2009qs,Malace:2014uea,Dutta:2010pg}.

\item 
The above  difference for the nuclear correction in $u_v$ and $d_v$ appears at the
level of the bound proton PDFs. When we construct a physical combination representing
the full nuclear PDF, $f^{A} = \frac{Z}{A}f^{p/A} + \frac{A-Z}{A}f^{n/A}$,
such as lead in Fig.~\ref{fig:nCTEQ15g_lead}, the combination yields  net corrections
for $u_v$ and $d_v$ which are close to each other and similar to those in the literature. 
We will discuss this in more detail in Sec.~\ref{sec:other_sets}.

Once more data are included, e.g. from the LHC, neutrino DIS experiments and a future $eA$ collider,
it should be possible to relax some of the assumptions.
\end{itemize}

In the following section, we will investigate the 
impact of these nPDFs and the corresponding uncertainty bands on 
the physical observables. 

%%%%%%%%%%%%%%%%%%%%%%%%%%%%%%%%%%%%%%%%%%%%%%%%%%%%%%%%%%%%%%%%%%%%%%%%%%%%%%%%%%%%%%%%%%%%%%
%
% Data over Theory - Fa/Fd
%
%%%%%%%%%%%%%%%%%%%%%%%%%%%%%
\subsection{Comparison with data}
\label{sec:dataComp}
%%%%%%%%%%%%%%%%%%%%%%%%%%%%%
%
%%%%%%%%%%%%%%%%%%%%%%%%%%%%%%%%%%%%%%%%%%%%%%%%%%%%%%%%%%%%%%%%%
% Data over Theory - Fa/D vs. X
\begin{figure*}[th!]
\begin{center}
\includegraphics[width=0.8\textwidth]{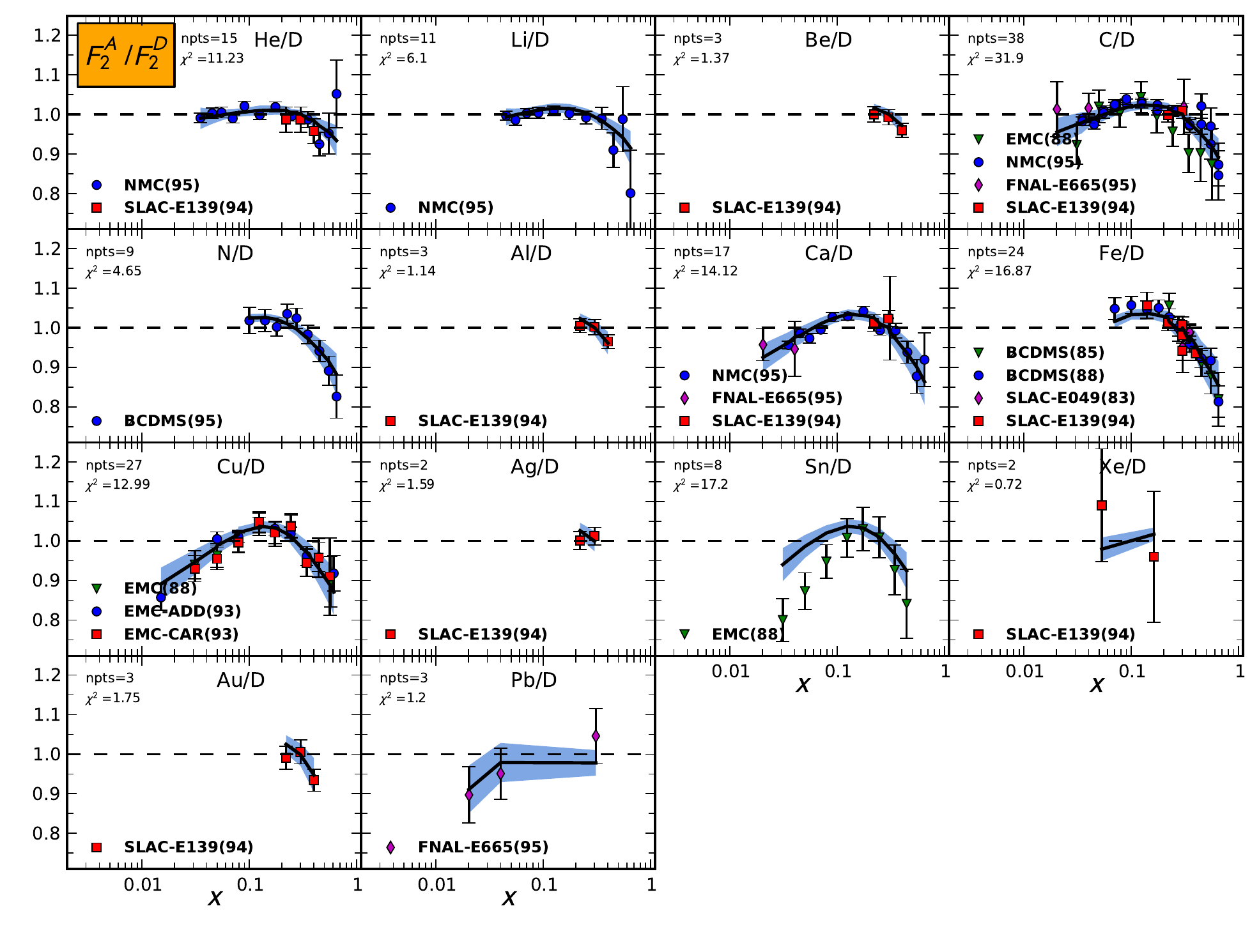}
\vspace*{-0.5cm}
\caption{Comparison of the \ncteqfit\ NLO theory predictions for $R=F_2^A(x,Q^2)/F_2^D(x,Q^2)$ as a function of $x$
with  nuclear target data. The theory predictions have been calculated at the $Q^2$ values of the corresponding data points.
The bands show the  uncertainty from the nuclear PDFs.
}
\label{fig:FaFd}
\vspace*{5ex}
\includegraphics[width=0.8\textwidth]{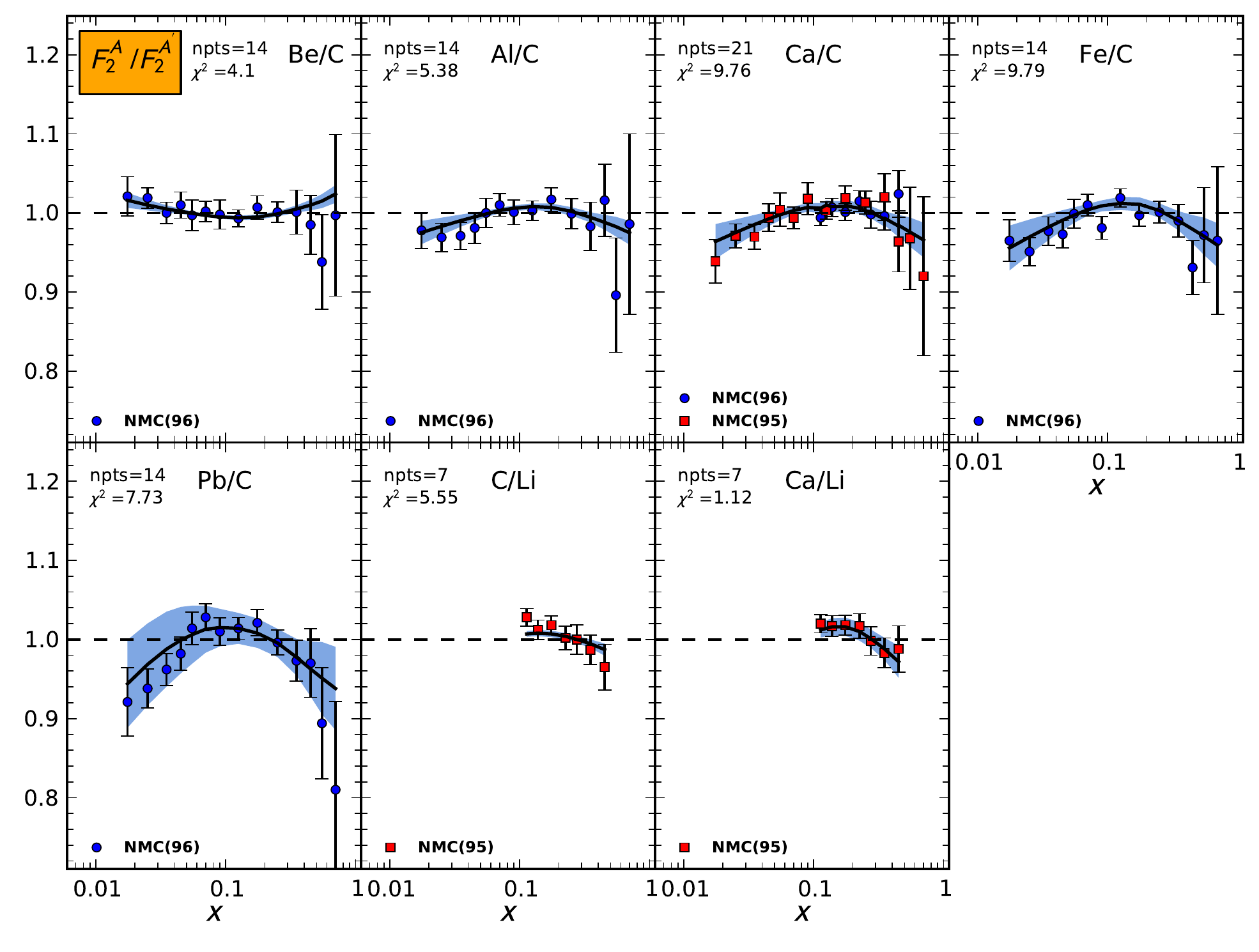}
\vspace*{-0.5cm}
\caption{Same as in Fig.~\ref{fig:FaFd} for $R=F_2^A(x,Q^2)/F_2^{A'}(x,Q^2)$.}
\label{fig:F2ratioVsX}
\end{center}
\end{figure*}
%%%%%%%%%%%%%%%%%%%%%%%%%%%%%%%%%%%%%%%%%%%%%%%%%%%%%%%%%%%%%%%%%

%%%%%%%%%%%%%%%%%%%%%%%%%%%%%%%%%%%%%%%%%%%%%%%%%%%%%%%%%%%%%%%%%
%--------------
%
% Data over Theory - Sn/C vs Q2
%--------------
\begin{figure*}[th]
\begin{center}
\includegraphics[width=0.8\textwidth,bb= 15 8 561 424]{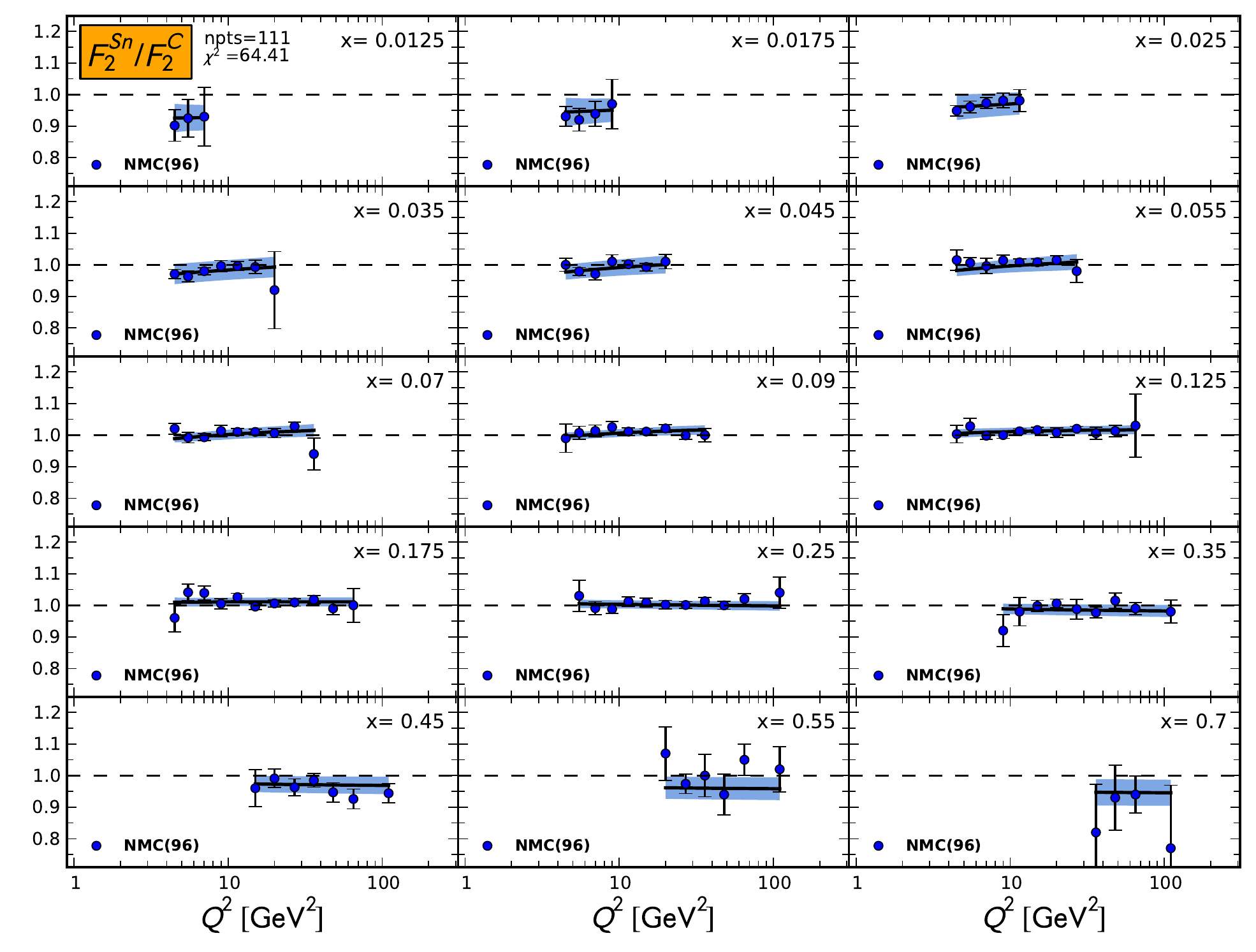}
\caption{Comparison of the \ncteqfit\ NLO theory predictions for $R=F_2^{Sn}/F_2^{C}$ 
as a function of $Q^2$ with  nuclear target data from the NMC collaboration. 
The bands show the  uncertainty from the nuclear PDFs.}
\label{fig:F2ratioVsQ}
\end{center}
\end{figure*}
%%%%%%%%%%%%%%%%%%%%%%%%%%%%%%%%%%%%%%%%%%%%%%%%%%%%%%%%%%%%%%%%%

While the $\chi^2/{\rm dof}$ is one measure of the quality of the fit
this alone obviously does not capture all the relevant characteristics.
To investigate the \ncteqfit\ result in more detail  we compare it to the most important and  
constraining data sets and consider strengths and limitations of both the fit and the available data sets. 

%
%%%%%%%%%%%%%%%%%%%%%%%%%%%%%
\subsubsection{DIS  data sets}
\label{sec:dataComp_DIS}
%%%%%%%%%%%%%%%%%%%%%%%%%%%%%
%
%%%%%%%%%%%%%%%%%%%%%%%%%%%%%%%%%%%%%%%%%%%%%%%%%%%%%%%%%%%%%%%%%
\begin{figure*}[th]
\begin{center}
%\subfloat[$Q^2=5$ GeV$^2$]
%
{
\includegraphics[width=0.47\textwidth,bb=7 10 422 277]{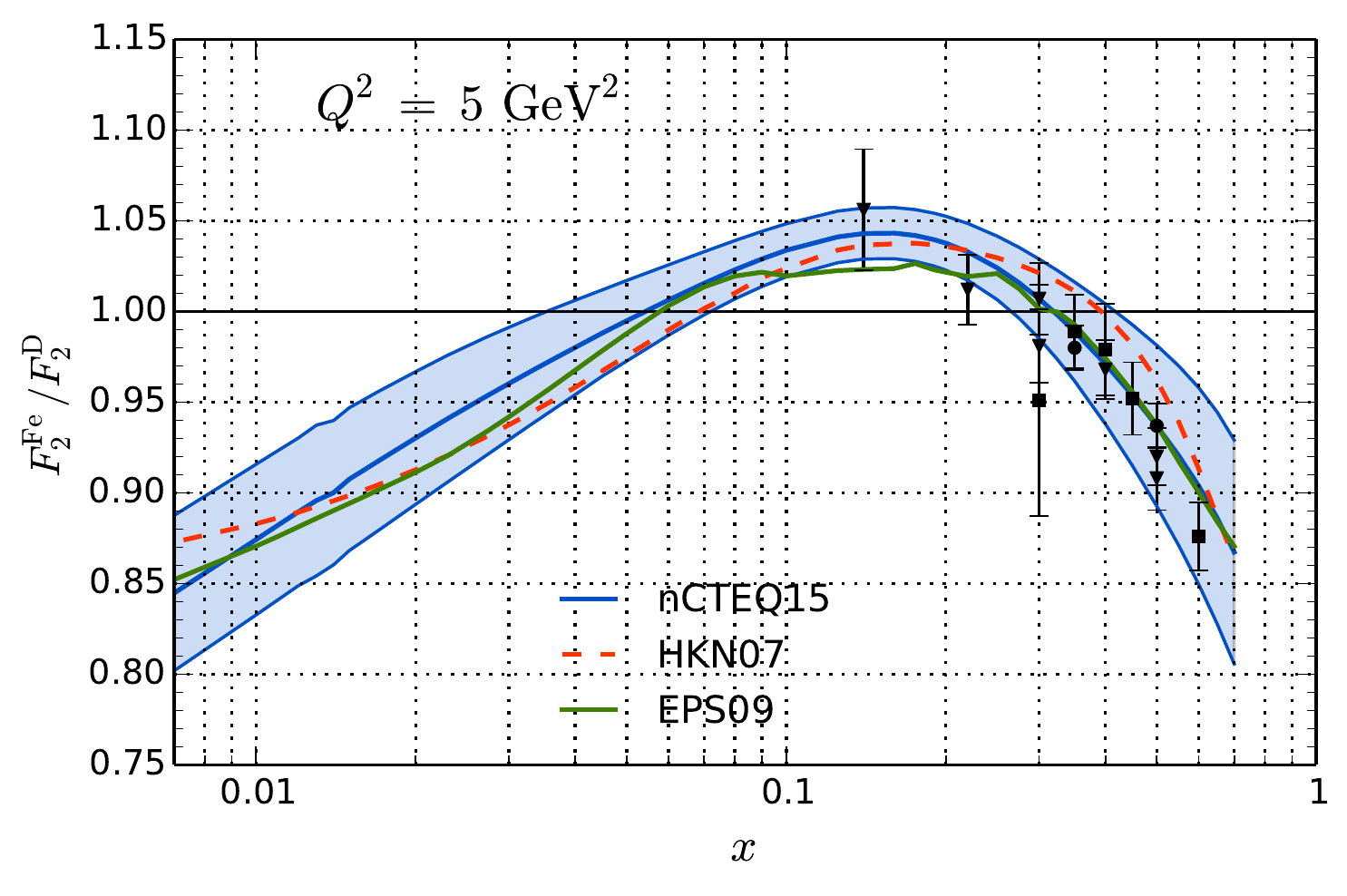}
\label{subfig:F2Fe_Q5}
}
\quad
%\subfloat[$Q^2=20$ GeV$^2$]
{
\includegraphics[width=0.47\textwidth,bb=7 10 422 277]{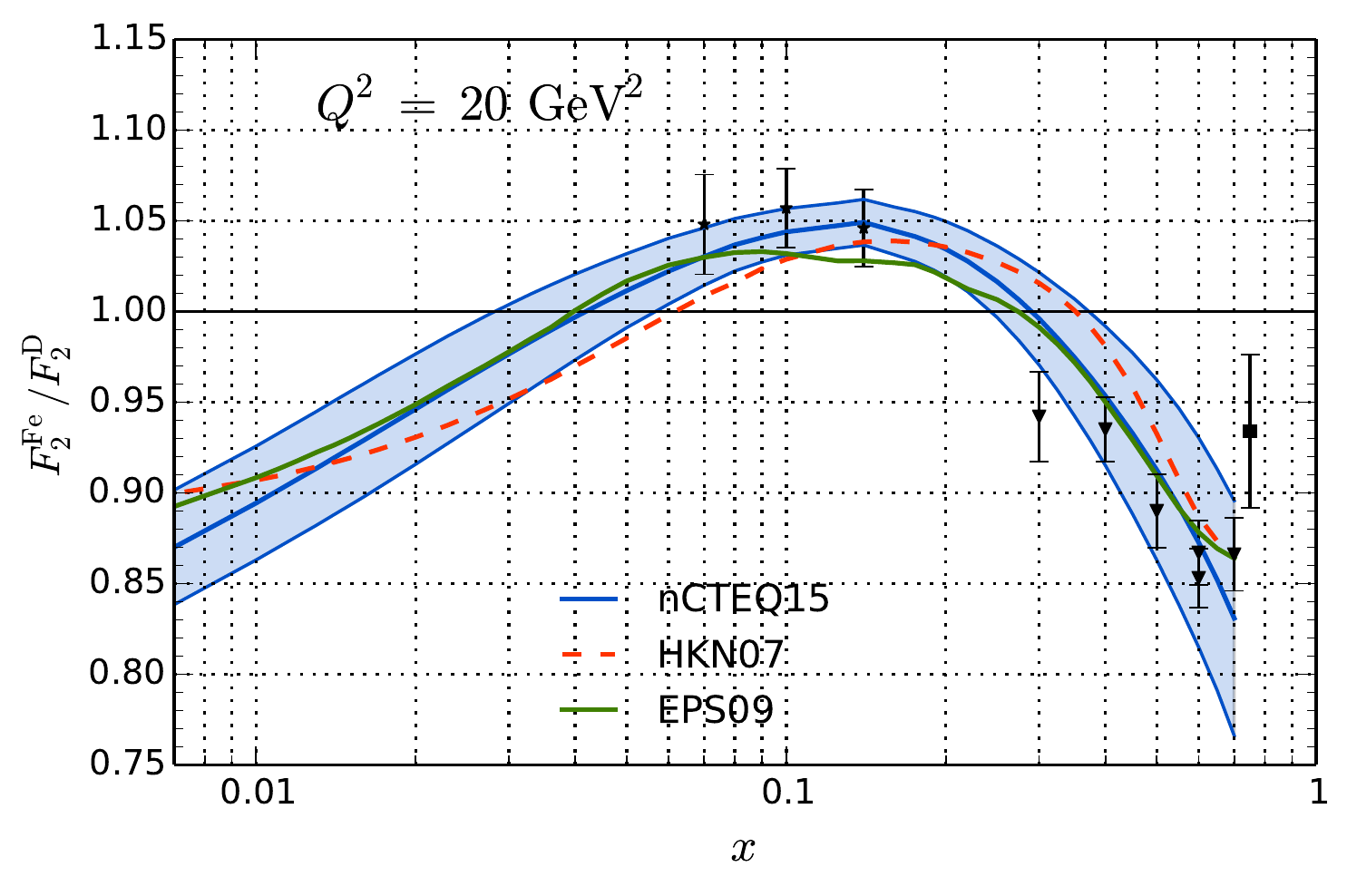}
\label{subfig:F2Fe_Q20}
}
\caption{Ratio of the $F_2$ structure functions for iron and deuteron
calculated with the \ncteqfit\ fit at 
(a)~$Q^2=5~\GeV^2$ and 
(b)~$Q^2=20~\GeV^2$.
This is compared with the fitted data from
SLAC-E049~\cite{Bodek:1983ec}
SLAC-E139~\cite{Gomez:1993ri}
SLAC-E140~\cite{Dasu:1993vk}
BCDMS-85~\cite{Bari:1985ga}
BCDMS-87~\cite{Benvenuti:1987az}
experiments and results from EPS09 and HKN07.
(The data points shown are within 50\% of the nominal $Q^2$ value.)
}
\label{fig:F2Fe-data}
\end{center}
\end{figure*}
%%%%%%%%%%%%%%%%%%%%%%%%%%%%%%%%%%%%%%%%%%%%%%%%%%%%%%%%%%%%%%%%%

%%%%%%%%%%%%%%%%%%%%%%%%%%%%%%%%%%%%%%%%%%%%%%%%%%%%%%%%%%%%%%%%%
%
% Data over Theory - DY
%--------------
\begin{figure*}[th]
\begin{center}
{
\includegraphics[width=0.49\textwidth]{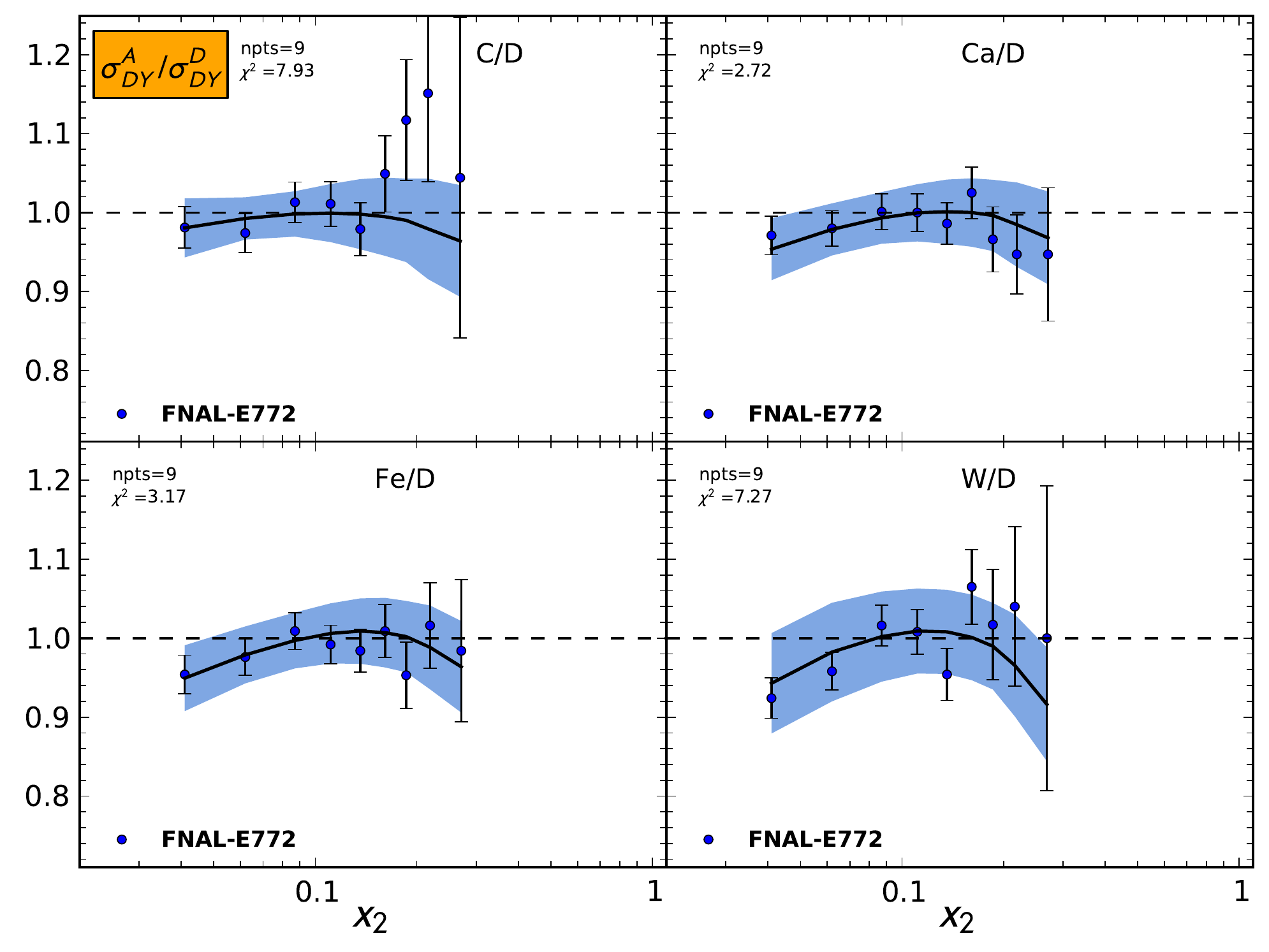}
}
{
\includegraphics[width=0.49\textwidth]{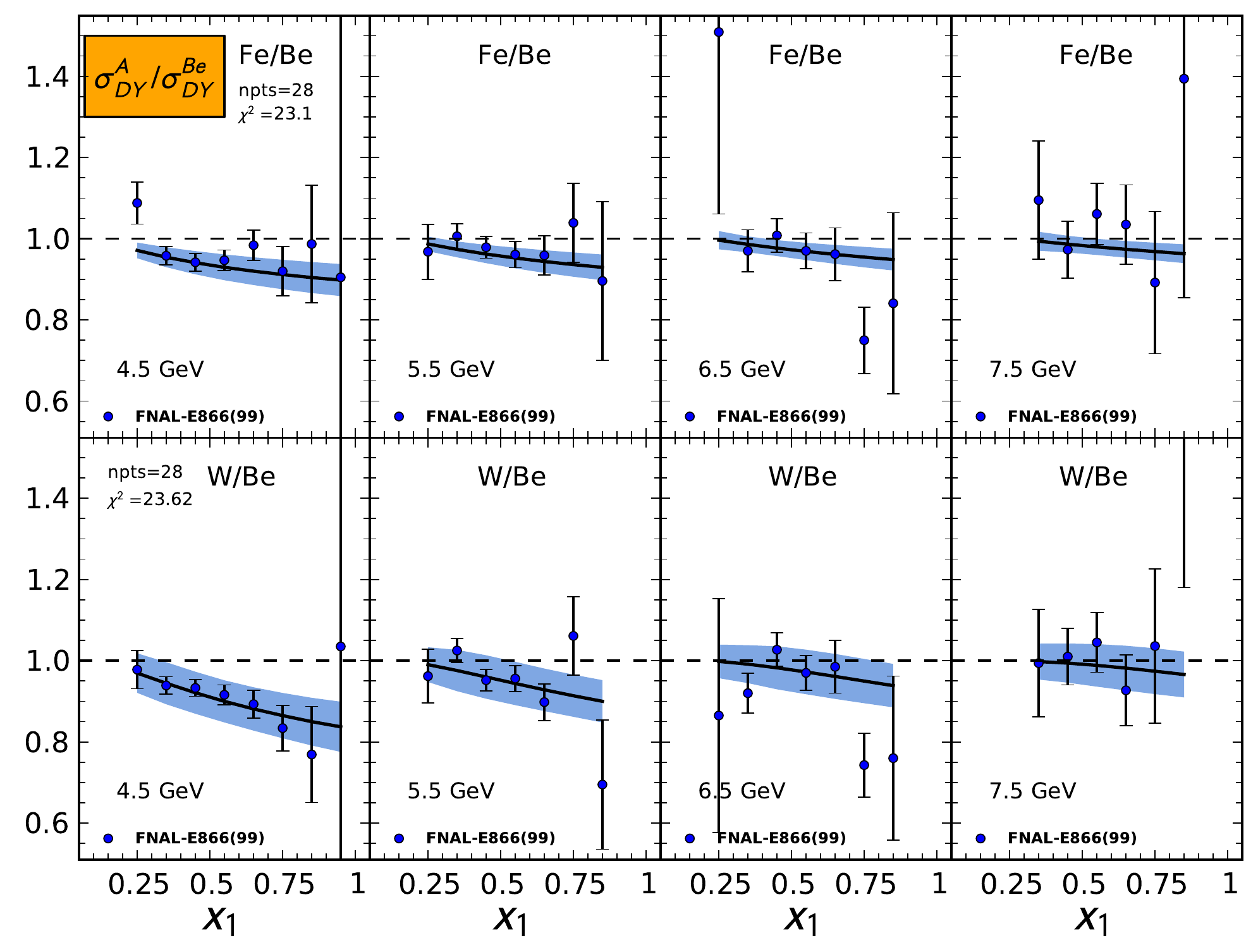}
}
\caption{Comparison of the \ncteqfit\ NLO theory predictions for $R=\sigma^{A}_{\rm DY}/\sigma^{A'}_{\rm DY}$ 
with data for several nuclear targets from the Fermilab experiments E772 (left) and E866 (right).
The error bands show the  uncertainty from the nuclear PDFs.
}
\label{fig:DY}
\end{center}
\end{figure*}
%%--------------
%%%%%%%%%%%%%%%%%%%%%%%%%%%%%%%%%%%%%%%%%%%%%%%%%%%%%%%%%%%%%%%%%

%%%%%%%%%%%%%%%%%%%%%%%%%%%%%%%%%%%%%%%%%%%%%%%%%%%%%%%%%%%%%%%%%
%INTRODUCTION:
%
The data from the deep-inelastic scattering 
experiments are by far the most numerous 
and provide the dominant contribution to the total $\chi^2$. 
These experiments are performed on a variety of nuclei which allow us to constrain the $A$ dependence of our parameters. 
Most of the data are extracted as a  ratio of $F_2$ structure functions $R=F_2^{A_1}/F_2^{A_2}$ for two different
targets $A_1$ and $A_2$. 
Note that in the present study we do not fit data from the very high $x$ region $x \gsim 0.7$
since they do not pass our kinematic cuts. As already mentioned the high $x$ region is theoretically
challenging due to a host of effects (higher twist, target mass corrections, large $x$ resummation,
deuteron wave-function, nuclear off-shell effects).
Some of these effects in the large $x$ and low $Q^2$ area have been investigated extensively
in the proton case by the CTEQ-CJ collaboration~\cite{Accardi:2009br,Accardi:2011fa}.
The nuclear case is even more challenging due to enhanced higher twist and
Fermi motion effects which lead to a steep rise of the structure function ratios in the limit $x\to1$.
For these reasons we avoid fitting the high $x$ region for the time being.

%%%%%%%%%%%%%%%%%%%%%%%%%%%%%%%%%%%%%%
%F2 RATIO AS A FUNCTION OF X:
%
The comparison of our fit to the DIS $F_2$ ratio data is shown in Figs.~\ref{fig:FaFd}
and \ref{fig:F2ratioVsX} as a function of $x$. 
Note, in these figures the data for different $Q^2$ are combined into a single plot 
as the scaling violations (discussed later) occur on a logarithmic scale and largely cancel
out in the ratios.

%%%%%%%%%%%%%%%%%%%%%%%%%%%%%%%%%%%%%%%%%%%%%%%%%%%%%%%%%%%%%%%%%
\begin{figure*}[th]
\begin{center}
%%%
\subfloat[ Comparison of the \ncteqfit\  fit with the data. 
The error bands are computed by adding the uncertainties in quadrature.
]{
\includegraphics[width=0.47\textwidth]{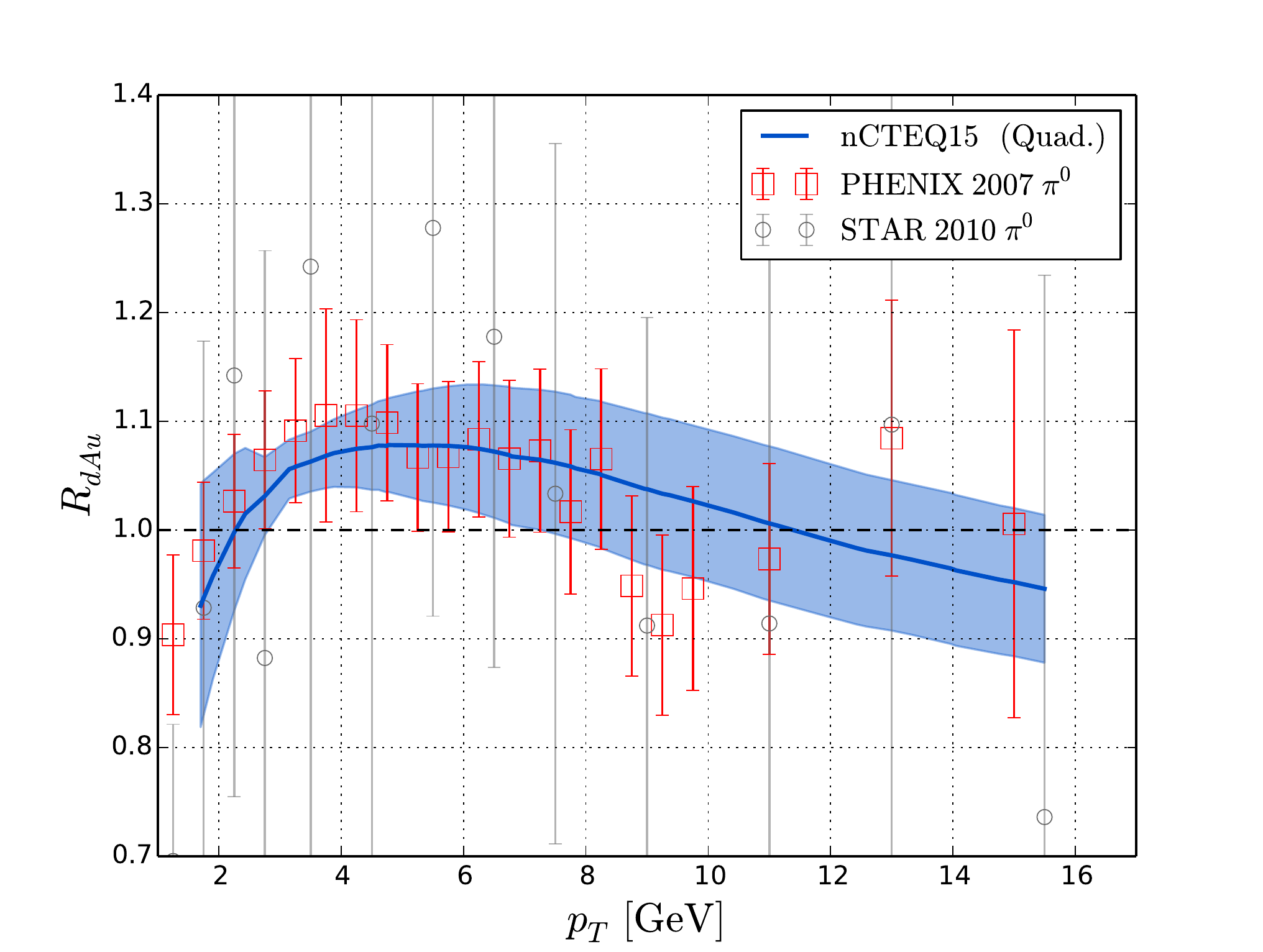}
\label{subfig:a}
}
\quad
%%%
\subfloat[ Comparison of the \ncteqfit\ and EPS09  fits with the data. 
The \ncteqfit\ error bands are computed using asymmetric uncertainties (MAX) to match EPS09.
]{
\includegraphics[width=0.47\textwidth]{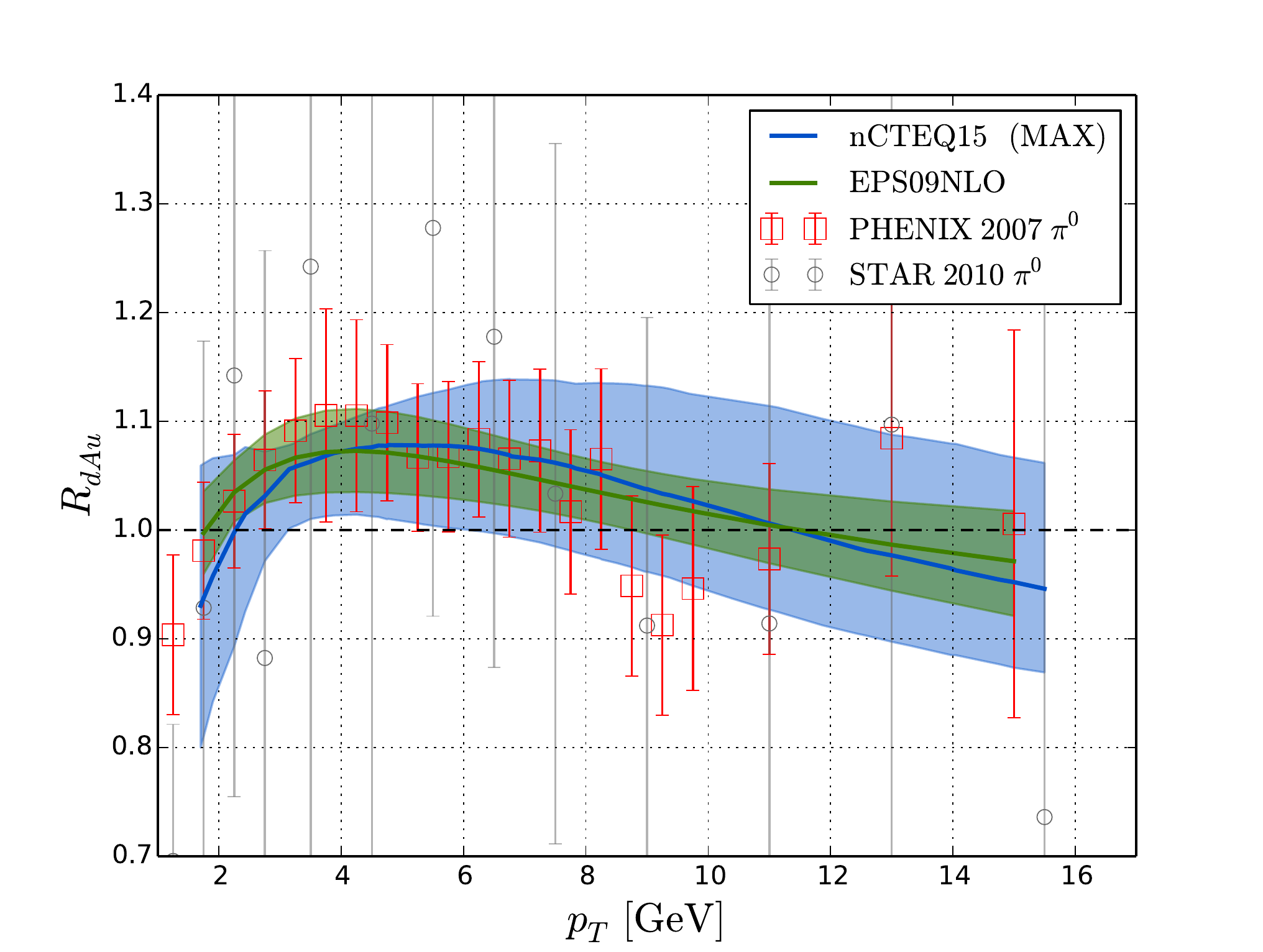}
\label{subfig:b}
}
\caption{
We display the comparison of the \ncteqfit\ and EPS09  fits  with the 
 PHENIX~\cite{Adler:2006wg} and STAR~\cite{Abelev:2009hx} data 
for  the ratio $R_{\text{dAu}}^{\pi}$.
The plotted PHENIX and STAR data are shifted by our fitted normalization. 
}\label{fig:pionEPS}
%%%
\subfloat[ Comparison of the \ncteqfit\  fit  
using the default BKK (blue) and the KKP fragmentation (violet) functions 
for the calculation of  $R_{\text{dAu}}^{\pi}$.
]{
\includegraphics[width=0.47\textwidth]{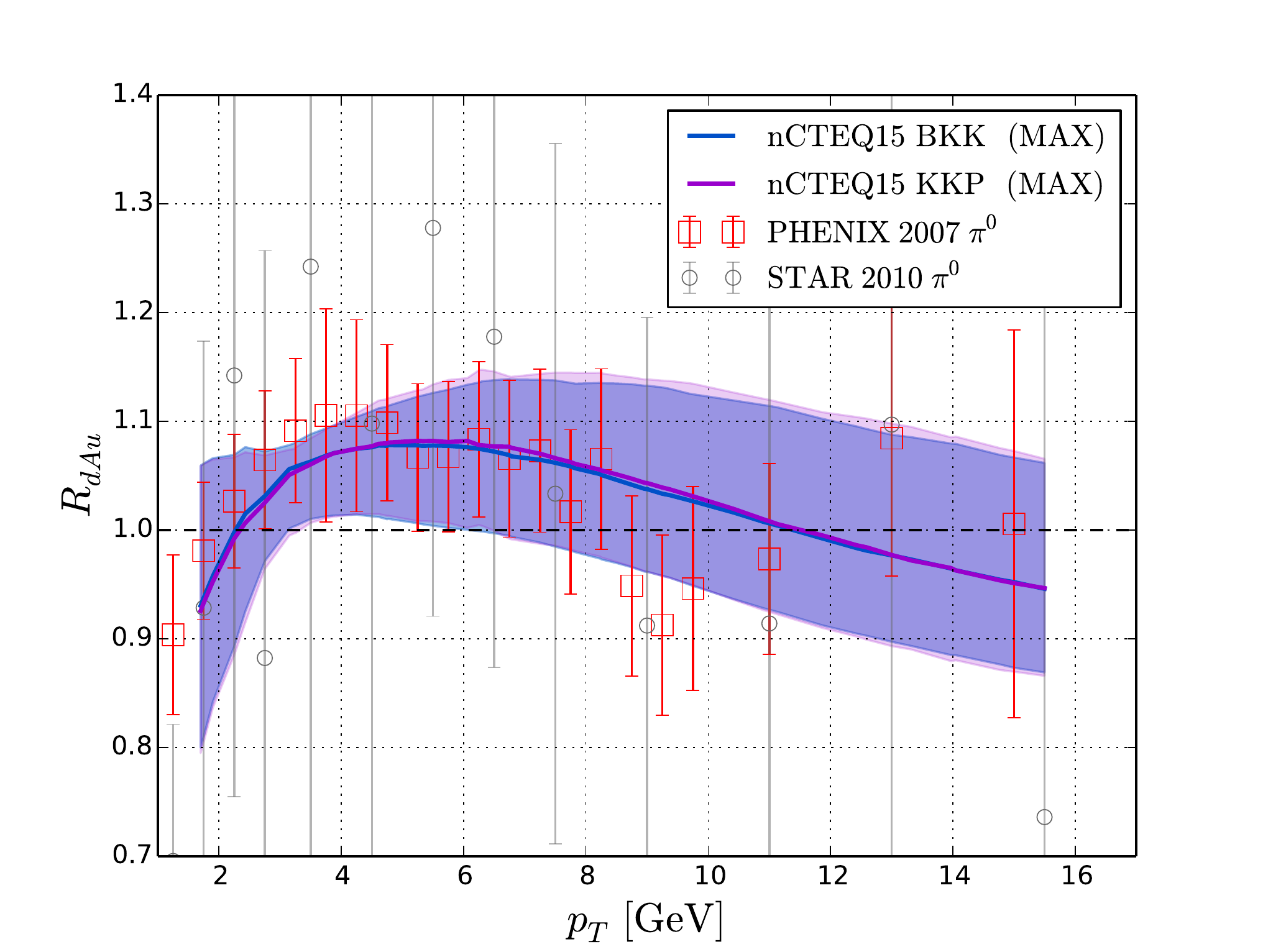}
\label{subfig:ff:a}
}
\quad
%%%
\subfloat[
Same as previous figure, but with a full re-analysis using the 
 BKK (blue) and the KKP fragmentation (violet) functions throughout the fitting procedure. 
]{
\includegraphics[width=0.47\textwidth]{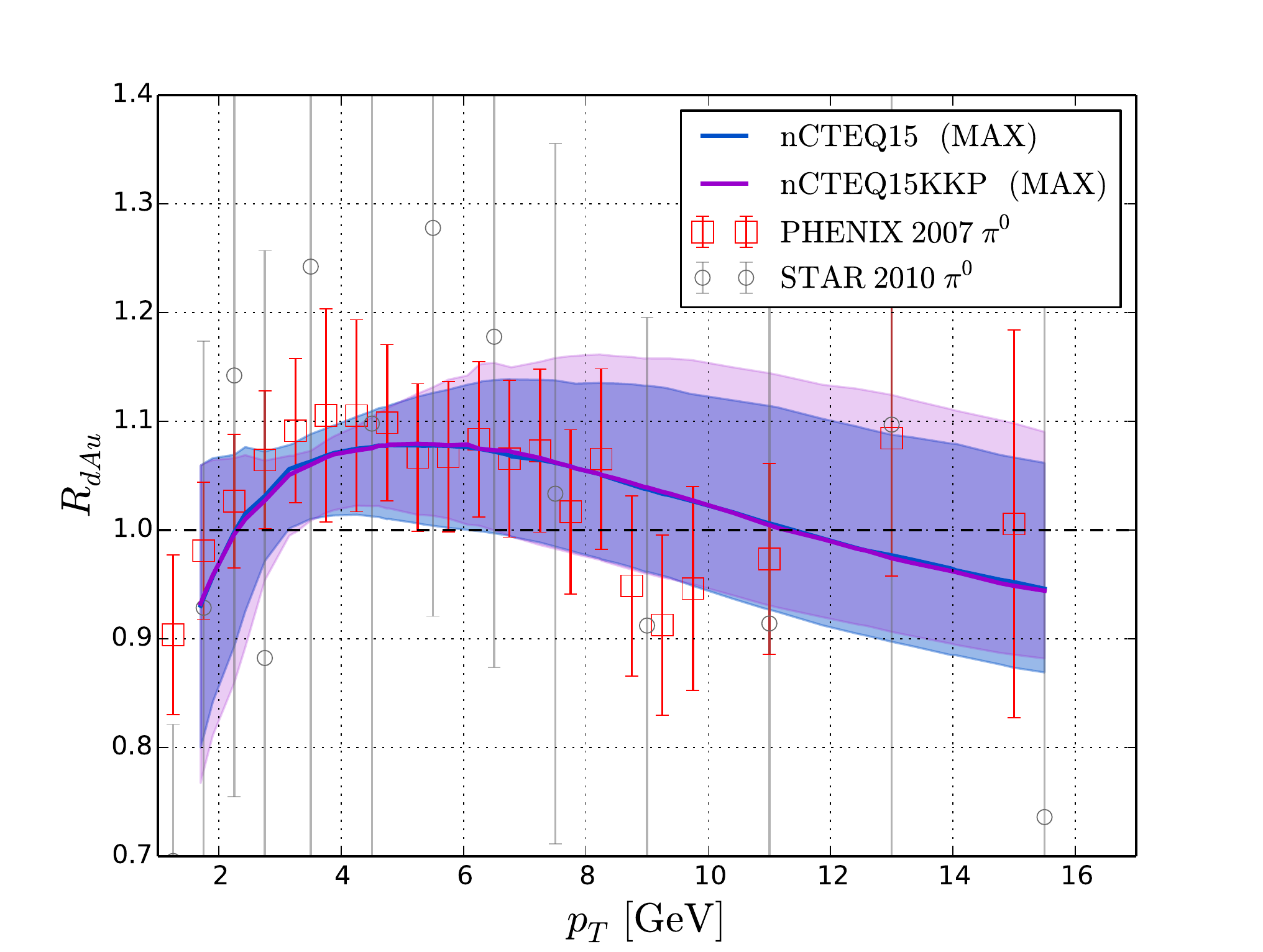}
\label{subfig:ff:b}
}
\caption{
We compare the impact of different fragmentation functions
on the observable $R_{\text{dAu}}^{\pi}$.
The \ncteqfit\ error bands are computed using asymmetric uncertainties to match EPS09.
}
\end{center}
\end{figure*}
%%%%%%%%%%%%%%%%%%%%%%%%%%%%%%%%%%%%%%%%%%%%%%%%%%%%%%%%%%%%%%%%%
%
%%%%%%%%%%%%%%%%%%%%%%%%%%%%%%%%%%%%%%%%%%%%%%%%%%%
% ncteq15 vs. ncteq15wp
%----------------
\begin{figure*}[th!]
\centering{}
\includegraphics[clip,width=0.48\textwidth]{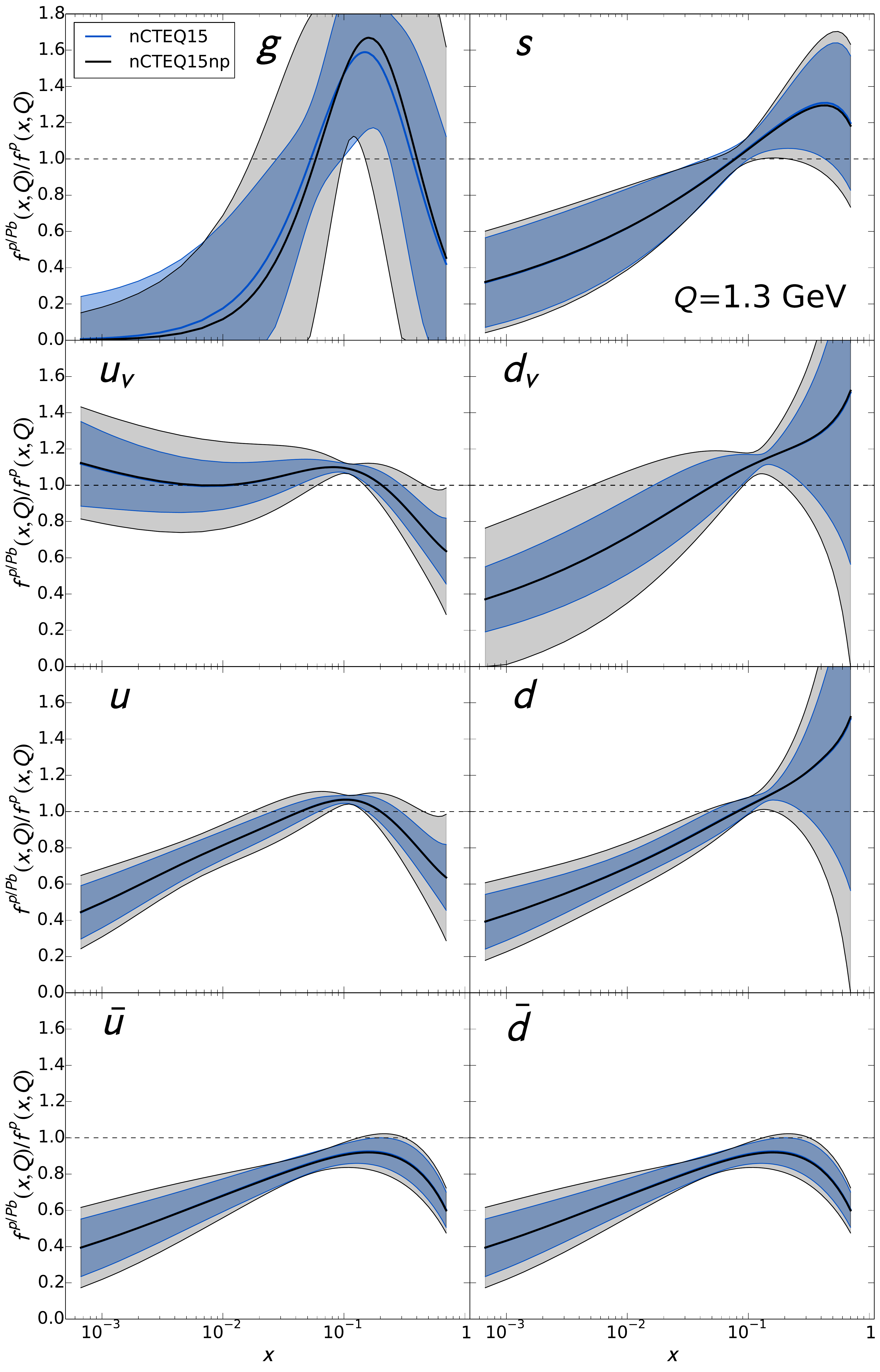}
\quad{}
\includegraphics[width=0.48\textwidth]{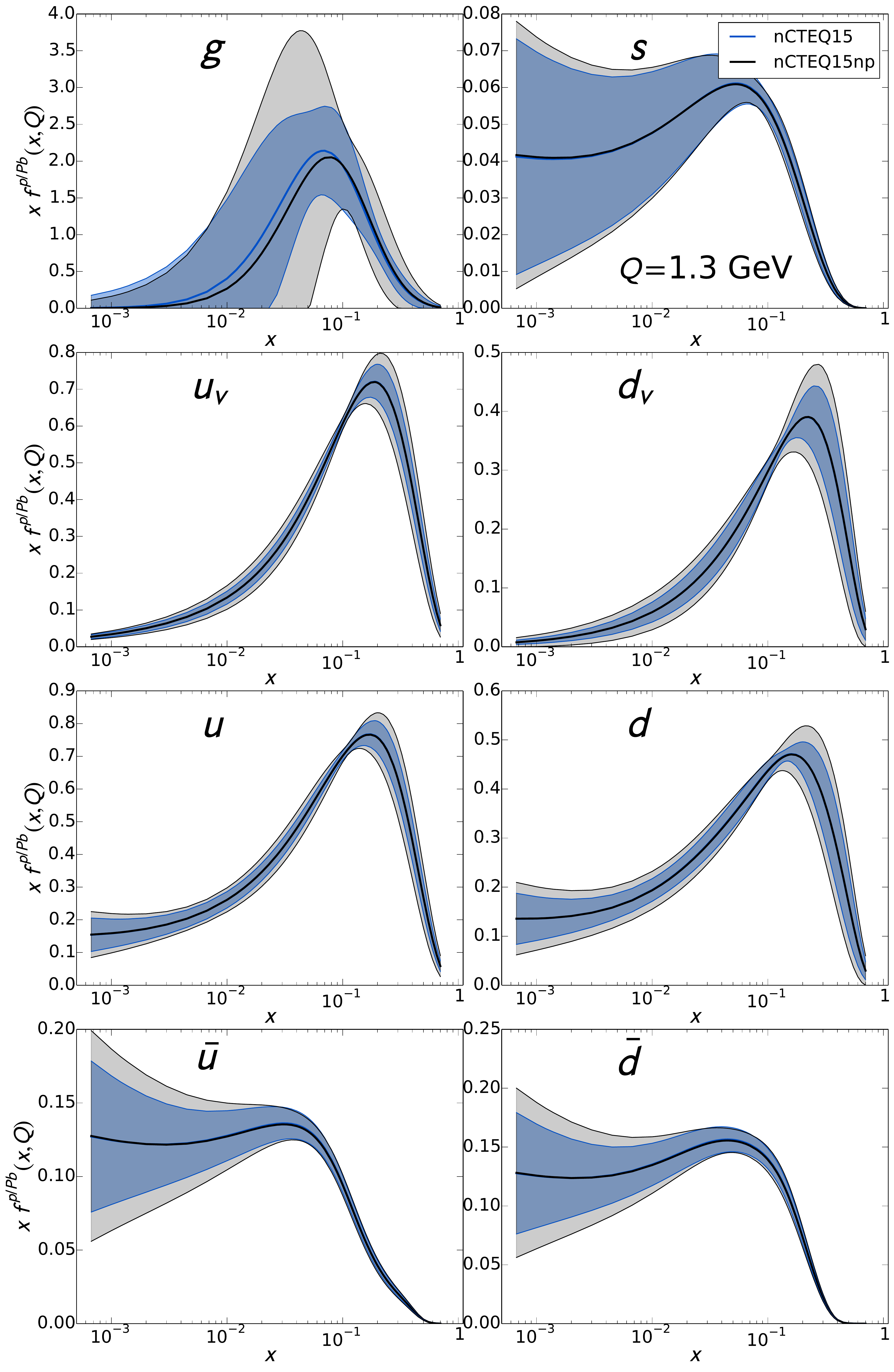}
\caption{Comparison of the \ncteqfit\ fit (blue)
with the
\ncteqnp\ fit without pion data (gray).
On the left we show nuclear modification factors defined as ratios of proton PDFs
bound in lead to the corresponding free proton PDFs, and on the right we show the
actual bound proton PDFs for lead. In both cases scale is equal to $Q=1.3$ GeV.}
\label{fig:nCTEQ15-vs-nCTEQ15np}
\end{figure*}
%%%%%%%%%%%%%%%%%%%%%%%%%%%%%%%%%%%%%%%%%%%%%%%%%%%

%%%%%%%%%%%%%%%%%%%%%%%%%%%%%%%%%%%%%%
%F2 RATIO FOR D AS A FUNCTION OF X:
%
Fig.~\ref{fig:FaFd} shows the ratio $F_2^A(x,Q^2)/F_2^D(x,Q^2)$ for a variety of experiments. 
The overall agreement of the fit with the data is excellent for a majority of the nuclei.
The discrepancy which can be seen for the EMC data taken on tin (Sn/D) is the same discrepancy 
we have pointed out in Sec.~\ref{sec:reschi2} when we investigated the $\chi^2$ of the individual experiments.
As already mentioned, this problem has been also encountered in previous analyses \cite{Hirai:2007sx,deFlorian:2011fp}
and we are unable to reconcile it with our fit. 

%%%%%%%%%%%%%%%%%%%%%%%%%%%%%%%%%%%%%%
%F2 RATIO FOR MISC A AS A FUNCTION OF X:
%
Similarly, Fig.~\ref{fig:F2ratioVsX} shows the structure function ratio $F_2^A(x,Q^2)/F_2^{A'}(x,Q^2)$ in comparison to NMC data for a variety of nuclear targets. 
These high-statistics data  are also  well described  by the results of the \ncteqfit\ fit.

%%%%%%%%%%%%%%%%%%%%%%%%%%%%%%%%%%%%%%
%F2 RATIO SN/C AS A FUNCTION OF Q2:
%
The  NMC data taken on tin and carbon  ($R=F_2^{Sn}/F_2^{C}$) cover a wider range in $Q^2$,
and we display these in Fig.~\ref{fig:F2ratioVsQ} as a function of $Q^2$ binned in $x$.  
As is well know, the logarithmic $Q^2$ scaling violations of the structure functions provide constraints on the low $x$
gluon distribution. Of course, compared to the very precise HERA data on the proton $F_2$ structure function
which extends over a very wide range of $Q^2$ values the NMC data have a much smaller $Q^2$ lever arm.
As a consequence the NMC data provide relatively weaker constraints on the nuclear
gluon PDF in the $x$ range of $(0.05,0.1)$.
We will discuss data constraints on gluon in more detail in Sec.~\ref{subsec:constraints}.

%%%%%%%%%%%%%%%%%%%%%%%%%%%%%%%%%%%%%%
%F2 RATIO FOR IRON VS. X FOR 2 Q2 CHOICES:
%
In Fig.~\ref{fig:F2Fe-data} we plot the nuclear correction  $R=F_2^{Fe}/F_2^D$ 
for iron vs. $x$ for two $Q^2$ values and compare the results with experimental data and with results from different nPDF groups. 
Comparing these two figures, we again see that there is a rather weak $Q^2$-dependence of the structure function ratio
between $Q^2=5~\GeV^2$ and $Q^2=20~\GeV^2$. As discussed above due to our strict kinematic cuts we do not extend our 
predictions to the high $x$ region $(x\gtrsim 0.7)$.

Taking into account both the nPDF uncertainty (represented by the error bands) and the experimental error bars, 
the data are generally compatible with the  \ncteqfit\  fit. 
In addition to comparing with data, we compare our predictions with those of HKN
\cite{Hirai:2007sx} and EPS \cite{Eskola:2009uj} and find a good
agreement within the errors of our analysis.

%%%%%%%%%%%%%%%%%%%%%%%%%%%%%
\subsubsection{Drell-Yan  data sets}
\label{sec:dataComp_DY}
%%%%%%%%%%%%%%%%%%%%%%%%%%%%%

%%%%%%%%%%%%%%%%%%%%%%%%%%%%%%%%%%%%%%
%DRELL YAN DATA 

We now turn to the Drell-Yan muon pair production process $p+A\to \mu^+ + \mu^- + X$.
In Fig.~\ref{fig:DY}~(a), we display the differential cross section ratio,
$R=(d\sigma^{pA}_{\rm DY}/dx_2dM)/(d\sigma^{pD}_{\rm DY}/dx_2dM)$, measured by the Fermilab
experiment E772, where $x_2$ is the momentum fraction of the parton inside the nucleus and
the invariant mass of the produced muon pair, $M$, covers the range $\sim(4.5, 13)$ GeV
(excluding the charmonium and bottonium resonances).
These data have been taken for large Feynman $x_F\sim x_1-x_2$ corresponding to smallish $x_2$ values.

Similarly, in Fig.~\ref{fig:DY}~(b), we present a comparison of our predictions with  large $x_F$ data
from the E866 experiment for the ratio $R=(d\sigma^{pA}_{\rm DY}/dx_1dM)/(d\sigma^{pD}_{\rm DY}/dx_1dM)$.
The data are arranged in four bins of the invariant mass ($M=\{4.5, 5.5, 6.5, 7.5\}$ GeV) and are 
presented as a function of the proton momentum fraction $x_1$.

As can be seen, the theory predictions describe the data quite well, 
except for some isolated points (generally those with large error bars). 

%%%%%%%%%%%%%%%%%%%%%%%%%%%%%
\subsubsection{Pion production data sets}
\label{sec:dataComp_pi}
%%%%%%%%%%%%%%%%%%%%%%%%%%%%%
%
The newest addition to the current analysis as compared to Ref.~\cite{Schienbein:2009kk} 
are the ratios of  double differential cross-sections for single inclusive pion data from 
the STAR and PHENIX experiments at RHIC. 
Specifically, we fit the ratio
\begin{equation}
\label{pion:ratio}
R_{\text{dAu}}^{\pi} = \frac{\tfrac{1}{2A}d^2\sigma_{\pi}^{\text{dAu}}/dp_T dy}{d^2\sigma_{\pi}^{\text{pp}}/dp_T dy}\,,
\end{equation}
and we include only the data  measured at central rapidity to exclude potential final-state effects 
(this criterion excludes any data from BRAHMS). 
Additionally, we fit the normalizations of the RHIC data and obtain 1.031 and 0.962 for PHENIX and STAR, respectively.
These values are within the experimental uncertainty.\footnote{We note that the EPS09 analysis obtained similar normalizations.}
Fitting  the single inclusive pion production has the added complication that it depends on the fragmentation functions (FFs).
As mentioned in Sec.~\ref{sec:framework}, pre-computed grids of convolutions with the free 
deuterium PDFs and a set of FFs are used to speed up the NLO calculation.

In Fig.~\ref{subfig:a}, PHENIX and STAR data are compared with predictions from the \ncteqfit\ fit using  the
Binnewies-Kniehl-Kramer (BKK) fragmentation functions~\cite{Binnewies:1994ju}. 
As the PHENIX data are more precise than the STAR data, the former will have a correspondingly larger impact on the resulting fit.

The EPS09 analysis \cite{Eskola:2009uj} also used this 
data
and we compare with their result 
in Fig.~\ref{subfig:b}.
Our central prediction for $R_{\text{dAu}}^{\pi}$  differs from EPS09 but lies within their uncertainty band; 
however,  our estimate of the PDF uncertainties differs substantially from EPS09.\footnote{The 
EPS09 analysis uses a different asymmetric definition of uncertainties given by
\begin{eqnarray*}
	(\Delta X^{+})^2 &=& \sum_k \left[ \max\left\{ X(S_k^+)-X(S_k^0), X(S_k^-)-X(S_k^0), 0 \right\} \right]^2,\\
	(\Delta X^{-})^2 &=& \sum_k \left[ \max\left\{ X(S_k^0)-X(S_k^+), X(S_k^0)-X(S_k^-), 0 \right\} \right]^2.
\end{eqnarray*}
To make this comparison 
consistent, we adopt the same definition when comparing with the EPS09 prediction.
} 
The main reason for this difference is the fact that EPS09 
chooses to include the single inclusive pion data with a large weight ($\times$20) to enhance its importance, 
and this  choice leads to the  suppression of the corresponding uncertainties.

Another source of difference can arise from  the choice of the fragmentation functions. 
The EPS09  analysis uses the  Kniehl-Kramer-P\"{o}tter (KKP) fragmentation functions \cite{Kniehl:2000fe}
whereas the \ncteqfit\  fit is based on the BKK FFs. 
To  investigate the effect of different fragmentation functions, 
we have calculated  $R_{\text{dAu}}^{\pi}$ 
using the KKP FFs {\em but} still using the \ncteqfit\  
nPDFs obtained employing the BKK FFs (see Fig.~\ref{subfig:ff:a}). 
As can be seen, the choice of different fragmentation functions 
yields only minor differences.

In a second step, we have also performed a complete  
reanalysis of the nuclear PDFs using the KKP fragmentation functions in both 
the fit and also for the calculation of  $R_{\text{dAu}}^{\pi}$
and this is shown in  Fig.~\ref{subfig:ff:b}.
The use of the KKP FFs does not change the central prediction for $R_{\text{dAu}}^{\pi}$
but slightly changes the nPDF uncertainties in the high-$p_T$ region.

In summary the use of two different sets of fragmentation functions, BKK and KKP,
has only a minor effect on the resulting nPDFs. This does not exclude a possibility 
that a larger effect on nPDFs is possible if other fragmentation functions are used \cite{Enterria:2013vba}.

%%%%%%%%%%%%%%%%%%%%%%%%%%%%%
\subsection{Fit without inclusive pion data (\ncteqnp)}
%%%%%%%%%%%%%%%%%%%%%%%%%%%%%
%
%%%%%%%%%%%%%%%%%%%%%%%%%%%%%%%%%%%%%%%%%%%%%%%%%%%%
\begin{figure}[th]
\begin{center}
\includegraphics[width=0.48\textwidth]{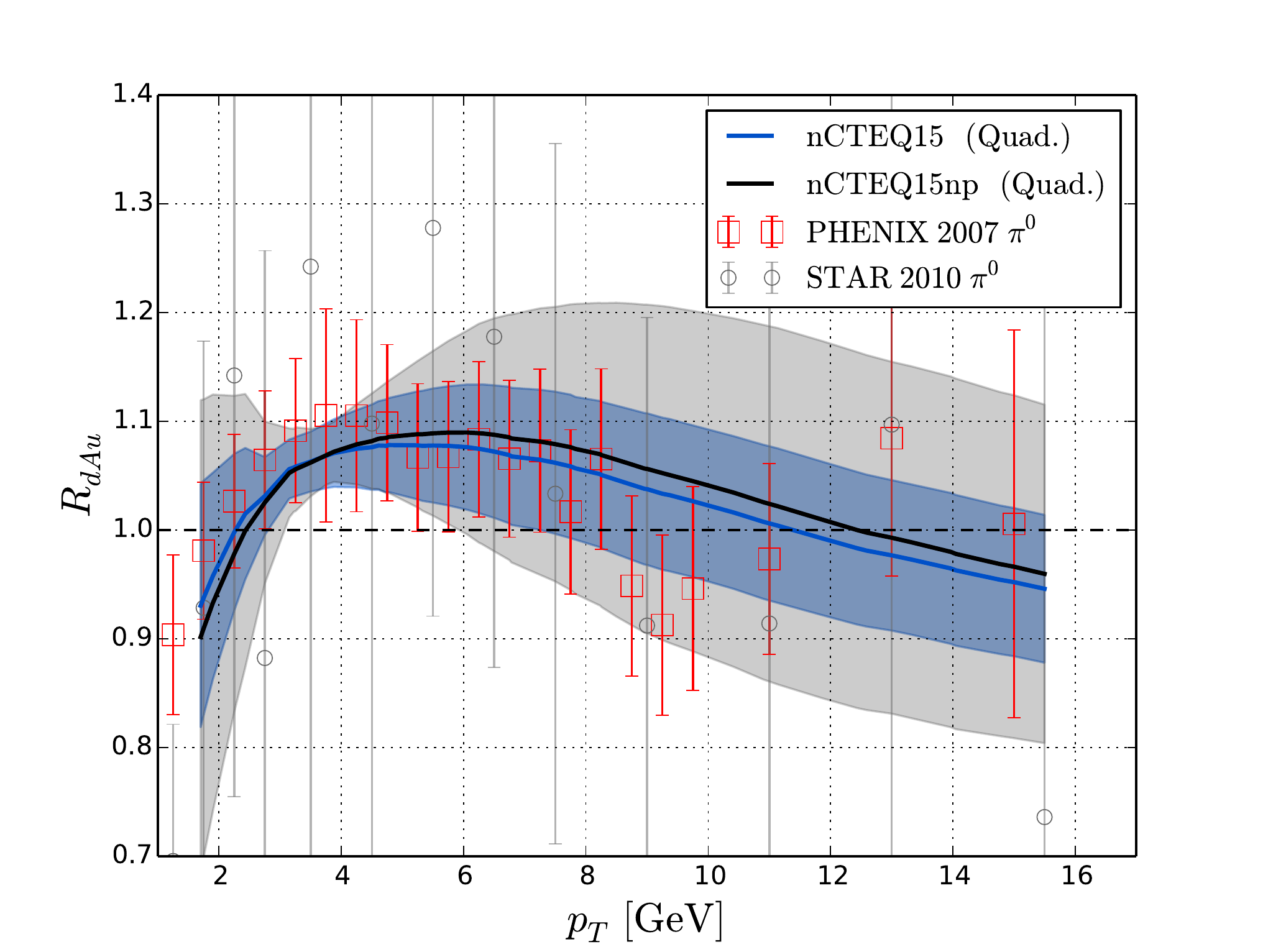}
\caption{
Comparison of the predictions of the \ncteqfit\ (solid blue) and \ncteqnp\ (dashed gray) fits to inclusive 
pion production data from PHENIX and STAR
demonstrating the effect of including these data sets.
Note that, the dark blue area is the overlap between the blue and gray bands.
}
\label{fig:fig9_e11NLOa-vs-decut4}
\end{center}
\end{figure}
%%%%%%%%%%%%%%%%%%%%%%%%%%%%%%%%%%%%%%%%%%%%%%%%%%%
%
To further analyze the impact of the newly added inclusive pion data and because the pion data introduce an unwanted 
dependence on fragmentation functions, we performed an alternative analysis which {\em does not} include the 
RHIC inclusive pion data (\ncteqnp).

In Fig.~\ref{fig:nCTEQ15-vs-nCTEQ15np}, we compare the results of the \ncteqfit\ fit 
with the ones of the alternative analysis \ncteqnp.
When examining the nuclear correction factors (left panels) we see the 
pion data have an impact on the gluon PDF
and to a lesser extent on the valence and sea quark distributions. 
For the central prediction, the inclusion of the pion data decreases the lead gluon PDF at large $x$
and increases it for smaller $x$; the two gluon distributions cross each other at $x\sim0.08$.
%%%%%%%%%%%%%%%%%%%%
Throughout most of the $x$-range the error bands are reduced with the exception of $x\sim0.1$
(and very small $x$ values) where they stay more or less unchanged.
This is precisely the range that is sensitive to the DIS Sn/C (and DY) data.
For most of the other PDF flavors, the change in the central value is minimal
(except for a few cases at high-$x$ where the magnitude of the PDFs are small). 
For these other PDFs, the inclusion of the pion data generally decreases the size of the error band.

In Fig.~\ref{fig:fig9_e11NLOa-vs-decut4} the predictions of the
\ncteqfit\ and  \ncteqnp\  fits are compared to the RHIC pion production data. 
The effect of the pion data is to increase $R_{dAu}^{\pi}$ for small $p_T$
and decrease it at larger $p_T$ by up to 5\%. 
The two central predictions cross each other at $p_T\sim4$ GeV. This can be connected
to the crossing of the gluon distributions in Fig.~\ref{fig:nCTEQ15-vs-nCTEQ15np}
(at $x\sim0.08$) which is in line with the kinematic mapping in Fig.~\ref{fig:pionXval}.

%%%%%%%%%%%%%%%%%%%%%%%%%%%%%%%%%%%%%%%%%%%%%%
\begin{figure*}[th!]
\begin{center}
\includegraphics[width=0.32\textwidth]{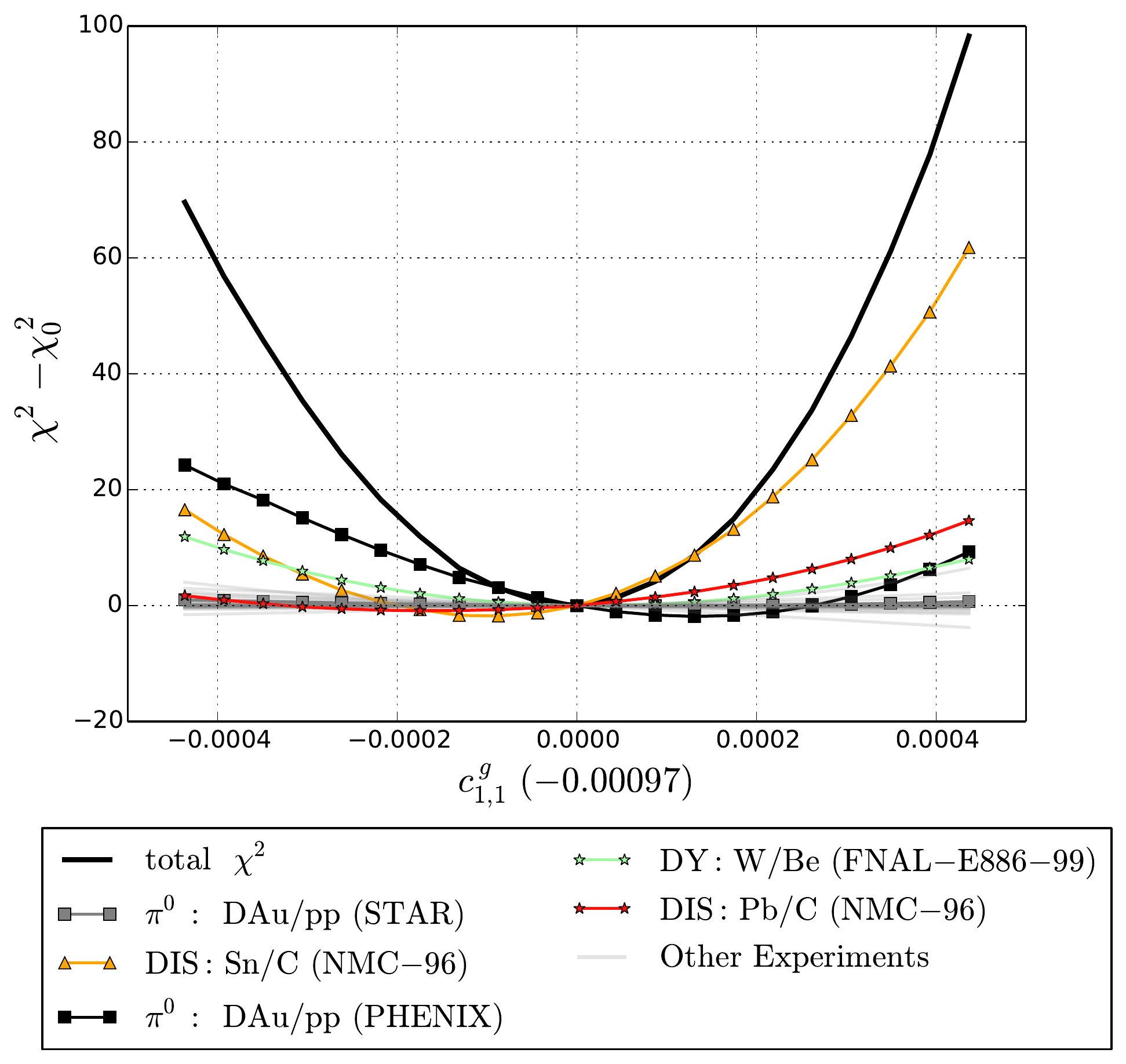}
\includegraphics[width=0.32\textwidth]{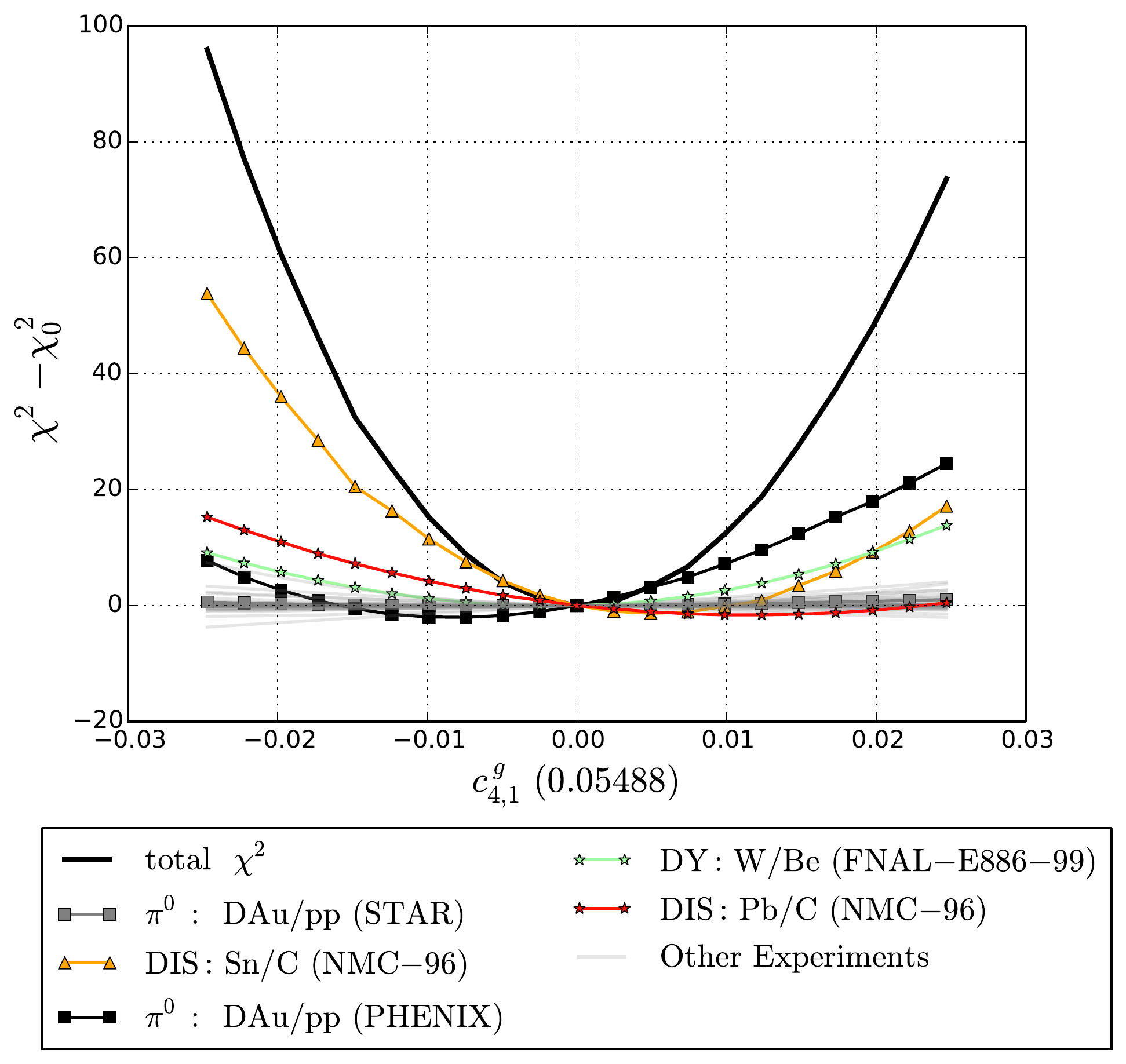}
\includegraphics[width=0.325\textwidth]{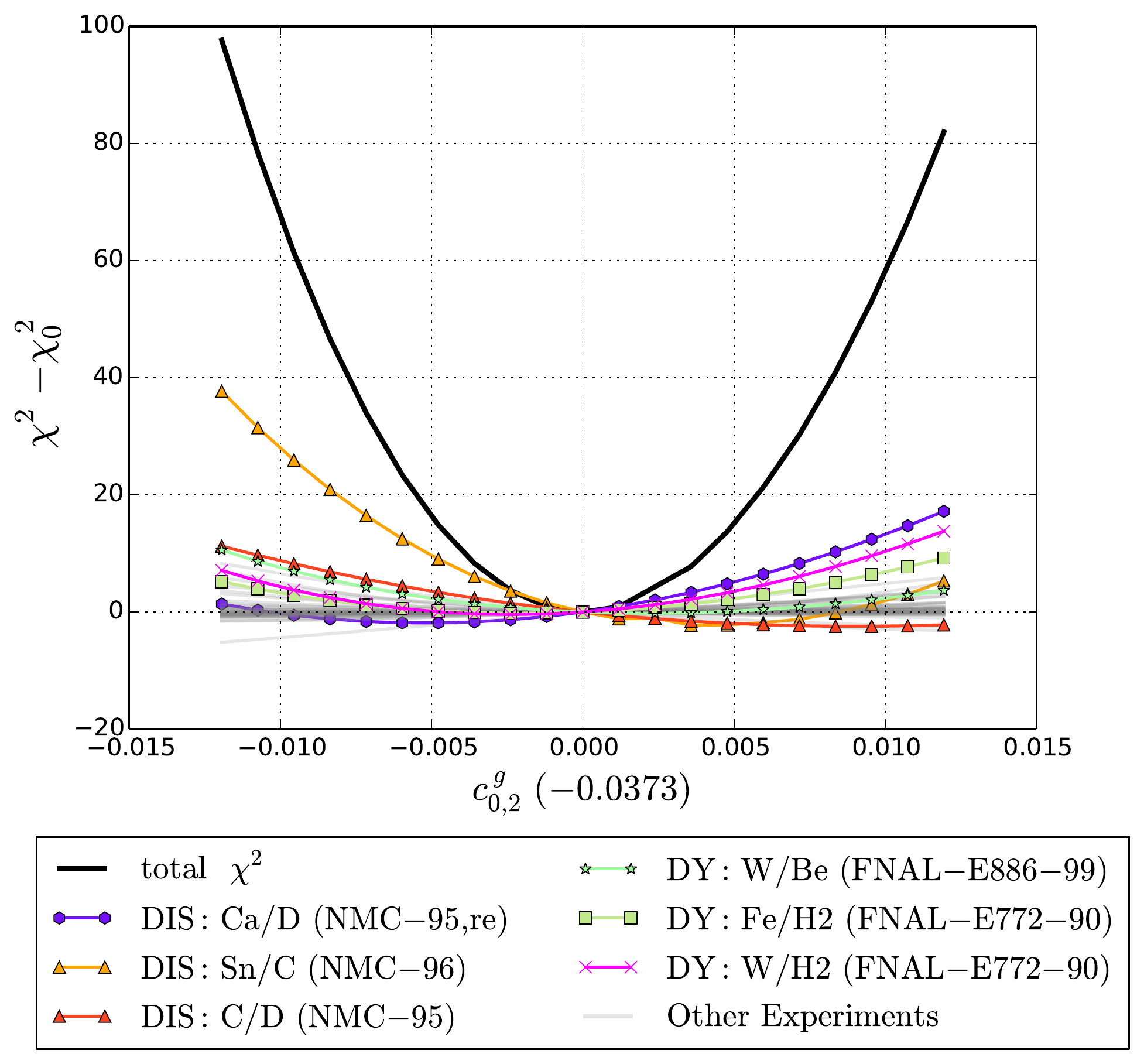}
\caption{
Contribution of different experiments to the total $\Delta\chi^2=\chi^2-\chi_0^2$
function (solid black line) for a  selection of gluon parameters 
(a) $c_{1,1}^g$, (b) $c_{4,1}^g$, (c) $c_{0,2}^g$.
On the $x$-axis we show the shift from the best fit value (indicated in the parenthesis, cf. Table \protect\ref{tab:params_nCTEQ15g}).
}
\label{fig:indivChi2}
\end{center}
\end{figure*}
%%%%%%%%%%%%%%%%%%%%%%%%%%%%%%%%%%%%%%%%%%%%%%

%%%%%%%%%%%%%%%%%%%%%%%%%%%%%%%%%%%%%%%%%%%%%%%%%%%%%%%%%%%%
%----------------------------
\subsection{Constraining the PDF flavors with data}
\label{subsec:constraints}
%----------------------------
%
Global analyses of PDFs necessarily include data from a wide variety of experiments which are 
differently sensitive to various PDF flavors. 
Examining the leading order expressions for DIS, DY, or $\pi$-production provides a simple estimate of 
which observable can constrain which PDF flavor combination. 
Additionally, we have to take into account the number of data points and their statistical and systematic uncertainties.
All of these factors contribute to the $\chi^2$-function; hence, 
we start with this measure to evaluate the impact of different experiments upon the PDF flavors. 

%%%%%%%%%%%%%%%%%%%%%%%%%%%%%%%%%%%%%%%%%%%%%%%%%%%%%%
\subsubsection{$\chi^2$ {\it vs.} the gluon parameters}
In Fig.~\ref{fig:indivChi2}, we compare the change of the global 
$\chi^2$ and the contributions from individual experiments to this change as a function of the shift of selected  gluon parameters 
\{$c_{1,1}^g$, $c_{4,1}^g$, $c_{0,2}^g$\} from the respective best fit values. 
Recall that the parameters \{$c_{1,1}^g$, $c_{4,1}^g$\} 
control the shape of the gluon PDF whereas \{$c_{0,2}^g$\} controls the $A$-dependence of the normalization.
The remaining gluon parameters behave in a similar manner as $c_{1,1}^g$ and $c_{4,1}^g$.

One feature that is immediately apparent is that the \ncteqfit\ minimum is not necessarily 
a minimum for all the experiments individually. 
For example, we see that the PHENIX experiment would prefer to shift $c_{1,1}^g$ to larger 
values ($\sim$0.002) while some of the DIS experiments (e.g., ID=5116, NMC-96 Pb/C) prefer a 
lower value for $c_{1,1}^g$  ($\sim$-0.002).
Therefore, 
the obtained 
fit is a  compromise that depends on the relative weight of the various data sets. 
This observation is part of the reason we consider a $\Delta\chi^2=1$ tolerance criterion impractical 
and choose $\Delta\chi^2=35$ (see Appendix~\ref{app:chidist}).
Moreover, for some experiments there may not even be a  
local minimum in the vicinity of the  \ncteqfit\ solution.
Thus, these figures highlight some of the tensions between the individual data sets
that the global fit must accommodate. 

On top of that, Fig.~\ref{fig:indivChi2} shows which experiments are most sensitive to the change of the underlying gluon parameters.
In turn the same experiments are the ones which have the largest impact when constraining the gluon PDF. Perhaps in contrast 
with expectations, the parameters analyzed in Fig.~\ref{fig:indivChi2} are mostly constrained by the NMC Sn/C data and data from 
several other DIS experiments. 
We also see that the inclusive pion production from PHENIX is sensitive to the gluon
shape parameters ($c_{1,1}^g$, $c_{4,1}^g$) but not to its normalization ($c_{0,2}^g$).
% We see that the inclusive pion production from PHENIX is also sensitive to the gluon parameters but is 
% by far not the leading contribution.

%%%%%%%%%%%%%%%%%%%%%%%%%%%%%%%%%%%%%%%%%%%%%%%%%%%%%%
%
\subsubsection{Correlations between data sets and PDFs}

Looking at the dependence of the $\chi^2$-function on only three gluon PDF parameters cannot give a complete picture and neither would 
inspecting the behavior for all gluon parameters because the momentum sum rule connect in fact all PDF flavors together. 
Therefore in the following we use different methods to study
the impact of individual experiments on different PDF flavors.

We introduce two quantities which will help us analyze the impact individual experiments have on constraining given PDF flavors.
The first quantity is the cosine of the correlation angle between two observables $X$ and $Y$
which was used in~\cite{Nadolsky:2008zw, Pumplin:2001ct} and can be defined as
\begin{widetext}
\begin{equation}
\label{eq:cosPhi}
\cos\phi[X,Y]
=   \frac{\sum_{i_{pdf}}\left(X_{i_{pdf}}^{(+)}-X_{i_{pdf}}^{(-)}\right)\left(Y_{i_{pdf}}^{(+)}-Y_{i_{pdf}}^{(-)}\right)}
{\sqrt{\sum_{i_{pdf}'}\left(X_{i_{pdf}'}^{(+)}-X_{i_{pdf}'}^{(-)}\right)^2}\sqrt{\sum_{i_{pdf}''}\left(Y_{i_{pdf}''}^{(+)}-Y_{i_{pdf}''}^{(-)}\right)^2}}\,,
\end{equation}
where the indices $i_{pdf}$ run over the 16 $z_{i_{pdf}}$ eigenvector directions.

In the following we will use the cosine of the correlation angle to investigate the correlations between 
the $\chi^2$ functions of the individual experiments and a single PDF. For example, in the case of the
gluon PDF the cosine of the correlation angle has the form
$\cos\phi[g(x,Q),\chi^2(j_{exp})]$.
This correlation cosine depends on $x$ and $Q$ through the gluon PDF, $g(x,Q)$, 
and on the particular experiment through $\chi^2(j_{exp})$.

Even though the cosine of the correlation angle is a useful quantity, it doesn't highlight the experiments with more data or smaller 
errors. It turns out that the normalization factors in Eq.~\eqref{eq:cosPhi} strongly reduce any sensitivity to the number of data points 
or to the size of the errors of an experimental data set. Therefore we introduce an alternate measure, the effective $\chi^2$ for an 
experiment $j_{exp}$, defined as 
\begin{equation}
\Delta\chi^2_{\text{eff}}(j_{exp},X) = \sum_{i_{pdf}}
  \frac{1}{2}\left(
                   \left|\chi^{2\;(+)}_{i_{pdf}}(j_{exp}) - \chi^{2\;(0)}_{i_{pdf}}(j_{exp})\right|
                 + \left|\chi^{2\;(-)}_{i_{pdf}}(j_{exp}) - \chi^{2\;(0)}_{i_{pdf}}(j_{exp})\right|
            \right)
  \left(\frac{X_{i_{pdf}}^{(+)}-X_{i_{pdf}}^{(-)}}
             {\sqrt{\sum\limits_{i_{pdf}'}\left(X_{i_{pdf}'}^{(+)}-X_{i_{pdf}'}^{(-)}\right)^2}}
  \right)^2 .
\label{eq:chi2Eff_g_chi2_eSTARnor11}
\end{equation}

\end{widetext}
As before, the  index $i_{pdf}$ runs over the 16 $z_i$ eigenvector directions. 

$\Delta\chi^2_{\text{eff}}$ is positive definite and comparing the definitions \eqref{eq:cosPhi} and 
\eqref{eq:chi2Eff_g_chi2_eSTARnor11} it is missing the normalization factor for the $\chi^2$ function which allows it
to be more sensitive to experiments with more data or smaller errors, i.e., experiments which have a larger impact in constraining 
single PDF flavors.

%--------------
\begin{figure*}[th]
\begin{center}
%%%%%%%%%%%%%%% gluon lead
\subfloat[\textbf{gluon}: $\Delta\chi^2_{\text{eff}}(g)$ at $Q=10$ GeV for lead]
{
\includegraphics[width=0.41\textwidth]{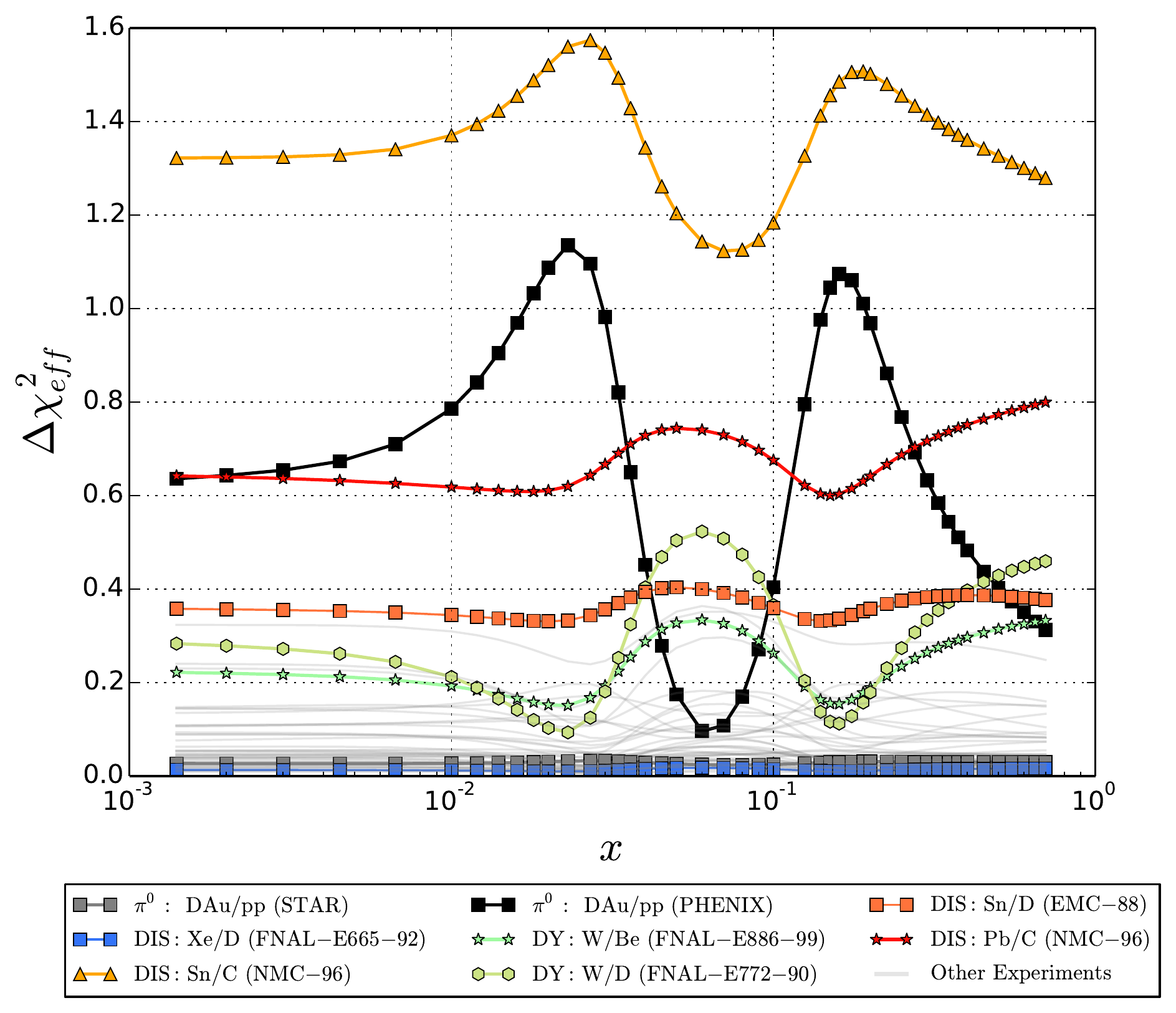}
\label{subfig:g-chi2eff_lead}
}
\qquad
\subfloat[\textbf{gluon}: $\cos\phi(g,\chi^2)$ at $Q=10$ GeV for lead]
{
\includegraphics[width=0.42\textwidth]{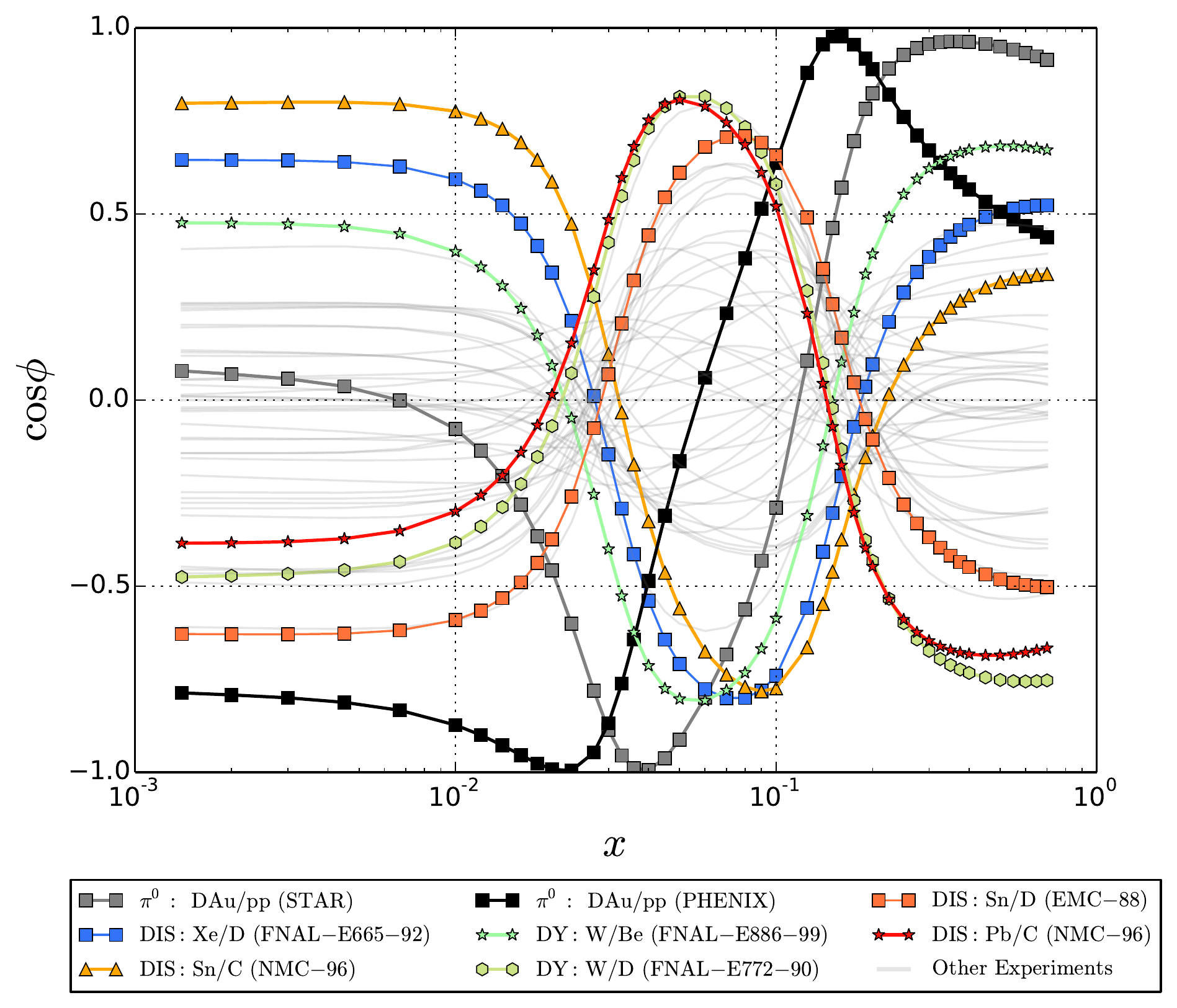}
\label{subfig:g-cosphi_lead}
}
\newline
%%%%%%%%%%%%%%% gluon carbon 10GeV
\subfloat[\textbf{gluon}: $\Delta\chi^2_{\text{eff}}(g)$ at $Q=10$ GeV for carbon]
{
\includegraphics[width=0.42\textwidth]{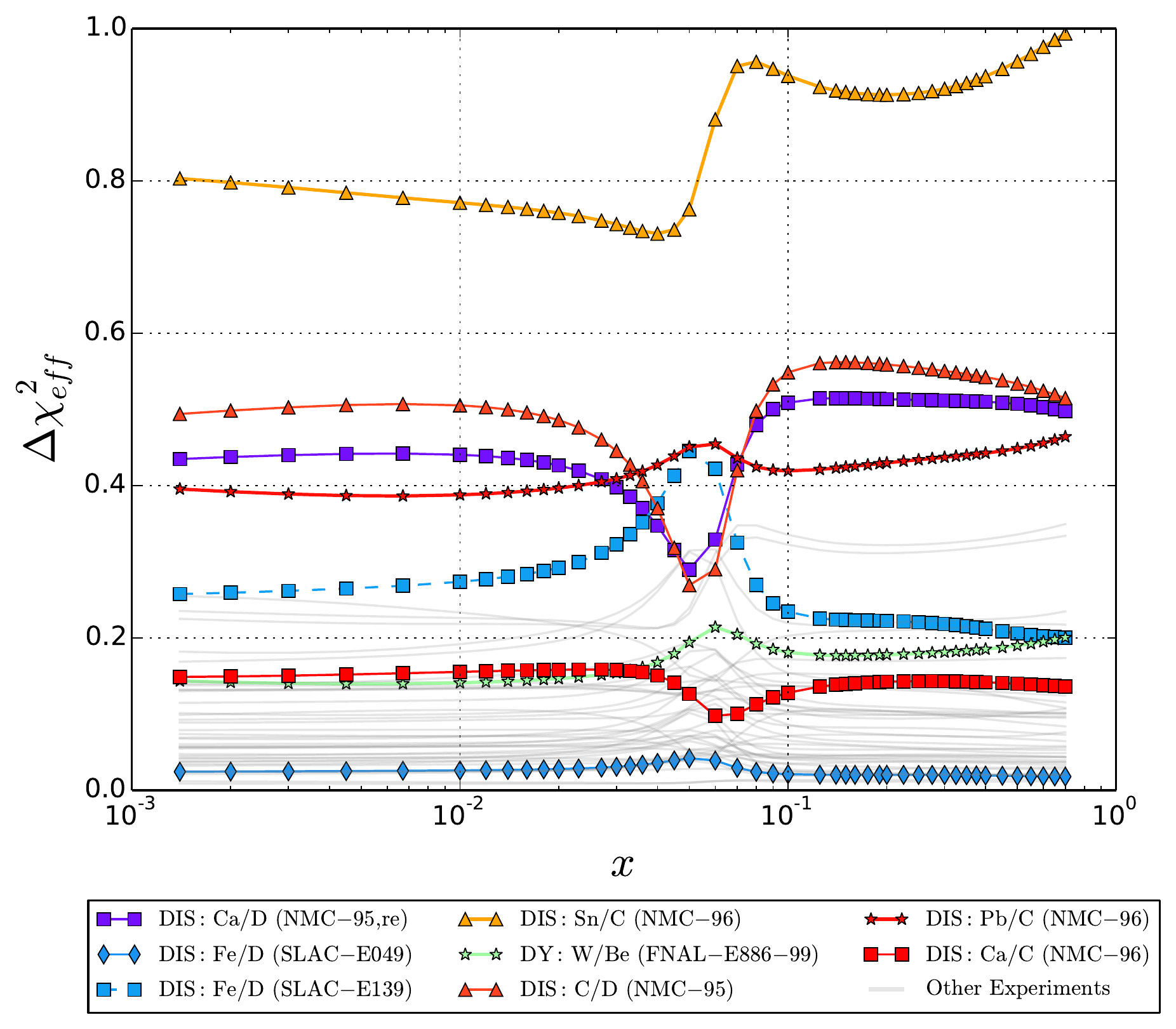}
\label{subfig:g-chi2eff_carbon}
}
\qquad
\subfloat[\textbf{gluon}: $\cos\phi(g,\chi^2)$ at $Q=10$ GeV for carbon]
{
\includegraphics[width=0.42\textwidth]{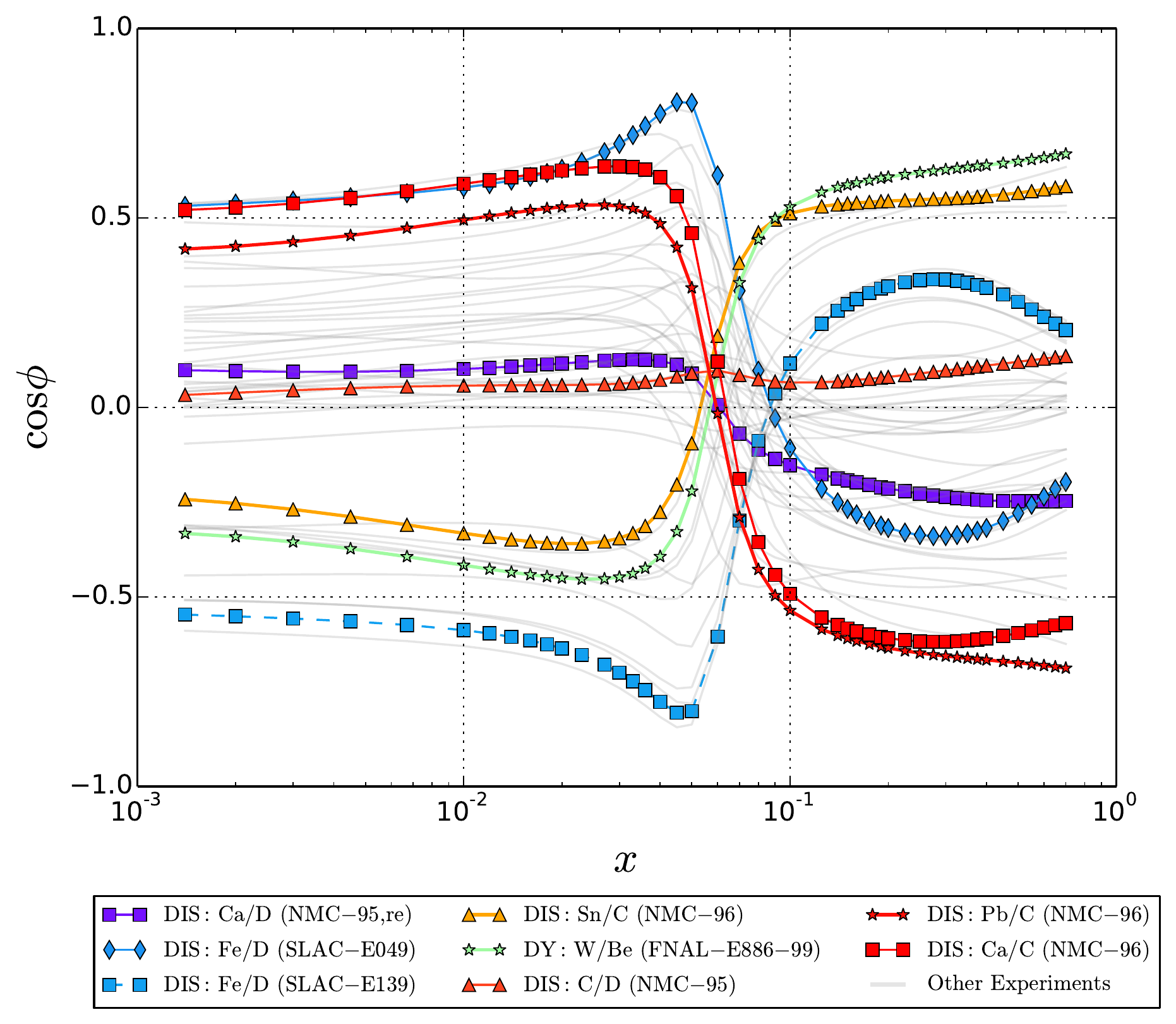}
\label{subfig:g-cosphi_carbon}
}
\newline
%%%%%%%%%%%%%
\caption{
Correlation measures for lead and carbon at $Q=10$ GeV 
for the gluon of the \ncteqfit\ fit.
The left panels display the  effective $\chi^2$ 
and the right panels display the  correlation cosine
as  a function of $x$.
}
\label{fig:correlationsGlue}
\end{center}
\end{figure*}
%--------------
%%%%%%%%%%%%%%%%%%%%%%%%%%%%%%%%%%%%%%%%%%%%%%

%%%%%%%%%%%%%%%%%%%%%%%%%%%%%%%%%%%%%%%%%%%%%%
%--------------
\begin{figure*}[th]
\begin{center}
%%%%%%%%%%%%%%% u-quark
%
\subfloat[\textbf{$u$-quark}: $\Delta\chi^2_{\text{eff}}(u)$ at $Q=10$ GeV for lead]
{
\includegraphics[width=0.42\textwidth]{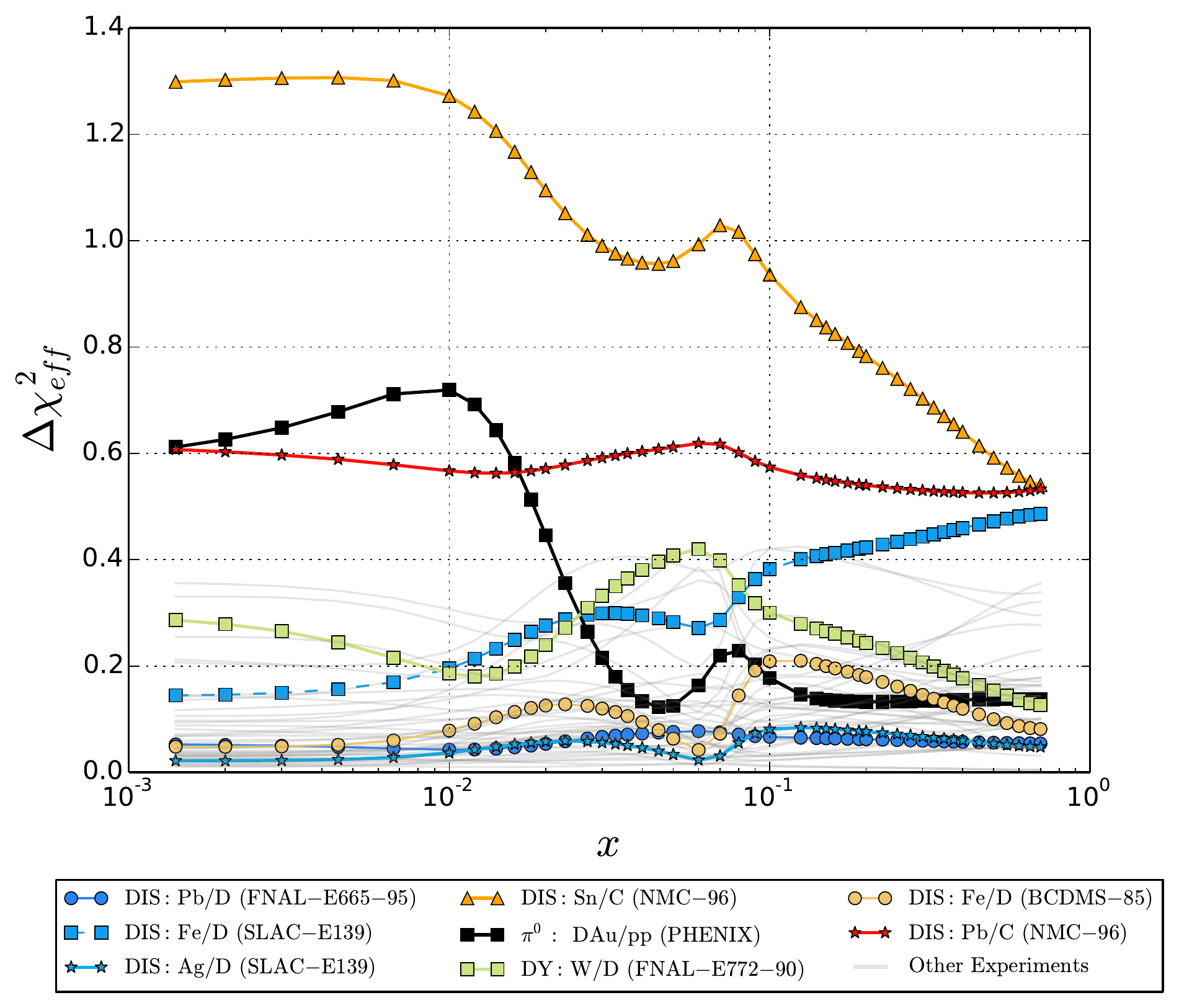}
\label{subfig:uval-chi2eff}
}
\qquad
\subfloat[\textbf{$u$-quark}: $\cos\phi(u,\chi^2)$ at $Q=10$ GeV for lead]
{
\includegraphics[width=0.42\textwidth]{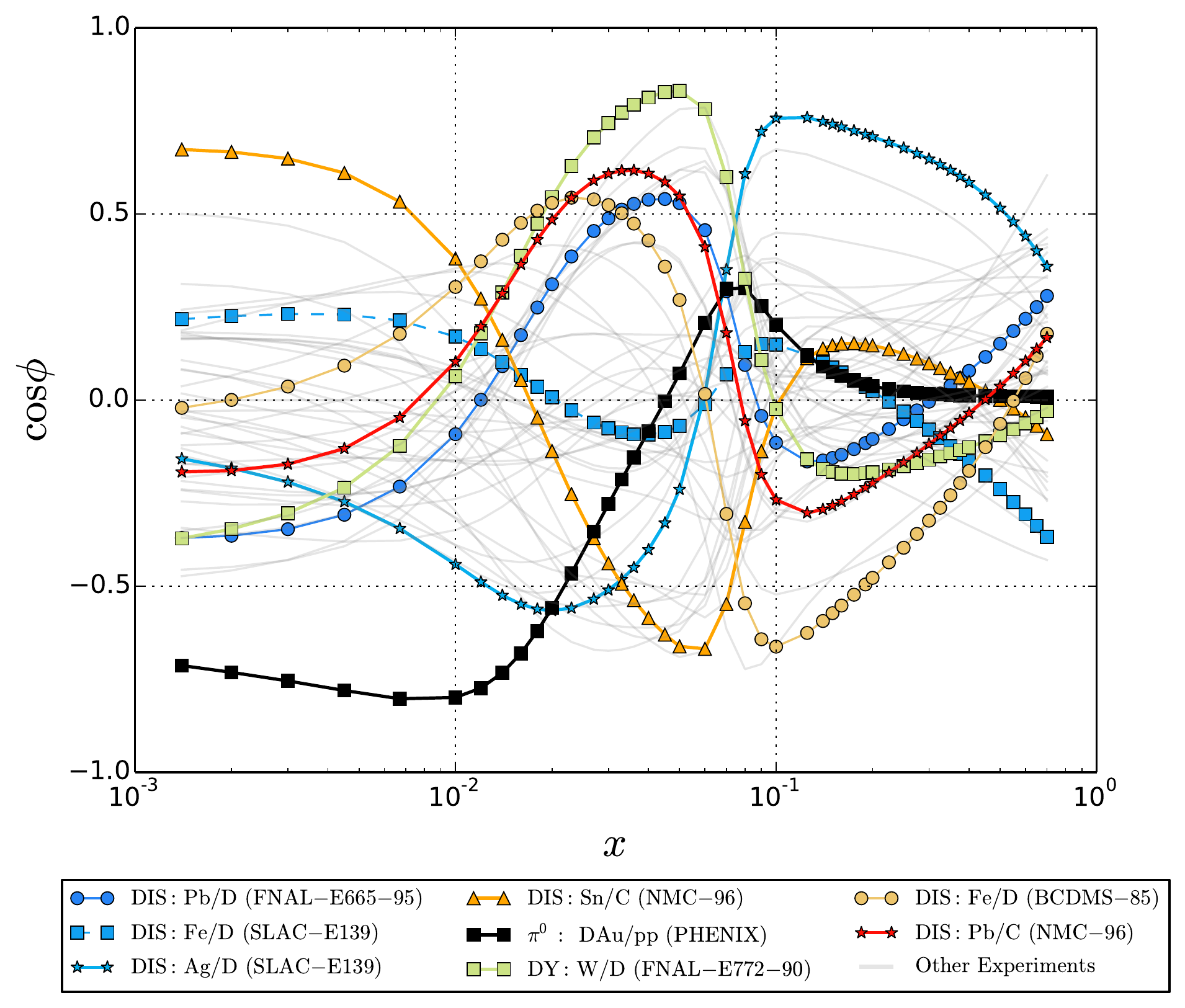}
\label{subfig:uval-cosphi}
}
\newline
%%%%%%%%%%%%%%% d-quark
%
\subfloat[\textbf{$d$-quark}: $\Delta\chi^2_{\text{eff}}(d)$ at $Q=10$ GeV for lead]
{
\includegraphics[width=0.42\textwidth]{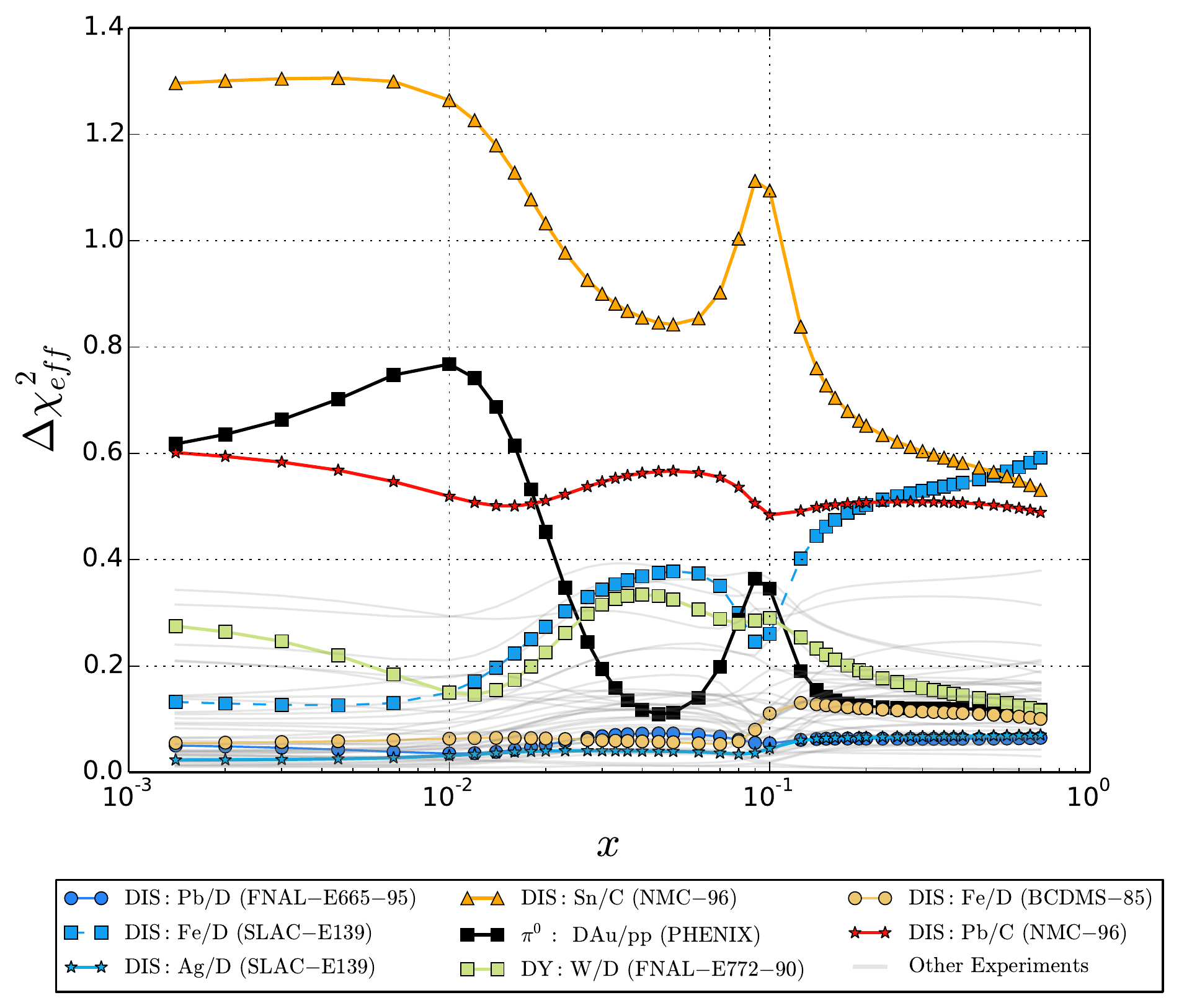}
\label{subfig:dval-chi2eff}
}
\qquad
\subfloat[\textbf{$d$-quark}: $\cos\phi(d,\chi^2)$ at $Q=10$ GeV for lead]
{
\includegraphics[width=0.42\textwidth]{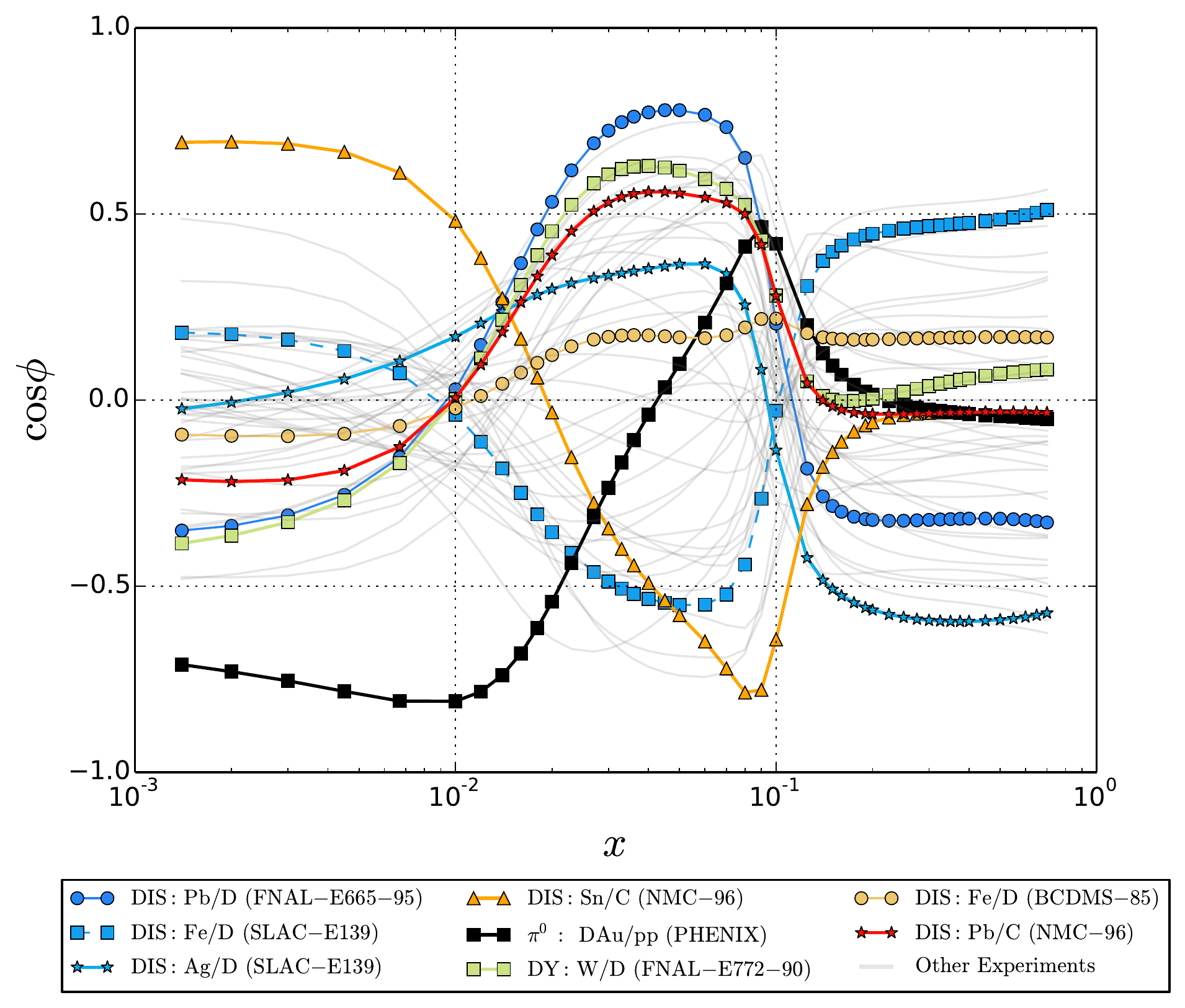}
\label{subfig:dval-cosphi}
}
\caption{
Correlation measures for lead at $Q=10$ GeV 
for the  $u$-quark and $d$-quark distributions
 of the \ncteqfit\ fit.
The left panels display the  effective $\chi^2$ 
and the right panels display the  correlation cosine
as  a function of $x$.
}
\label{fig:correlationsVal}
\end{center}
\end{figure*}
%--------------
%%%%%%%%%%%%%%%%%%%%%%%%%%%%%%%%%%%%%%%%%%%%%%

In Figs.~\ref{fig:correlationsGlue} 
and \ref{fig:correlationsVal} 
we display both the   $\Delta\chi^2_{\text{eff}}$
and correlation cosine 
as a function of $x$.
These plots do not exhibit a strong $Q$ dependence,
so we only display them for one value of $Q=10~\GeV$.

%%%%%%%%%%%%%%%%%%%%%%%%%%%%%%%%%%
%%%%%%%%%%%%%%%%%%%%%%%%%%%%%%%%%%
%%%%%%%%%%%%%%%%%%%%%%%%%%%%%%%%%%

%%%%%%%%%%%%%%%%%%%%%%%%%%%%%%%%%%
%GLUON: LEAD: EFF-CHI2

We now examine the $\Delta\chi^2_{\text{eff}}$ results for the gluon PDF in lead ($A$=207) as show in
Fig.~\ref{subfig:g-chi2eff_lead}. 
For readability, we primarily show the data sets which have the  largest impact on 
$\Delta\chi^2_{\text{eff}}$; these are generally the data sets which involve the 
heaviest targets.
The strong influence of the DIS  Sn/C set reflects a combination of the large $Q^2$ coverage of the 
data and the small errors. 
The DIS Pb/C data, and to a lesser extent  the DIS Sn/D data, also provide constraints for the 
gluon PDF in lead. 
The PHENIX pion production data contributes strongly in the central $x$ region;
conversely, the effect of the STAR data is negligible due to the larger uncertainties.
Additionally,  the DY data on heavy targets 
(W tungsten with Be and D) 
also play a role in determining the gluon lead PDF;
this is due to the fact that the DY data cover a range $\sim(20,170)$ GeV$^2$ in the invariant 
mass of the muon pair, which creates some sensitivity to the gluon PDF via scale evolution.

%%%%%%%%%%%%%%%%%%%%%%%%%%%%%%%%%%
%GLUON: LEAD: CORRELATION COSINE

In Fig.~\ref{subfig:g-cosphi_lead}  we show the correlation cosine for the gluon PDF in lead.
The DIS Sn/C and DY W/Be data sets have positive correlations at large and small $x$, and 
a negative dip in the middle. 
Contrary to this, 
the DIS Pb/C, Sn/D and DY W/D data sets have  the opposite behavior.
Hence, these data sets are anti-correlated which indicates
that they pull against each other in the fit. 
This is precisely what we have observed in Fig.~\ref{fig:indivChi2} for the gluon parameters.
Also, the PHENIX data have a separate $x$-dependence (arising from a separate production mechanism),
and this will further help us separate the PDF flavor components. 

Finally, there are two data sets (STAR and DIS Xe/D) 
that have relatively large correlation cosines, 
but do not have a large influence on the $\Delta\chi^2_{\text{eff}}$ of
Fig.~\ref{subfig:g-chi2eff_lead}; thus, we need to take care when interpreting the results of 
the correlation cosine plots and use this in combination with  $\Delta\chi^2_{\text{eff}}$.

%%%%%%%%%%%%%%%%%%%%%%%%%%%%%%%%%%
%GLUON: CARBON: EFF CHI2: 

We now consider the gluon PDF in carbon  ($A$=12) to see if the general observations above 
apply in the case of a lighter nuclei. 
In Fig.~\ref{subfig:g-chi2eff_carbon} we see the primary data sets constraining 
 $\Delta\chi^2_{\text{eff}}$ are the DIS sets involving ratios of carbon (Sn/C, C/D, Pb/C)
or other comparable nuclei (Ca/D). 
Note the DY data on heavy tungsten (W) and the pion production data on gold (Au) 
are not shown as they do
not contribute significantly to  $\Delta\chi^2_{\text{eff}}$ for carbon.

%%%%%%%%%%%%%%%%%%%%%%%%%%%%%%%%%%
%GLUON: CARBON: CORRELATION: 

The correlation cosines for the gluon PDF in carbon are shown in
 Figs.~\ref{subfig:g-cosphi_carbon}.
We see the DIS Pb/C data have a positive correlation cosine at small $x$ and 
a negative correlation cosine at large $x$.
The DIS Sn/C data shows the opposite behavior; hence, these data sets will 
pull against each other in the fit. 
The DIS C/D and Ca/D data generally have a small correlation cosine throughout the  
$x$ range.
As in the case of the gluon in lead, we see there are a number of data sets 
(such as DIS Fe/D) 
that have a large correlation cosine but yield a small contribution 
to the $\Delta\chi^2_{\text{eff}}$; thus, we need to use 
both the  $\Delta\chi^2_{\text{eff}}$ and $\cos\phi$ information together when drawing our conclusions.

%%%%%%%%%%%%%%%%%%%%%%%%%%%%%%%%%%
%U/D: LEAD

We now turn our attention to 
 $u^{p/Pb}$ and $d^{p/Pb}$ distributions for lead at $Q=10~\GeV$ 
as shown in  Fig.~\ref{fig:correlationsVal}.
For these PDFs, not only is the $Q$-dependence rather mild, 
but the differences between heavy and light nuclei are also not 
as pronounced as in the gluon case.
The $\Delta\chi^2_{\text{eff}}$ for the  $u$ and $d$ PDFs depends almost 
exclusively on the DIS data from heavy targets (Sn/C, Pb/C, Fe/D), 
with some contributions from PHENIX pion production data at small-$x$, and 
a minimal contribution from the DY W/D data.

Turning to the $\cos\phi$ plots, we see the 
DIS Sn/C and the Fe/D data start with a positive correlation cosine at small $x$ and 
moves negative for increasing $x$, while the DIS Pb/C and the DY W/D data do the opposite;
hence, these sets are anti-correlated in this region. 
For small to medium $x$ values, the general pattern is similar between the $u$ and $d$ 
correlation plots, but they differ some at large $x$ where we see, for example, the 
DIS Fe/D data has a positive correlation cosine for $u$ but a negative one for $d$;
this will be useful in differentiating $u$ and $d$ PDFs at large $x$. 
As with the gluon correlation plots, there are  a number of data sets 
(such as the DIS Ag/D) which have large correlation cosines but small contributions to $\Delta\chi^2_{\text{eff}}$;
thus, they have minimal effect constraining the PDFs.

%%%%%%%%%%%%%%%%%%%%%%%%%%%%%
\subsection{Comparison with different global analyses}
\label{sec:other_sets}
%%%%%%%%%%%%%%%%%%%%%%%%%%%%%

%
%%%%%%%%%%%%%%%%%%%%%%%%%%%%%%%%%%%%%%%%%%%%%%%%%%%%%
% nCTEQ15g vs. EPS vs. DSSZ vs. HKN @ Q=2.0GeV
%----------------
\begin{figure*}[th!]
\centering{}
\includegraphics[clip,width=0.48\textwidth]{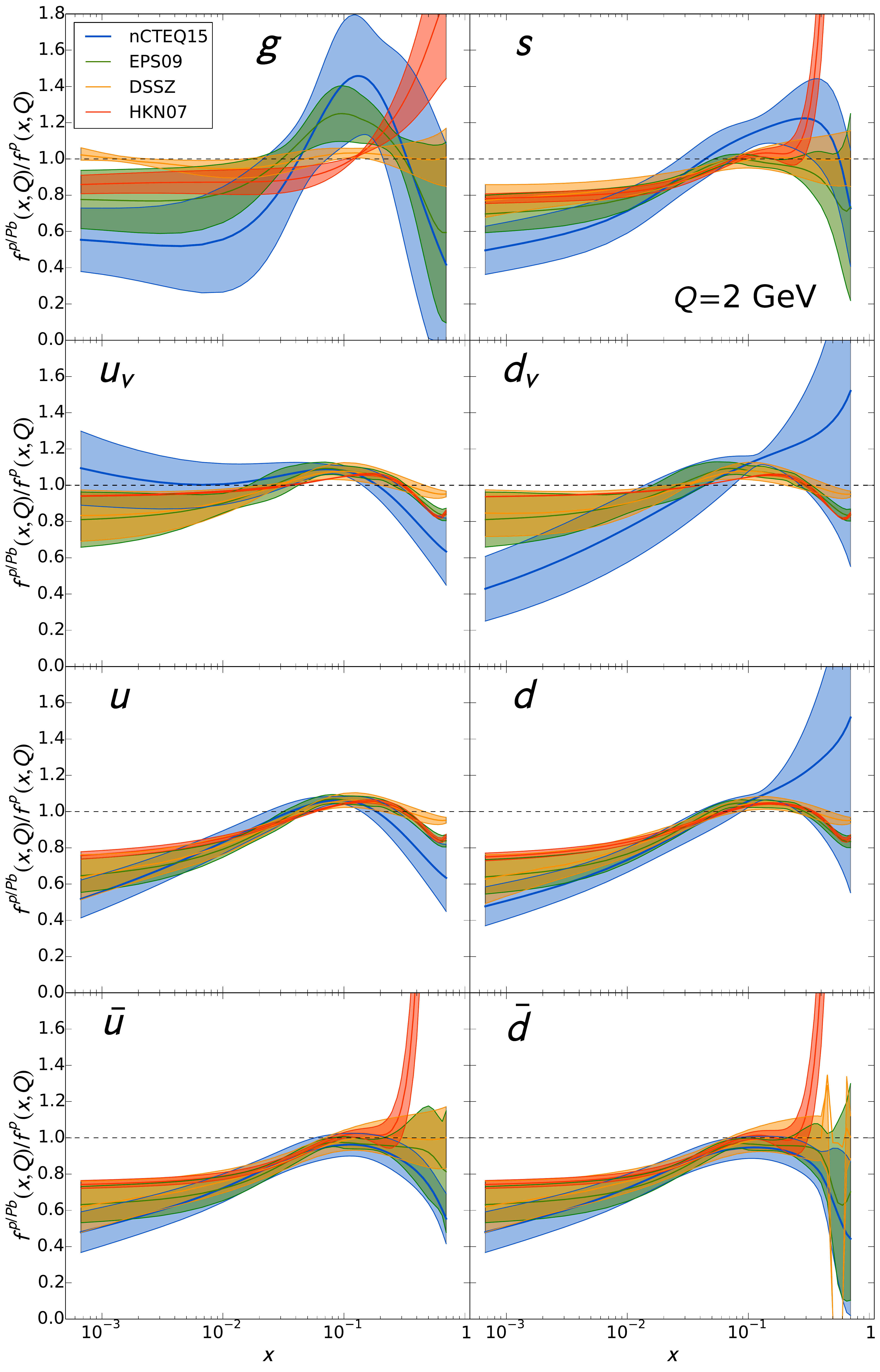}
\quad{}
\includegraphics[width=0.48\textwidth]{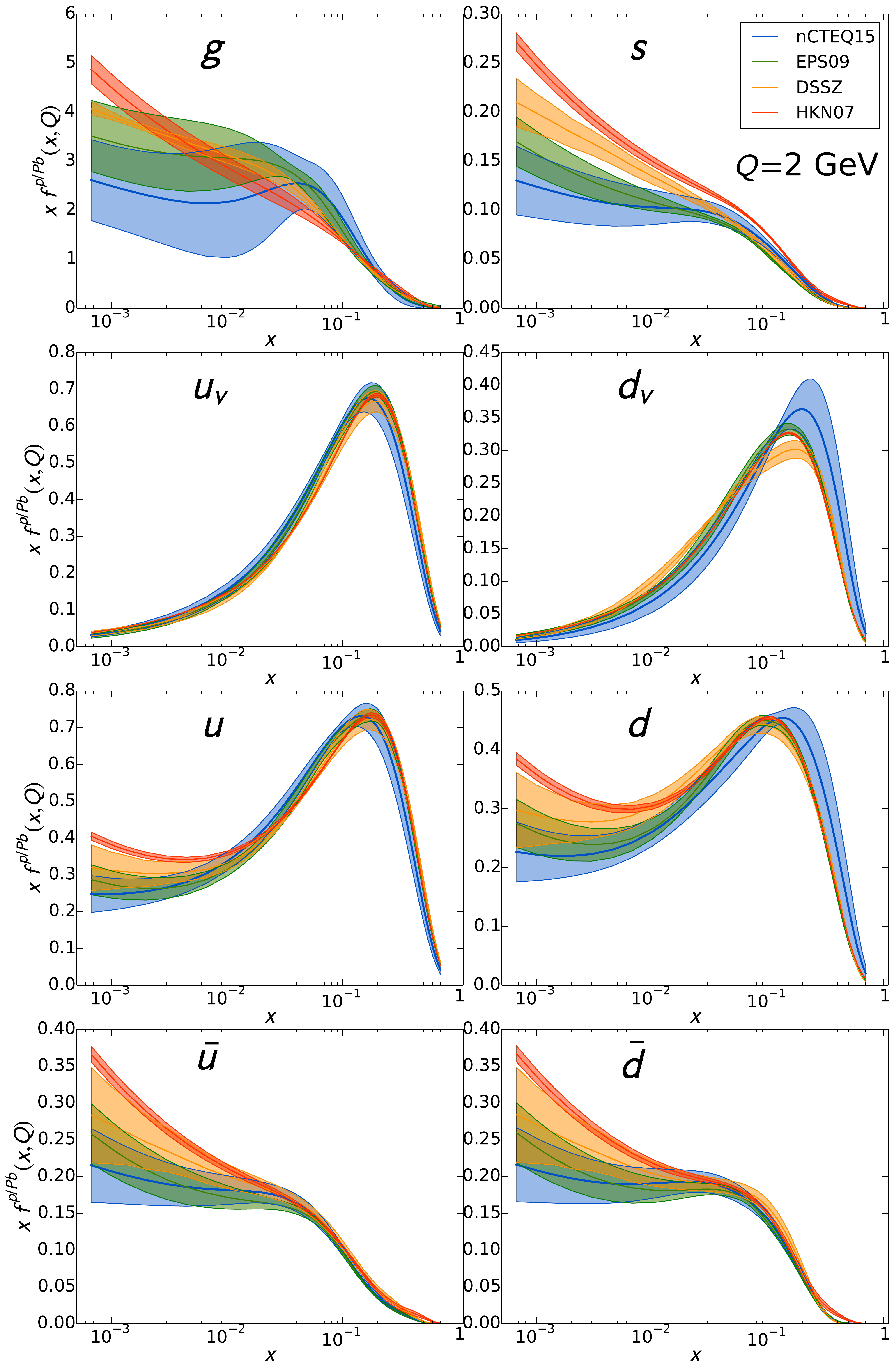}
\caption{Comparison of  the \ncteqfit\ fit (blue)
with results from other groups:
EPS09~\cite{Eskola:2009uj} (green),
DSSZ~\cite{deFlorian:2011fp} (orange),
HKN07~\cite{Hirai:2007sx} (red).
The left panel shows nuclear modification factors for lead,
 and the right panel the actual
PDFs of a proton bound in \emph{lead}. 
The scale is $Q=2~\GeV$.
The wide spread of the ratios at large $x$ are an unphysical artifact
due to the vanishing of the PDFs in this region.}
\label{fig:compar_PDFs_Q1}
\end{figure*}
%%%%%%%%%%%%%%%%%%%%%%%%%%%%%%%%%%%%%%%%%%%%%%%%%%%%%

%%%%%%%%%%%%%%%%%%%%%%%%%%%%%%%%%%%%%%%%%%%%%%%%%%%%%
% nCTEQ15g vs. EPS vs. DSSZ vs. HKN @ Q=10GeV
%----------------
\begin{figure*}[th]
\centering{}
\includegraphics[clip,width=0.48\textwidth]{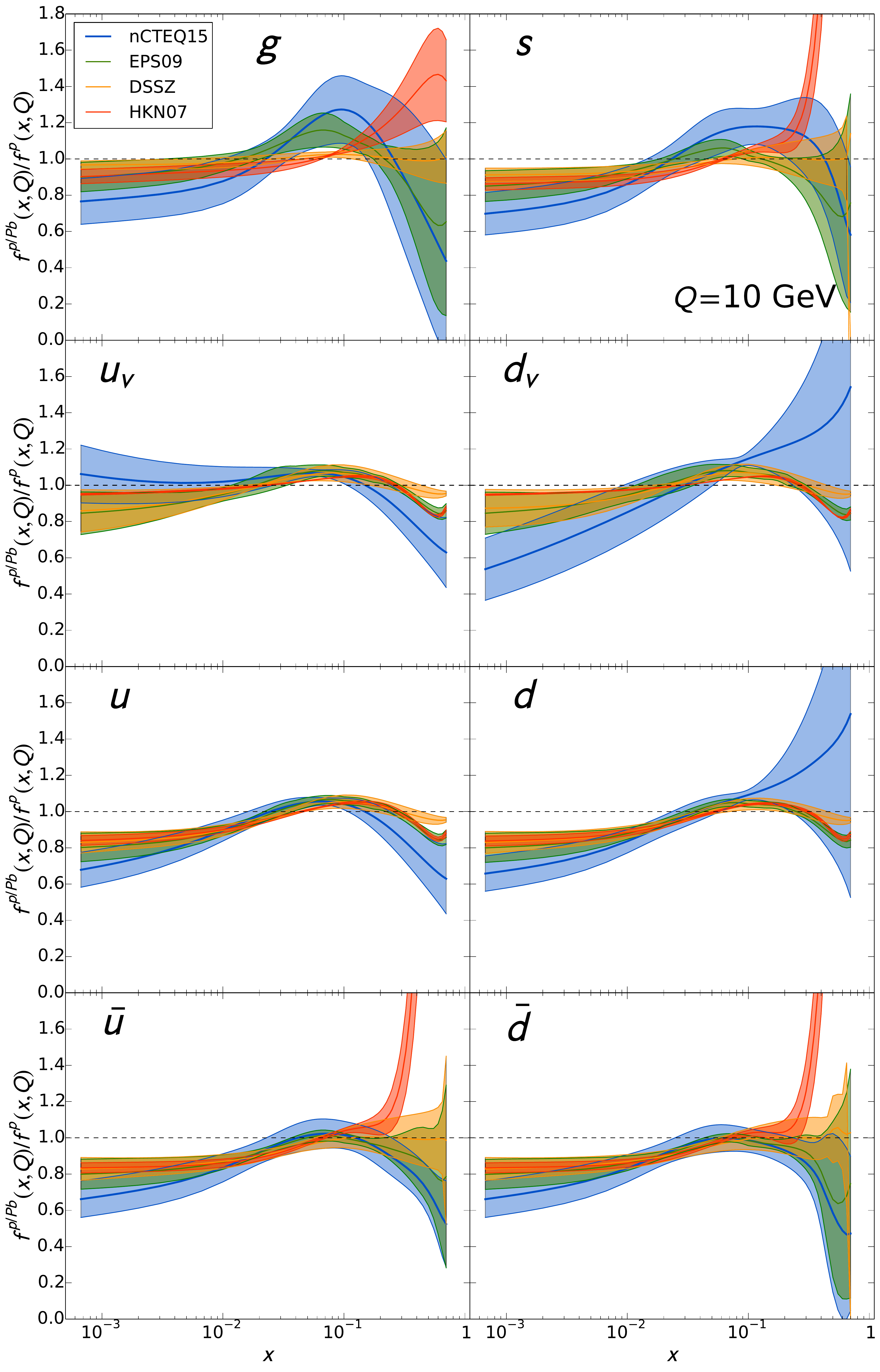}
\quad{}
\includegraphics[width=0.48\textwidth]{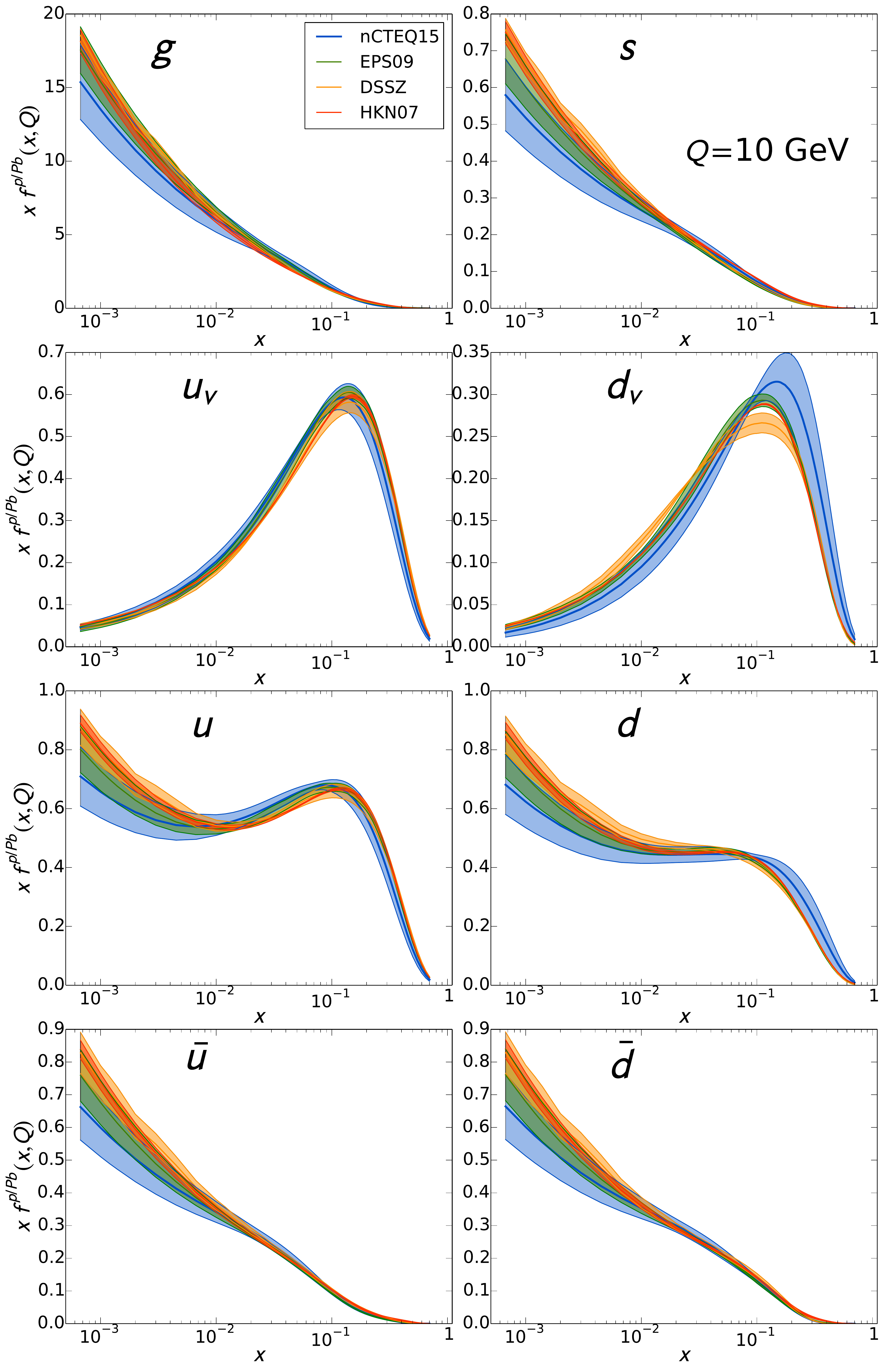}
\caption{
Same  as Fig.~\protect\ref{fig:compar_PDFs_Q1}, with $Q= 10~\GeV$.}
\label{fig:compar_PDFs_Q10}
\end{figure*}
%%%%%%%%%%%%%%%%%%%%%%%%%%%%%%%%%%%%%%%%%%%%%%%%%%%%%
%
%
We now compare our \ncteqfit\ PDFs with other recent 
nuclear parton distributions in the literature. 
Specifically, we will consider 
DSSZ~\cite{deFlorian:2011fp},
EPS09~\cite{Eskola:2009uj},
and
HKN07~\cite{Hirai:2007sx}.%
    \footnote{Note that there is also a very recent global nPDF analysis performed
    at NNLO level~\cite{Khanpour:2016pph}.}
Our data set selection and technical aspects of our analysis are
closest to that of EPS09.
In Figs.~\ref{fig:compar_PDFs_Q1} and \ref{fig:compar_PDFs_Q10},
we plot nuclear
modifications for the PDFs of a proton bound in lead, $f^{p/Pb}/f^p$ (left),
as well as the bound proton PDFs themselves, $f^{p/Pb}$ (right),
for different flavors for a selection of $Q$ scales.

%%%%%%%%%%%%%%%%%%%%%%%%%%%%%%%%%%
%U-BAR, D-BAR, AND S

For the $\bar{u}$ and $\bar{d}$ PDFs at $Q=2~\GeV$, \ncteqfit\ has significant overlap with 
the other sets through much of the $x$ range with a stronger shadowing  at small $x$. 
Our results at $x<10^{-2}$ are extrapolated since they are not constrained by data due to the cut $Q>2$ GeV 
which was imposed in order to reduce higher twist contributions.
Therefore, it is likely that the uncertainty band at $x<10^{-2}$ underestimates the true PDF uncertainties.
While this trend repeats itself for the strange quark PDF, the spread at small $x$ is 
slightly increased.%
\footnote{%
In this analysis the $s$-quark nuclear effects are completely determined by the $\bar{u}$ and $\bar{d}$ 
nuclear PDFs and by the gluon nuclear PDF through evolution. Due to these constraints the error of
the $s$-quark nuclear PDF is underestimated. A comprehensive analysis would require including 
the charged-current $\nu$-DIS data as in \cite{Kovarik:2010uv} along with using a proton PDF baseline
where the strange distribution was determined from different data such as the $W+c$ production at the LHC.
}
In fact, at $Q=2$ GeV the small $x$ behavior of the strange PDF of all four fits is quite distinct
with little overlap between the uncertainty bands (see Fig.~\ref{fig:compar_PDFs_Q1}b).
As  we move to higher $Q$ values, the DGLAP evolution tends to bring the various PDF sets into 
closer agreement, particularly at small $x$ values. 
For example, already at $Q=10~\GeV$ the  \ncteqfit\ bands overlap the other PDFs across a much broader $x$ range than at 
low $Q$ values. 

%%%%%%%%%%%%%%%%%%%%%%%%%%%%%%%%%%
%GLUON 

In the case of the gluon, there is considerable variation among the different PDF sets at $Q$ close to the initial scale. 
Again, the \ncteqfit\ exhibits a stronger shadowing suppression along with a
larger enhancement in the anti-shadowing region ($x \sim 0.1$).
In addition, the uncertainty band for $x\gtrsim 0.02$ is considerably larger than
the uncertainty bands of the other groups. 
The \ncteqfit\ result is largely compatible with the result of EPS09 even though the shape of the central prediction
is more suppressed in the shadowing region and enhanced in the anti-shadowing region. 
We have less overlap with the  HKN07 and DSSZ  bands, in part, due to their smaller uncertainty bands. 
Moving to larger $Q$ values, the DGLAP evolution again causes the different PDFs to converge,.

\begin{figure*}[th]
\centering{}
\includegraphics[clip,width=0.9\textwidth]{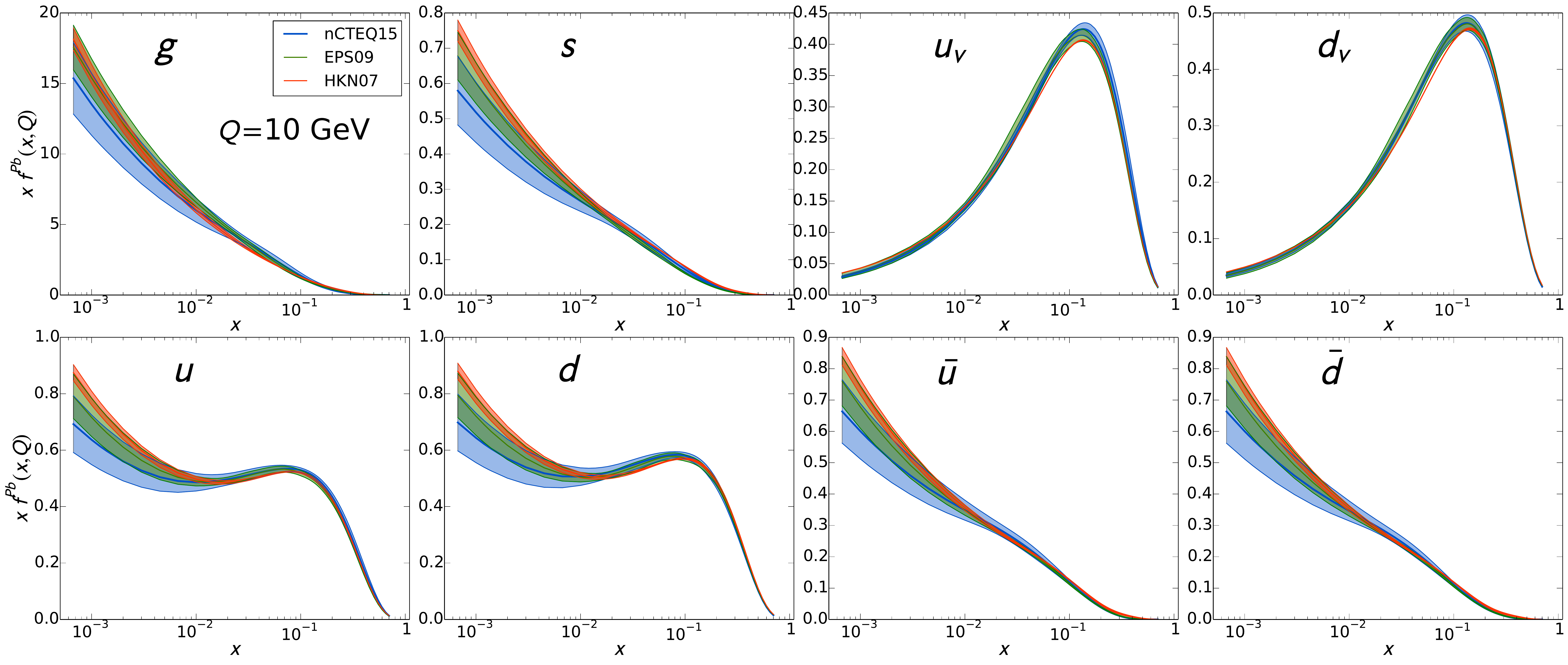}
\\
\includegraphics[width=0.9\textwidth]{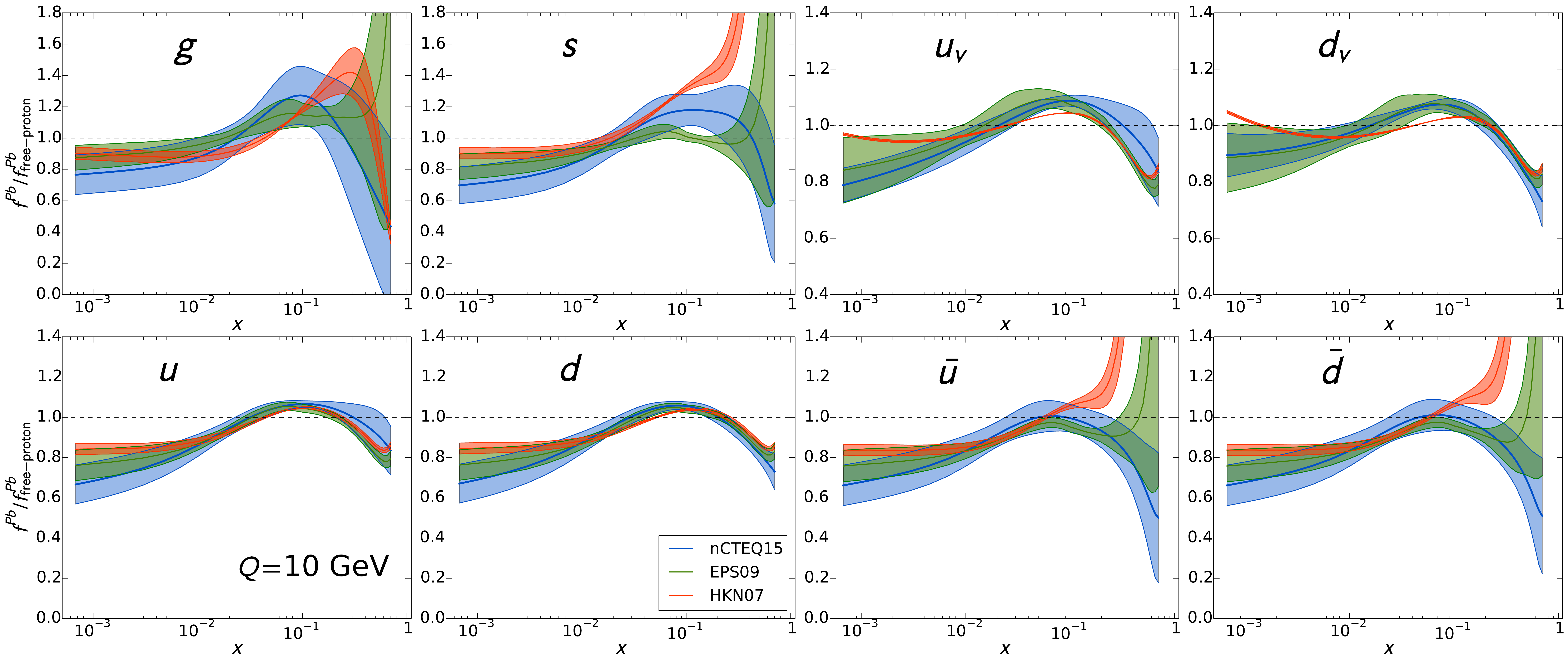}
\caption{
(upper panel) Comparison of the full nuclear lead distributions,
$f^{Pb} = \frac{82}{207}f^{p/Pb} + \frac{207-82}{207}f^{n/Pb}$,
for \ncteqfit\ (blue), EPS09 (green) and HKN07 (red) at $Q= 10~\GeV$.
Lower panel shows the same distributions compared to the lead PDF, $f^{Pb}$,
constructed of free proton distributions.
The wide spread of the ratios at large $x$ are an unphysical artifact
due to the vanishing of the PDFs in this region.}
\label{fig:compar_FullPDFs_Q10}
\end{figure*}

Note that  the ratio plots of Figs.~\ref{fig:compar_PDFs_Q1} and \ref{fig:compar_PDFs_Q10} 
have quite a wide spread at large $x$ values. 
This unphysical behavior is an artifact due to the vanishing of the PDFs in this region. 
The spread is largest for those PDFs with minimal support at  large $x$--specifically
$g,s, \bar{u},\bar{d}$.
Also, these effects are reduced when we construct the full nuclear lead distribution 
as shown in Fig.~\ref{fig:compar_FullPDFs_Q10}. 
%

%%%%%%%%%%%%%%%%%%%%%%%%%%%%%%%%%%
%U-VAL AND D-VAL 
%
Examining the $u$- and $d$-valence distributions, 
one can see that  PDF sets \{HKN07, EPS09, DSSZ\} 
agree quite closely with each other throughout the $x$ range.
While the \ncteqfit\ fit uncertainty bands generally overlap the other
sets, we see on average the $u_v$ distribution is softer and the $d_v$
distribution is harder.
These differences reflect the  fact that the  HKN07, EPS09, and DSSZ fits assume that the nuclear
corrections $R_{u_v}$ and $R_{d_v}$ are the same, while the \ncteqfit\ fit allows 
them to vary independently.
Clearly, there is no physical reason to assume that $u_v$ and $d_v$ must have a universal nuclear
correction factor, and there exist models in the literature~\cite{Cloet:2009qs,Malace:2014uea,Dutta:2010pg}
which indeed predict non-universal modifications. 

The obvious question is whether the additional freedom to decouple the
$R_{u_v}$ and $R_{d_v}$ nuclear corrections yields a substantial
improvement in the fit.
To shed more light on this issue, we have generated a modified fit
where we have forced the $u_v$ and $d_v$ nuclear
corrections to be similar to the EPS09 PDF set.%
    \footnote{As we are fitting directly the nuclear PDFs 
    $f^{p/A}(x,Q)$ and not  the ratios $f^{p/A}(x,Q)/f^{p}(x,Q)$, 
    it is non-trivial to force  the nuclear corrections to be exactly
    the same if the underlying proton PDFs differ.
    We are able to find an approximate solution by equating the
    $u_v$ and $d_v$ coefficients $c_{i,j}$ for 
    $\{ij\}=\{11,12,21,22,31,32,51,52\}$ and refitting the PDFs.}
We find that the $\chi^2$/dof for this modified fit is comparable ($\Delta\chi^2\lsim 5$) to our 
original \ncteqfit\ at a level well below our tolerance criteria of $\Delta\chi^2=35$.
Therefore, we conclude that the current data sets are not sufficiently sensitive to 
distinguish the $u_v$ and $d_v$ nuclear corrections to a good degree. 
Hence, the advantage  of independent $R_{u_v}$ and $R_{d_v}$ correction factors 
is currently limited, which however will change with more data 
(e.g. from the LHC).\footnote{In an earlier study we did find
    an {\it apparent} difference due to independent $R_{u_v}$ and $R_{d_v}$ nuclear corrections.
    The present updated analysis additionally includes: 
    i) an improved treatment of the $\{A,Z\}$ isoscalar corrections,
    ii) QED radiative corrections for DIS data sets, 
    iii) use of full theory (instead of K-factors) to obtain the final minimum, 
    and
    iv) improved numerical precision for the DY process.
    With these improvements, the $\chi^2$ of the modified fit is now comparable to \ncteqfit.
}

%%%%%%%%%%%%%%%%%%%%%%%%%%%%%%%%%%%%%%%%%%%%%%%%%%%%%
% Full Nuclear PDF: nCTEQ15g vs. EPS vs. DSSZ vs. HKN @ Q=10GeV
%----------------
%%%%%%%%%%%%%%%%%%%%%%%%%%%%%%%%%%%%%%%%%%%%%%%%%%%%%
%

%
To better understand this result, we observe in Figs.~\ref{fig:compar_PDFs_Q1} and \ref{fig:compar_PDFs_Q10},
that the $u_v$ and $d_v$ ratios exhibit opposite $x$-dependence as compared with the \{HKN07, EPS09, DSSZ\}
sets. That is the  $u_v$  ratio is below the other sets at large $x$ and above at small $x$; the  $d_v$ ratio does the opposite. 
As the nuclear data sets probe a linear combination of $u_v$ and $d_v$, 
this raises the question as to whether the above 
differences might cancel when combined.

In Fig.~\ref{fig:compar_FullPDFs_Q10}  we now compare the full nuclear
lead PDFs from the different groups. The upper panel shows
the PDFs themselves, and the lower one shows their ratio compared to the
nuclear combination constructed out of the free proton -- the full nuclear correction. 
From this comparison we can clearly see that the
large differences in the effective bound proton distributions of valence quarks 
(Figs.~\ref{fig:compar_PDFs_Q1}, \ref{fig:compar_PDFs_Q10})
translate
into much smaller differences in the full nuclear PDFs that actually enter the
calculation of observables.%
    \footnote{Note that only up and down distributions differ between the
    full nuclear PDFs and the PDFs of the bound proton; the gluon and strange
    distributions are the same.}
In particular, we see that $u_v$ and $d_v$ distributions of the \ncteqfit\ fit are 
in very good agreement with the EPS09 results, and have substantial (but not complete) 
overlap with HKN07.\footnote{The DSSZ set (not show) is similar to HKN07 in that 
it has  substantial (but not complete) overlap.}

Of course, as the data can only constrain the full nuclear PDF in the combination 
$f^{A} = \frac{Z}{A}f^{p/A} + \frac{A-Z}{A}f^{n/A}$, 
we conclude that better separation of $u_v$ and $d_v$ distributions require
more data on non-isoscalar targets.
We also note that the currently available DIS data use a number of non-isoscalar
targets and would have the potential to partially distinguish $u_v$ and $d_v$
distributions; unfortunately many of these data sets have been corrected for the
neutron excess and in turn lost this ability.
%
%%%%%%%%%%%%%%%%%%%%%%%%%%%%%%%%%%%%%%%%%%%%%%%%%%%%%

%!TEX root=paper.tex
%+++++++++++++++++++++++++++++++++++++++++++++++++++++++
\section{Summary and conclusions}
\label{sec:sum}
%+++++++++++++++++++++++++++++++++++++++++++++++++++++++

In this paper we have presented the first complete analysis of nuclear PDFs with errors in the \cteq framework.
The resulting fit, \ncteqfit, uses 
the available charged lepton DIS, DY and inclusive pion data taken on a variety of nuclear
targets. The uncertainty of this analysis is presented in the form of error PDFs which are constructed using an adapted
Hessian method.

Within our framework we are able to obtain a good fit to all data. The output of the \ncteqfit\ analysis is a complete set 
of nuclear PDFs with uncertainties for any $A=\{1, ... , 208\}$. 
A selection of nuclear PDFs for the most common nuclei are made publicly 
available,\footnote{The nPDF sets for the current \ncteqfit\ analysis as well as for the alternative \ncteqnp\ analysis
are available for download at {\tt http://ncteq.hepforge.org} as well as on the LHAPDF website.}
but custom nPDFs can be generated for any $\{A,Z \}$ combination.

In comparison to our previous analysis~\cite{Schienbein:2009kk}, we have included the data from the inclusive pion production
at RHIC. The new data provide additional constraints mostly for the nuclear gluon PDF but the description of the data relies 
on the fragmentation functions.
Therefore we also provide an alternative conservative result \ncteqnp\ which does not include the inclusive
pion data and is hence fragmentation function independent.

Compared to other global analyses (HKN07, EPS09, and DSSZ) there are a number of important differences:
\begin{itemize}
 \item In contrast to the other analyses, we parameterize the nuclear PDFs directly instead of the nuclear corrections factors.
 \item In addition, our $u$- and $d$-valence distributions are parametrized independently.
 \item Other differences arise from the selection of data points used in the fit. In particular we impose more
 conservative kinematic cuts in order to minimize effects from higher twists and target mass corrections.
\end{itemize}

Overall our results are compatible with the other nPDFs but after a detailed look we see distinct differences
(see Fig.~\ref{fig:compar_FullPDFs_Q10}).
\begin{itemize}
 \item[(i)]
 The \ncteqfit\ nuclear gluon PDF has a larger shadowing  at small-$x$ than the other global analyses.
 Our result is compatible with the result of EPS09 as the error bands are overlapping
 throughout the entire $x$ range. 
 The overlap in case of HKN07 and (especially) DSSZ is limited especially in the small-$x$ region
 where no data constraints are present (and uncertainties of HKN07 and DSSZ are very small).
 This highlights the fact that nPDF uncertainties, in particular for gluon,
 are underestimated and different gluon solutions are possible~\cite{Stavreva:2010mw}.
 \item[(ii)] 
%%%%%%%%%%%%%%%%%%%%%%%%%%%%%%%%%%%%%%%
Our valence distributions for a bound proton in lead differ  as we allow  separate nuclear corrections for $u_v$ and $d_v$.
Compared to  the other groups; 
 our $d$-valence PDF is harder and $u$-valence PDF is softer.
 However, when the full lead nucleus is constructed, these differences are substantially reduced
 and we observe a good agreement between all groups. 
%%%%%%%%%%%%%%%%%%%%%%%%%%%%%%%%
%
 \item[(iii)] The \ncteqfit\ light sea quark distributions are in very good agreement with the ones from the other groups for $x\gsim10^{-2}$.
 At smaller $x$ where there are no data constraints the individual error bands clearly underestimates the uncertainty.
 \item[(iv)] It should be also mentioned that strange distributions are currently not fitted in any of the nPDF analyses
 and are fixed by imposing additional assumptions; this leads to quite significant differences between different groups.
\end{itemize}

All in all we find relatively good agreement between different nPDFs. Most of the noticeable differences between them 
occur in regions without any constraints from data and so they can be attributed to different assumptions such as parameterization of 
the nuclear effects. 

In view of the differences, the true nPDF uncertainties should be obtained by combining the results of all 
analyses and their uncertainties. In particular, this is true for the gluon distribution where the small $x$ behavior
is basically unconstrained and every single nPDF analysis substantially underestimates it (see our earlier study~\cite{Stavreva:2010mw}).

The \ncteq\ framework used for the \ncteqfit\ fit can combine data from both proton and nuclear targets into a
single coherent analysis. Using \ncteqfit\ fit as a reference, it will be interesting to include the upcoming LHC data
as we continue to investigate the relations between the proton and the nuclear PDFs.

\begin{figure*}[th!]
\includegraphics[width=0.9\textwidth]{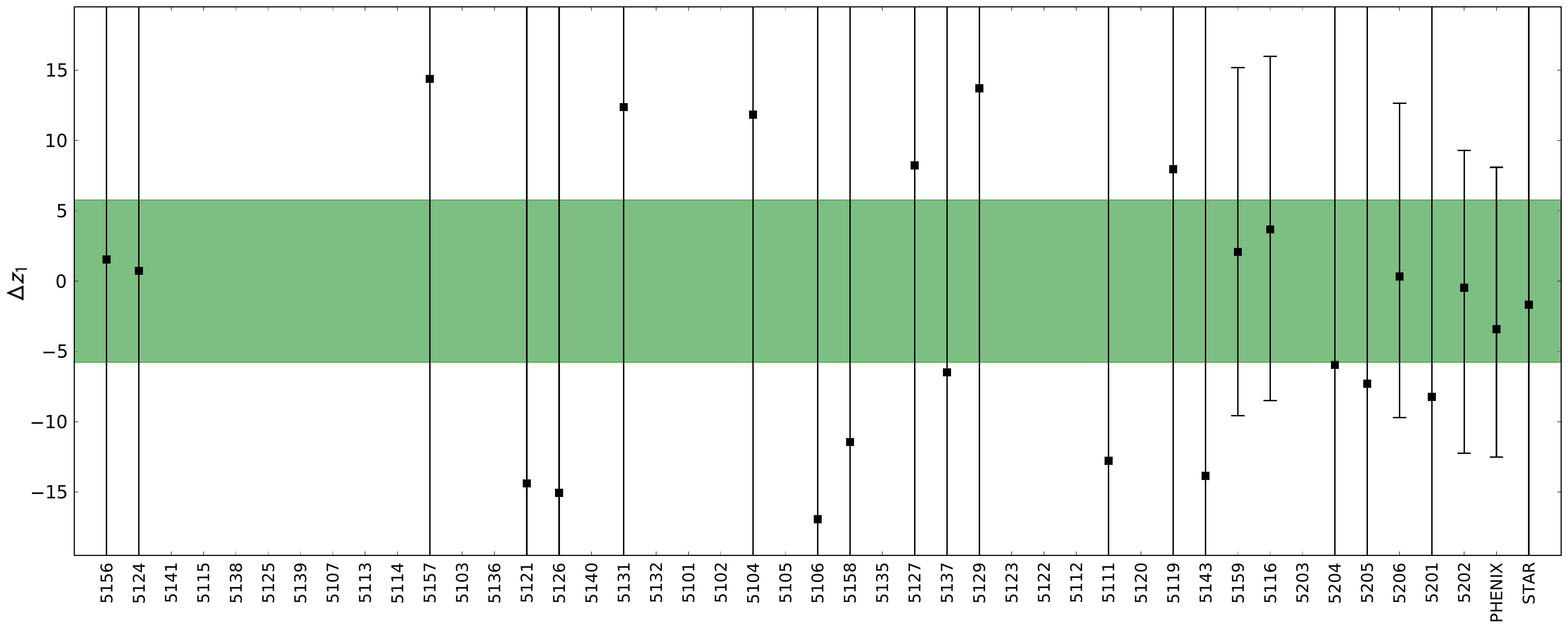}
\caption{The 90\% confidence level limits from different data sets in the eigenvector direction $\tilde{z}_1$. 
The $\chi^2$-minimum for each experiment is denoted by a black square, 
and the green band demonstrates the interval of the eigenvector
parameter corresponding to the final $\Delta\chi^2$.}
\label{fig:evex}
\end{figure*}
%

%!TEX root=paper.tex
%+++++++++++++++++++++++++++++++++++++++++++++++++++++++
\section*{Acknowledgments}
\label{sec:acks}
%+++++++++++++++++++++++++++++++++++++++++++++++++++++++
The authors would like to thank J. Gao for proposing the use of $\Delta\chi^2_{\text{eff}}$ as an alternative to the
correlation cosine and for other useful discussions. Furthermore, we would like to thank 
M. Botje, 
E. Godat,
S. Kumano, 
P. Nadolsky
and
V. Radescu
for valuable discussions.
We acknowledge the hospitality of CERN, DESY, and Fermilab where a portion of this work was performed. 
The work of J.F.O. was supported by the DOE under the grant DE-FG02-13ER41942. T. J. was partly supported by the
Research Executive Agency (REA) of the European Union under the Grant Agreement No. PITN-GA-2010-264564
(LHCPhenoNet).
This work was also partially supported by the 
U.S. Department of Energy under Grant No. DE-FG02-13ER41996, 
and by Projet international de cooperation scientifique PICS05854 between
France and the U.S.
Fermilab is operated by Fermi Research Alliance, LLC under Contract No. DE-AC02-07CH11359 with the United States Department of Energy.

\appendix
%!TEX root=paper.tex
%
%+++++++++++++++++++++++++++++++++++++++++++++++++++++++
\section{Determination of $\Delta\chi^2$ and Hessian rescaling}
\label{app:Hess}
%
%+++++++++++++++++++++++++++++++++++++++++++++++++++++++
\subsection{Determination of $\Delta\chi^2$}
\label{app:chidist}
%+++++++++++++++++++++++++++++++++++++++++++++++++++++++
%
In this appendix we discuss the details of the determination of $\Delta\chi^2$ which is 
motivated by the treatment presented in Refs.~\cite{Stump:2001gu,Martin:2009iq,Eskola:2009uj}.
We investigate how the global fit describes each experiment by examining
$\chi^2_k$ which is the  individual $\chi^2$-contribution
of experiment $k$  with $N_k$ data points. 
We can then see how  $\chi_k^2$ changes when varying PDF parameters along each eigenvector direction $\tilde{z}_i$  
of Eq.~(\ref{eq:zTOy}).

The probability distribution for the $\chi_k^{2}$ given that
the fit has $N_k$ degrees of freedom is: 
\begin{equation}
P(\chi_k^{2},N_k)=\frac{(\chi_k^{2})^{N_k/2-1}e^{-\chi_k^{2}/2}}{2^{N_k/2}\Gamma(N_k/2)}\,.
\label{eq:chi2dist}
\end{equation}
This allows us to define the percentiles $\xi_{p}$ via 
\begin{equation}\label{perc}
\int_{0}^{\xi_{p}}P(\chi^{2},N)d\chi^{2}=p\%\quad\;{\rm where}\quad p=\{50,90,99\}\,.
\end{equation}
Here, $\xi_{50}$ serves as an estimate of
the mean of the $\chi^{2}$ distribution and $\xi_{90}$, for example,
gives us the value where there is only a 10\% probability that a fit
with $\chi^{2}>\xi_{90}$ genuinely describes the given set of data.

Due to fluctuations in the data and possible 
incompatibilities between experiments, the global $\chi^2$ minimum does not necessarily coincide with $\chi^2$-minima of individual 
experiments. Moreover, for the same reason, the minimum $\chi^2$ for each experiment, $\chi^2_{k,0}$, can be far away from the expected 
minimum given by $\xi_{50}$. In order to use the percentiles defined in Eq.~(\ref{perc}) to define the 90\% confidence level 
(C.L.), we rescale the $\xi_{90}$ percentile to take into account the position of the minimum as
\begin{equation}
	\tilde{\xi}_{90}\rightarrow \xi_{90}\left(\frac{\chi^2_{k,0}}{\xi_{50}}\right)\,.
\end{equation}
%
% before rescaling
%----------------
\begin{figure*}[t]
\centering{}
\subfloat[$\chi^2$ function in parameter space.] 
{
\includegraphics[clip,width=0.48\textwidth]{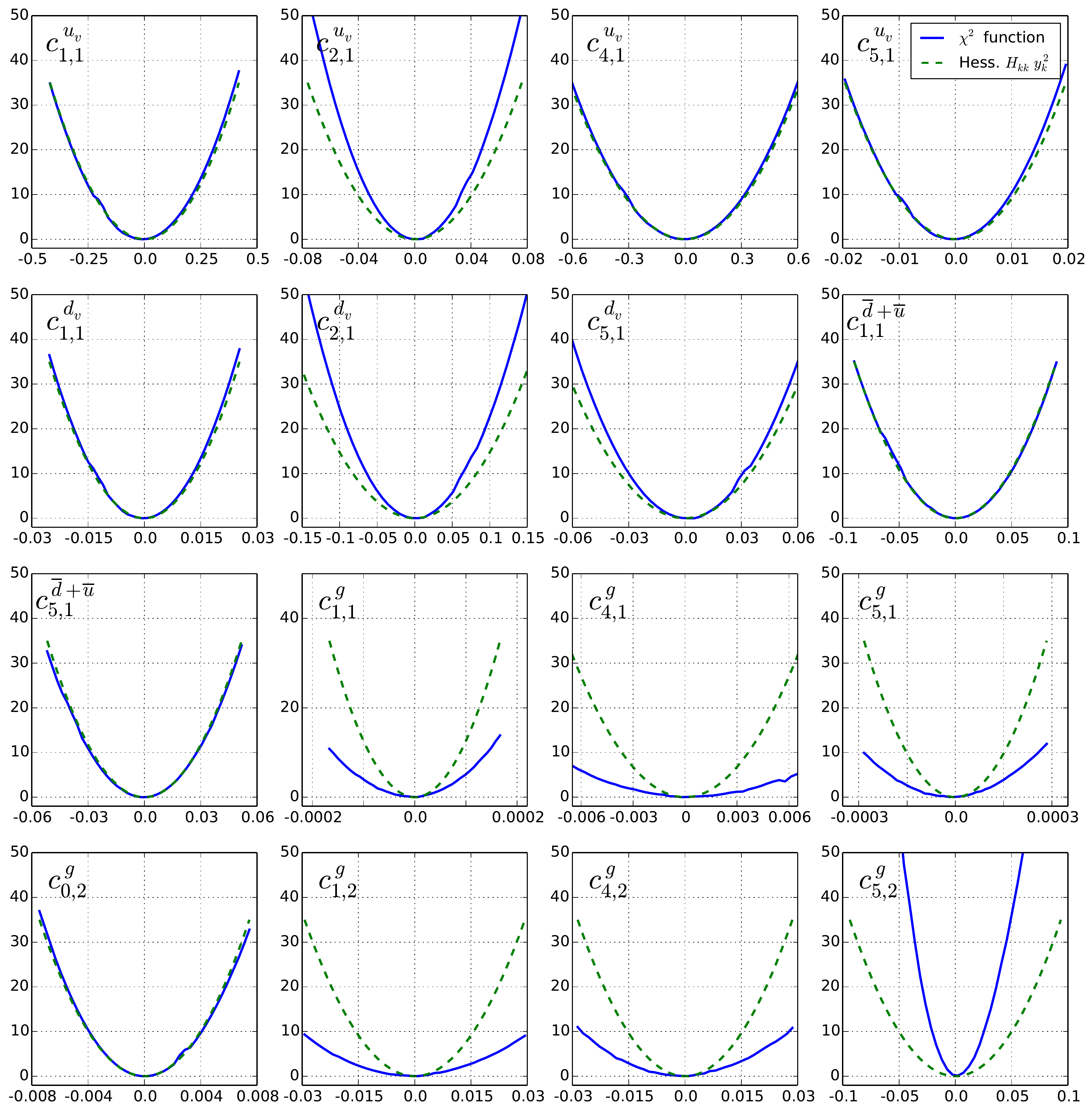}
\label{subfig:1Dscans_beforeRescaling_OP}
}
\quad{}
\subfloat[$\chi^2$ function in eigenvector space.] 
{
\includegraphics[width=0.47\textwidth]{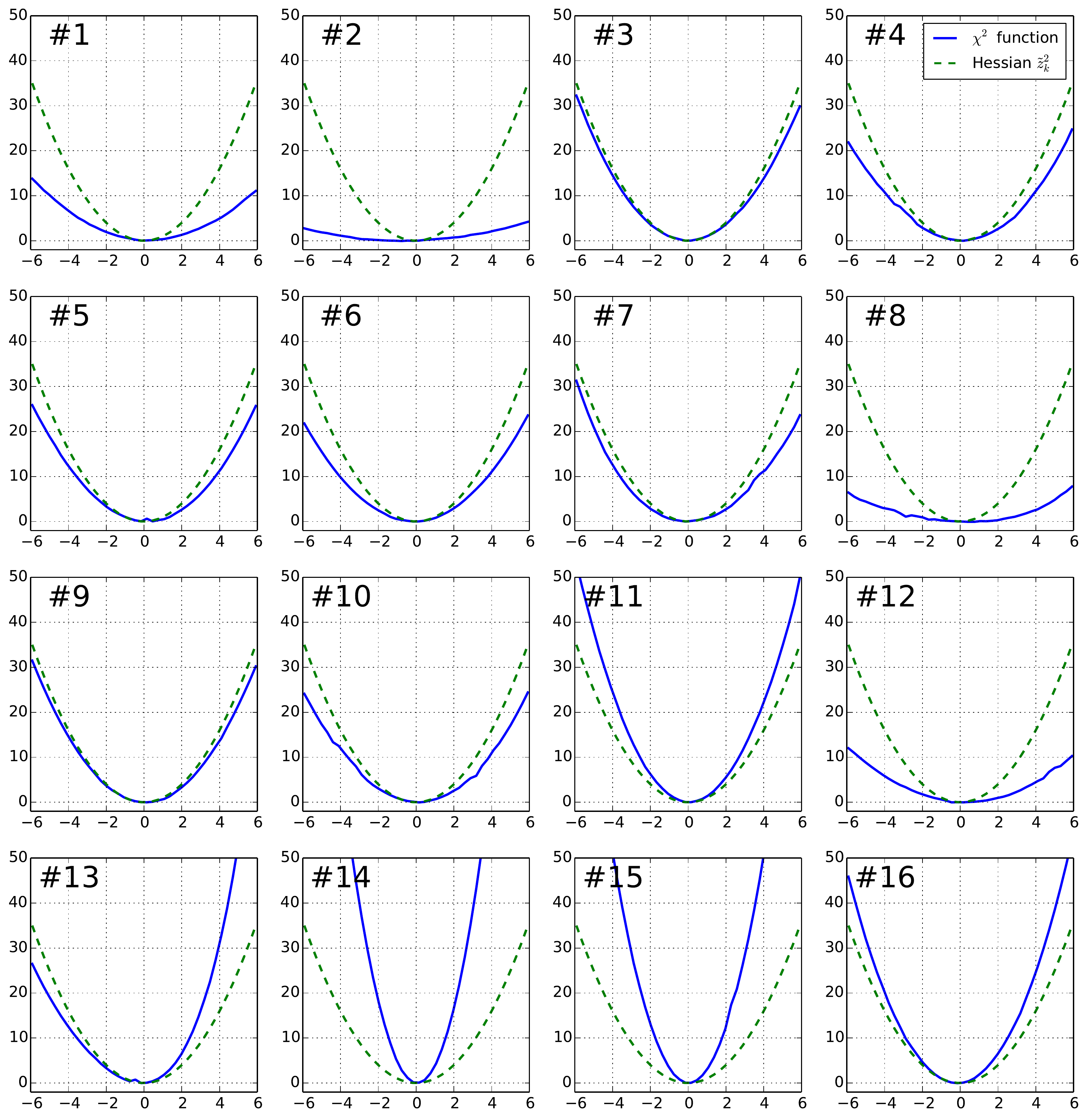}
\label{subfig:1Dscans_beforeRescaling_EV}
}
\caption{
These plots display the  Hessian {\em before} the ``rescaling'' procedure.
\\
$\chi^2$ function relative to its value at the minimum, $\Delta\chi^2=\chi^2-\chi^2_0$,
plotted along the 16 fitting parameters of the original space (left) 
and along the $z_i$ directions in the eigenvector space (right). 
The actual $\chi^2$ function is plotted with solid lines, 
and the Hessian approximation $\Delta\chi^2=\tilde{z}_i^2$  is shown with dashed lines.
}
\label{fig:1Dscans_beforeRescaling}
\end{figure*}
%----------------

For each eigenvector direction given by a variation of the parameter $\tilde{z}_i$ and every experiment, we define an interval
\begin{equation}
	z_i^{(k)-}\leq \tilde{z}_i\leq z_i^{(k)+},
\end{equation}
where the $\chi_k^2$ stays within the 90\% C.L. limit (i.e. $\chi_k^2<\tilde{\xi}_{90}$). 
For each eigenvector direction we then construct an interval 
$(z_i^-,z_i^+)$ where all experiments stay within the 90\% C.L. limit as
\begin{equation}
	(z_i^-,z_i^+)\equiv \bigcap_k\, (z_i^{(k)-},z_i^{(k)+})\,.
\end{equation}
These intervals can  obviously be different for each eigenvector, depending on the fact how well the experiments constrain the 
variations in this eigenvector direction. For $n$ free parameters we obtain $2n$ parameters 
$\{z_1^-,z_1^+,z_2^-,z_2^+,\ldots,z_n^-,z_n^+\}$ which we can use to define the global tolerance as
\begin{equation}
	\Delta\chi^2\equiv \sum_i\frac{(z_i^+)^2 + (z_i^-)^2}{2n}\,.
\end{equation}
Having performed the procedure described in this Section, we have arrived at $\Delta\chi^2 = 35$. One can compare how this choice of
global tolerance (the same for every eigenvector direction) agrees with the rescaled 90\% confidence level (C.L.) for each experiment
in every direction. In Fig.~\ref{fig:evex} we show this comparison for only one single eigenvector direction as all the others are rather 
similar.
%
%+++++++++++++++++++++++++++++++++++++++++++++++++++++++
\subsection{Hessian rescaling}
\label{app:Hess-rescaling}
%+++++++++++++++++++++++++++++++++++++++++++++++++++++++
%
Choosing a larger tolerance  $\Delta\chi^2=35$ as argued above might pose a problem for the Hessian approach
as it requires using information from a larger neighborhood of the global minimum which is
not necessarily well described in the quadratic approximation. 
Fig.~\ref{fig:1Dscans_beforeRescaling} confirms that this is the case for the {\ncteqfit} fit.
Both in the original parameter space, Fig.~\ref{subfig:1Dscans_beforeRescaling_OP}, and in the
eigenvector basis, Fig.~\ref{subfig:1Dscans_beforeRescaling_EV}, we can see directions where
$\chi^2$-function deviates substantially from the quadratic approximation when $\Delta\chi^2\sim 35$. 
This is a problem because  in the Hessian approach we use  
the eigenvector basis to determine the ranges of the normalized parameters $\tilde{z}_i$ 
Fig.~\ref{fig:1Dscans_beforeRescaling} shows that if we 
take $\Delta\tilde{z}_i = \sqrt{\Delta\chi^2} \approx \sqrt{35}$, then depending on the
specific eigen-direction we would largely overestimate or underestimate the error on our
parameters (see e.g. plots \#1, \#2 and \#14 in Fig.~\ref{subfig:1Dscans_beforeRescaling_EV}).

\begin{figure*}[t]
\centering{}
\subfloat[$\chi^2$ function in parameter space.]
{
\includegraphics[clip,width=0.48\textwidth]{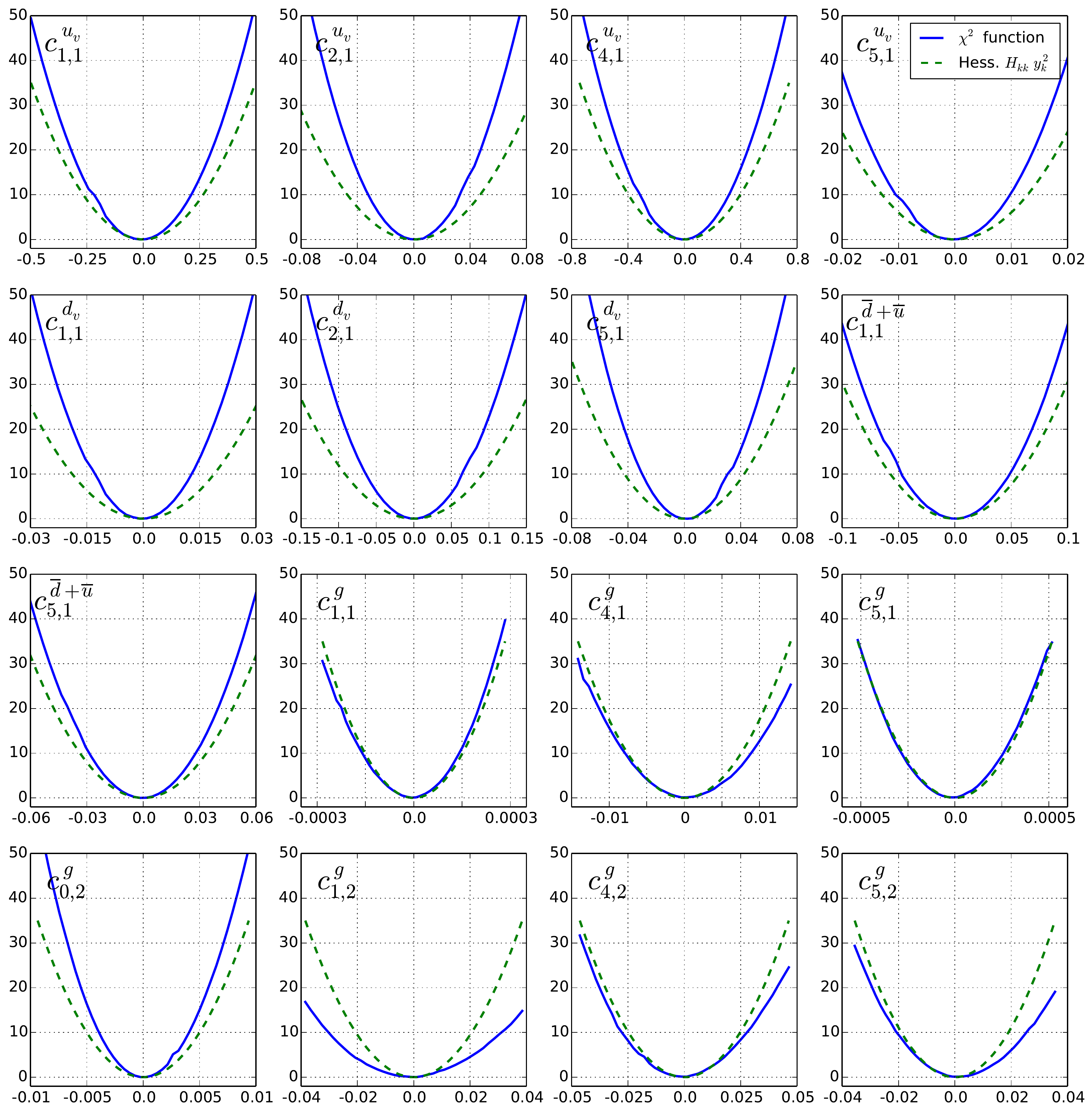}
\label{subfig:1Dscans_afterRescaling_OP}
}
\quad{}
\subfloat[$\chi^2$ function in eigenvector space.] 
{
\includegraphics[width=0.47\textwidth]{nCTEQ15_EV_4x4.pdf}
\label{subfig:1Dscans_afterRescaling_EV}
}
\caption{
These plots display the  Hessian {\em after} the ``rescaling'' procedure.
\\
$\chi^2$ function relative to its value at the minimum, $\Delta\chi^2=\chi^2-\chi^2_0$,
plotted along the 16 fitting parameters of the original space (left) 
and along the $z_i$ directions in the eigenvector space (right). 
The actual $\chi^2$ function is plotted with solid lines, 
and the Hessian approximation $\Delta\chi^2=\tilde{z}_i^2$  is shown with dashed lines.
}
\label{fig:1Dscans_afterRescaling}
\end{figure*}

To improve the constraints provided by the $\chi^2$-function, we redefine the Hessian which we use
to determine the error PDFs using the formalism described in Sec.~\ref{sec:hessian}. 
We keep the eigenvector information intact, 
but rescale the eigenvalues of the original Hessian (which corresponds to rescaling the parameters $\tilde{z}_i$) so that
the modified Hessian better describes the $\chi^2$-function not only in the minimum ($\Delta\chi^2 =0$) 
but also at $\Delta\chi^2 =35$.
For each eigenvector direction, we identify the parameter values $\tilde{z}_i^\pm$ where
\begin{equation}
	\Delta\chi^2(\tilde{z}_i^\pm)\equiv \chi^2(\tilde{z}_i^\pm) - \chi^2_0 = 35\,,
\end{equation}
where $\chi^2_0$ is the minimum of the $\chi^2$. Using the $\tilde{z}_i^\pm$, we rescale the corresponding eigenvalue as
\begin{equation}
	\lambda_i\mapsto \lambda_i'=\frac{|\tilde{z}_i^+|^2+|\tilde{z}_i^-|^2}{2\sqrt{\Delta\chi^2}}\,\lambda_i\ .
\end{equation}
The impact of the rescaling of the Hessian can be seen on Fig.~\ref{subfig:1Dscans_afterRescaling_EV} where one notices that
the description of the $\chi^2$-function in the eigenvector basis is highly improved, especially in region  where
$\Delta\chi^2 =35$. 
The description of the $\chi^2$-function in the original parameter space  
(Fig.~\ref{subfig:1Dscans_afterRescaling_OP}) is also improved but to a lesser extent. 
However, this is a secondary feature as we are working in the eigenvector space
when defining the error PDFs.
%
% after rescaling
%----------------
%----------------

%!TEX root=paper.tex
%+++++++++++++++++++++++++++++++++++++++++++++++++++++++
\section{Usage of \ncteq\ PDFs}
\label{app:usage}
%+++++++++++++++++++++++++++++++++++++++++++++++++++++++

We provide a set of PDF tables for the \ncteqfit\ and \ncteqnp\ fits at the {\tt nCTEQ} Hepforge 
website~\cite{nCTEQwebpage}. We provide the tables in the older CTEQ PDS format together with a
dedicated interface as well as in the new LHAPDF6 format~\cite{Buckley:2014ana}.
In the future the LHAPDF6 grids will be also available at the LHAPDF website~\cite{LHAPDFwebpage}.

We provide tables for both bound proton PDFs $f^{p/A}(x,Q)$ as well as grids for the resulting 
full nuclear PDFs $f^A = Z/A\ f^{p/A} + (A-Z)/A\ f^{n/A}$. The bound proton PDFs allow a direct 
comparison of the nPDFs for different $A$ values as displayed in 
Fig.~\ref{fig:compar_PDFs_diff-nuc}. On the other hand the full nuclear PDFs can be used directly to calculate 
cross-sections for the nuclear collisions.

At the moment we distribute grids for a selection of
nuclei that are commonly used in the high energy/nuclear
experiments. In particular, we provide girds for:
\{He, Li, Be, C, N, Al, Ca, Fe, Cu, Kr, Ag, Sn, Xe, W, Au, Pb\}.
Since our parametrization is continuous in A and Z it
allows us to generate PDFs for any nuclei or isotopes.
In case users are interested in having the \ncteqfit\ 
distributions for a nucleus that is not included in
our standard selection, we can generate the PDFs upon
request.

%%%%%%%%%%%%%%%%%%%%%%%%%%%%%
%\bibliographystyle{utphys}
%% \bibliographystyle{apsrev}
%\bibliography{biblio}

\begin{thebibliography}{10}

\bibitem{Collins:1985ue}
J.~C. Collins, D.~E. Soper, and G.~F. Sterman, ``{Factorization for Short
  Distance Hadron - Hadron Scattering},''
\href{http://dx.doi.org/10.1016/0550-3213(85)90565-6}{{\em Nucl.Phys.}
  {\bfseries B261} (1985) 104}.
%%CITATION = NUPHA,B261,104;%%.

\bibitem{Bodwin:1984hc}
G.~T. Bodwin, ``{Factorization of the Drell-Yan Cross-Section in Perturbation
  Theory},''
\href{http://dx.doi.org/10.1103/PhysRevD.34.3932,
  10.1103/PhysRevD.31.2616}{{\em Phys.Rev.} {\bfseries D31} (1985) 2616}.
%%CITATION = PHRVA,D31,2616;%%.

\bibitem{Collins:1998rz}
J.~C. Collins, ``{Hard scattering factorization with heavy quarks: A General
  treatment},'' \href{http://dx.doi.org/10.1103/PhysRevD.58.094002}{{\em
  Phys.Rev.} {\bfseries D58} (1998) 094002},
\href{http://arxiv.org/abs/hep-ph/9806259}{{\ttfamily arXiv:hep-ph/9806259
  [hep-ph]}}.
%%CITATION = HEP-PH/9806259;%%.

\bibitem{Gao:2013xoa}
J.~Gao, M.~Guzzi, J.~Huston, H.-L. Lai, Z.~Li, {\em et~al.}, ``{CT10
  next-to-next-to-leading order global analysis of QCD},''
  \href{http://dx.doi.org/10.1103/PhysRevD.89.033009}{{\em Phys.Rev.}
  {\bfseries D89} no.~3, (2014) 033009},
\href{http://arxiv.org/abs/1302.6246}{{\ttfamily arXiv:1302.6246 [hep-ph]}}.
%%CITATION = ARXIV:1302.6246;%%.

\bibitem{Ball:2014uwa}
{\bfseries NNPDF} Collaboration, R.~D. Ball {\em et~al.}, ``{Parton
  distributions for the LHC Run II},''
  \href{http://dx.doi.org/10.1007/JHEP04(2015)040}{{\em JHEP} {\bfseries 1504}
  (2015) 040},
\href{http://arxiv.org/abs/1410.8849}{{\ttfamily arXiv:1410.8849 [hep-ph]}}.
%%CITATION = ARXIV:1410.8849;%%.

\bibitem{Harland-Lang:2014zoa}
L.~Harland-Lang, A.~Martin, P.~Motylinski, and R.~Thorne, ``{Parton
  distributions in the LHC era: MMHT 2014 PDFs},''
  \href{http://dx.doi.org/10.1140/epjc/s10052-015-3397-6}{{\em Eur.Phys.J.}
  {\bfseries C75} no.~5, (2015) 204},
\href{http://arxiv.org/abs/1412.3989}{{\ttfamily arXiv:1412.3989 [hep-ph]}}.
%%CITATION = ARXIV:1412.3989;%%.

\bibitem{Alekhin:2013nda}
S.~Alekhin, J.~Bluemlein, and S.~Moch, ``{The ABM parton distributions tuned to
  LHC data},'' \href{http://dx.doi.org/10.1103/PhysRevD.89.054028}{{\em
  Phys.Rev.} {\bfseries D89} (2014) 054028},
\href{http://arxiv.org/abs/1310.3059}{{\ttfamily arXiv:1310.3059 [hep-ph]}}.
%%CITATION = ARXIV:1310.3059;%%.

\bibitem{Owens:2012bv}
J.~Owens, A.~Accardi, and W.~Melnitchouk, ``{Global parton distributions with
  nuclear and finite-$Q^2$ corrections},''
  \href{http://dx.doi.org/10.1103/PhysRevD.87.094012}{{\em Phys.Rev.}
  {\bfseries D87} no.~9, (2013) 094012},
\href{http://arxiv.org/abs/1212.1702}{{\ttfamily arXiv:1212.1702 [hep-ph]}}.
%%CITATION = ARXIV:1212.1702;%%.

\bibitem{Jimenez-Delgado:2014twa}
P.~Jimenez-Delgado and E.~Reya, ``{Delineating parton distributions and the
  strong coupling},'' \href{http://dx.doi.org/10.1103/PhysRevD.89.074049}{{\em
  Phys.Rev.} {\bfseries D89} no.~7, (2014) 074049},
\href{http://arxiv.org/abs/1403.1852}{{\ttfamily arXiv:1403.1852 [hep-ph]}}.
%%CITATION = ARXIV:1403.1852;%%.

\bibitem{Aubert:1983xm}
{\bfseries European Muon} Collaboration, J.~Aubert {\em et~al.}, ``{The ratio
  of the nucleon structure functions $F2_n$ for iron and deuterium},''
\href{http://dx.doi.org/10.1016/0370-2693(83)90437-9}{{\em Phys.Lett.}
  {\bfseries B123} (1983) 275}.
%%CITATION = PHLTA,B123,275;%%.

\bibitem{deFlorian:2011fp}
D.~de~Florian, R.~Sassot, P.~Zurita, and M.~Stratmann, ``{Global Analysis of
  Nuclear Parton Distributions},''
  \href{http://dx.doi.org/10.1103/PhysRevD.85.074028}{{\em Phys.Rev.}
  {\bfseries D85} (2012) 074028},
\href{http://arxiv.org/abs/1112.6324}{{\ttfamily arXiv:1112.6324 [hep-ph]}}.
%%CITATION = ARXIV:1112.6324;%%.

\bibitem{Eskola:2009uj}
K.~Eskola, H.~Paukkunen, and C.~Salgado, ``{EPS09: A New Generation of NLO and
  LO Nuclear Parton Distribution Functions},''
  \href{http://dx.doi.org/10.1088/1126-6708/2009/04/065}{{\em JHEP} {\bfseries
  0904} (2009) 065},
\href{http://arxiv.org/abs/0902.4154}{{\ttfamily arXiv:0902.4154 [hep-ph]}}.
%%CITATION = ARXIV:0902.4154;%%.

\bibitem{Schienbein:2009kk}
I.~Schienbein, J.~Yu, K.~Kovarik, C.~Keppel, J.~Morfin, {\em et~al.}, ``{PDF
  Nuclear Corrections for Charged and Neutral Current Processes},''
  \href{http://dx.doi.org/10.1103/PhysRevD.80.094004}{{\em Phys.Rev.}
  {\bfseries D80} (2009) 094004},
\href{http://arxiv.org/abs/0907.2357}{{\ttfamily arXiv:0907.2357 [hep-ph]}}.
%%CITATION = ARXIV:0907.2357;%%.

\bibitem{Hirai:2007sx}
M.~Hirai, S.~Kumano, and T.-H. Nagai, ``{Determination of nuclear parton
  distribution functions and their uncertainties in next-to-leading order},''
  \href{http://dx.doi.org/10.1103/PhysRevC.76.065207}{{\em Phys.Rev.}
  {\bfseries C76} (2007) 065207},
\href{http://arxiv.org/abs/0709.3038}{{\ttfamily arXiv:0709.3038 [hep-ph]}}.
%%CITATION = ARXIV:0709.3038;%%.

\bibitem{Qiu:2003cg}
J.-W. Qiu, ``{QCD factorization and rescattering in proton nucleus
  collisions},'' \href{http://arxiv.org/abs/hep-ph/0305161}{{\ttfamily
  arXiv:hep-ph/0305161 [hep-ph]}}.
{hep-ph/0305161}.
%%CITATION = HEP-PH/0305161;%%.

\bibitem{Accardi:2004be}
A.~Accardi, N.~Armesto, M.~Botje, S.~Brodsky, B.~Cole, {\em et~al.}, ``{Hard
  probes in heavy ion collisions at the LHC: PDFs, shadowing and $pA$
  collisions},'' \href{http://arxiv.org/abs/hep-ph/0308248}{{\ttfamily
  arXiv:hep-ph/0308248 [hep-ph]}}.
{hep-ph/0308248}.
%%CITATION = HEP-PH/0308248;%%.

\bibitem{Morfin:2012kn}
J.~G. Morfin, J.~Nieves, and J.~T. Sobczyk, ``{Recent Developments in
  Neutrino/Antineutrino - Nucleus Interactions},''
  \href{http://dx.doi.org/10.1155/2012/934597}{{\em Adv. High Energy Phys.}
  {\bfseries 2012} (2012) 934597},
  \href{http://arxiv.org/abs/1209.6586}{{\ttfamily arXiv:1209.6586 [hep-ex]}}.

\bibitem{Armesto:2006ph}
N.~Armesto, ``{Nuclear shadowing},''
  \href{http://dx.doi.org/10.1088/0954-3899/32/11/R01}{{\em J. Phys.}
  {\bfseries G32} (2006) R367--R394},
  \href{http://arxiv.org/abs/hep-ph/0604108}{{\ttfamily arXiv:hep-ph/0604108
  [hep-ph]}}.

\bibitem{Kopeliovich:2012kw}
B.~Z. Kopeliovich, J.~G. Morfin, and I.~Schmidt, ``{Nuclear Shadowing in
  Electro-Weak Interactions},''
  \href{http://dx.doi.org/10.1016/j.ppnp.2012.09.004}{{\em Prog. Part. Nucl.
  Phys.} {\bfseries 68} (2013) 314--372},
  \href{http://arxiv.org/abs/1208.6541}{{\ttfamily arXiv:1208.6541 [hep-ph]}}.

\bibitem{Schienbein:2007fs}
I.~Schienbein, J.~Yu, C.~Keppel, J.~Morfin, F.~Olness, {\em et~al.}, ``{Nuclear
  parton distribution functions from neutrino deep inelastic scattering},''
  \href{http://dx.doi.org/10.1103/PhysRevD.77.054013}{{\em Phys.Rev.}
  {\bfseries D77} (2008) 054013},
\href{http://arxiv.org/abs/0710.4897}{{\ttfamily arXiv:0710.4897 [hep-ph]}}.
%%CITATION = ARXIV:0710.4897;%%.

\bibitem{Kovarik:2010uv}
K.~Kovarik, I.~Schienbein, F.~Olness, J.~Yu, C.~Keppel, {\em et~al.},
  ``{Nuclear corrections in neutrino-nucleus DIS and their compatibility with
  global NPDF analyses},''
  \href{http://dx.doi.org/10.1103/PhysRevLett.106.122301}{{\em Phys.Rev.Lett.}
  {\bfseries 106} (2011) 122301},
\href{http://arxiv.org/abs/1012.0286}{{\ttfamily arXiv:1012.0286 [hep-ph]}}.
%%CITATION = ARXIV:1012.0286;%%.

\bibitem{Paukkunen:2013grz}
H.~Paukkunen and C.~A. Salgado, ``{Agreement of Neutrino Deep Inelastic
  Scattering Data with Global Fits of Parton Distributions},''
  \href{http://dx.doi.org/10.1103/PhysRevLett.110.212301}{{\em Phys.Rev.Lett.}
  {\bfseries 110} no.~21, (2013) 212301},
\href{http://arxiv.org/abs/1302.2001}{{\ttfamily arXiv:1302.2001 [hep-ph]}}.
%%CITATION = ARXIV:1302.2001;%%.

\bibitem{Aivazis:1993kh}
M.~Aivazis, F.~I. Olness, and W.-K. Tung, ``{Leptoproduction of heavy quarks.
  1. General formalism and kinematics of charged current and neutral current
  production processes},''
  \href{http://dx.doi.org/10.1103/PhysRevD.50.3085}{{\em Phys.Rev.} {\bfseries
  D50} (1994) 3085--3101},
\href{http://arxiv.org/abs/hep-ph/9312318}{{\ttfamily arXiv:hep-ph/9312318
  [hep-ph]}}.
%%CITATION = HEP-PH/9312318;%%.

\bibitem{Aivazis:1993pi}
M.~A.~G. Aivazis, J.~C. Collins, F.~I. Olness, and W.-K. Tung,
  ``Leptoproduction of heavy quarks. 2. a unified qcd formulation of charged
  and neutral current processes from fixed target to collider energies,'' {\em
  Phys. Rev.} {\bfseries D50} (1994) 3102--3118,
\href{http://arxiv.org/abs/hep-ph/9312319}{{\ttfamily hep-ph/9312319}}.
%%CITATION = HEP-PH 9312319;%%.

\bibitem{Kramer:2000hn}
.~Kramer, Michael, F.~I. Olness, and D.~E. Soper, ``{Treatment of heavy quarks
  in deeply inelastic scattering},''
  \href{http://dx.doi.org/10.1103/PhysRevD.62.096007}{{\em Phys.Rev.}
  {\bfseries D62} (2000) 096007},
\href{http://arxiv.org/abs/hep-ph/0003035}{{\ttfamily arXiv:hep-ph/0003035
  [hep-ph]}}.
%%CITATION = HEP-PH/0003035;%%.

\bibitem{Kretzer:1998ju}
S.~Kretzer and I.~Schienbein, ``Heavy quark initiated contributions to deep
  inelastic structure functions,'' {\em Phys. Rev.} {\bfseries D58} (1998)
  094035,
\href{http://arxiv.org/abs/hep-ph/9805233}{{\ttfamily hep-ph/9805233}}.
%%CITATION = PHRVA,D58,094035;%%.

\bibitem{Guzzi:2011ew}
M.~Guzzi, P.~M. Nadolsky, H.-L. Lai, and C.-P. Yuan, ``{General-Mass Treatment
  for Deep Inelastic Scattering at Two-Loop Accuracy},''
  \href{http://dx.doi.org/10.1103/PhysRevD.86.053005}{{\em Phys.Rev.}
  {\bfseries D86} (2012) 053005},
\href{http://arxiv.org/abs/1108.5112}{{\ttfamily arXiv:1108.5112 [hep-ph]}}.
%%CITATION = ARXIV:1108.5112;%%.

\bibitem{Stavreva:2012bs}
T.~Stavreva, F.~Olness, I.~Schienbein, T.~Jezo, A.~Kusina, {\em et~al.},
  ``{Heavy Quark Production in the ACOT Scheme at NNLO and N3LO},''
  \href{http://dx.doi.org/10.1103/PhysRevD.85.114014}{{\em Phys.Rev.}
  {\bfseries D85} (2012) 114014},
\href{http://arxiv.org/abs/1203.0282}{{\ttfamily arXiv:1203.0282 [hep-ph]}}.
%%CITATION = ARXIV:1203.0282;%%.

\bibitem{Botje:2010ay}
M.~Botje, ``{QCDNUM: Fast QCD Evolution and Convolution},''
  \href{http://dx.doi.org/10.1016/j.cpc.2010.10.020}{{\em Comput. Phys.
  Commun.} {\bfseries 182} (2011) 490--532},
\href{http://arxiv.org/abs/1005.1481}{{\ttfamily arXiv:1005.1481 [hep-ph]}}.
%%CITATION = 1005.1481;%%.

\bibitem{Schienbein:2007gr}
I.~Schienbein {\em et~al.}, ``{A Review of Target Mass Corrections},''
  \href{http://dx.doi.org/10.1088/0954-3899/35/5/053101}{{\em J. Phys.}
  {\bfseries G35} (2008) 053101},
\href{http://arxiv.org/abs/0709.1775}{{\ttfamily arXiv:0709.1775 [hep-ph]}}.
%%CITATION = 0709.1775;%%.

\bibitem{Aurenche:1999nz}
P.~Aurenche, M.~Fontannaz, J.~Guillet, B.~A. Kniehl, and M.~Werlen, ``{Large
  $p_T$ inclusive $\pi^0$ cross-sections and next-to-leading-order QCD
  predictions},'' \href{http://dx.doi.org/10.1007/s100520000309}{{\em
  Eur.Phys.J.} {\bfseries C13} (2000) 347--355},
\href{http://arxiv.org/abs/hep-ph/9910252}{{\ttfamily arXiv:hep-ph/9910252
  [hep-ph]}}.
%%CITATION = HEP-PH/9910252;%%.

\bibitem{incNLL-website}
``Incnlo version 1.4.'' \url{http://lapth.cnrs.fr/PHOX_FAMILY/readme_inc.html}.

\bibitem{Stavreva:2010mw}
T.~Stavreva, I.~Schienbein, F.~Arleo, K.~Kovarik, F.~Olness, {\em et~al.},
  ``{Probing gluon and heavy-quark nuclear PDFs with gamma + Q production in pA
  collisions},'' \href{http://dx.doi.org/10.1007/JHEP01(2011)152}{{\em JHEP}
  {\bfseries 1101} (2011) 152},
\href{http://arxiv.org/abs/1012.1178}{{\ttfamily arXiv:1012.1178 [hep-ph]}}.
%%CITATION = ARXIV:1012.1178;%%.

\bibitem{Owens:2007kp}
J.~Owens, J.~Huston, C.~Keppel, S.~Kuhlmann, J.~Morfin, {\em et~al.}, ``{The
  Impact of new neutrino DIS and Drell-Yan data on large-x parton
  distributions},'' \href{http://dx.doi.org/10.1103/PhysRevD.75.054030}{{\em
  Phys.Rev.} {\bfseries D75} (2007) 054030},
\href{http://arxiv.org/abs/hep-ph/0702159}{{\ttfamily arXiv:hep-ph/0702159
  [HEP-PH]}}.
%%CITATION = HEP-PH/0702159;%%.

\bibitem{Pumplin:2002vw}
J.~Pumplin {\em et~al.}, ``New generation of parton distributions with
  uncertainties from global qcd analysis,'' {\em JHEP} {\bfseries 07} (2002)
  012,
\href{http://arxiv.org/abs/hep-ph/0201195}{{\ttfamily hep-ph/0201195}}.
%%CITATION = HEP-PH 0201195;%%.

\bibitem{Stump:2003yu}
D.~Stump, J.~Huston, J.~Pumplin, W.-K. Tung, H.~Lai, {\em et~al.}, ``{Inclusive
  jet production, parton distributions, and the search for new physics},'' {\em
  JHEP} {\bfseries 0310} (2003) 046,
\href{http://arxiv.org/abs/hep-ph/0303013}{{\ttfamily arXiv:hep-ph/0303013
  [hep-ph]}}.
%%CITATION = HEP-PH/0303013;%%.

\bibitem{Niculescu:2005rh}
I.~Niculescu, J.~Arrington, R.~Ent, and C.~Keppel, ``{Moments of nuclear and
  nucleon structure functions at low Q**2 and the momentum sum rule},''
  \href{http://dx.doi.org/10.1103/PhysRevC.73.045206}{{\em Phys.Rev.}
  {\bfseries C73} (2006) 045206},
\href{http://arxiv.org/abs/hep-ph/0509241}{{\ttfamily arXiv:hep-ph/0509241
  [hep-ph]}}.
%%CITATION = HEP-PH/0509241;%%.

\bibitem{Fomin:2010ei}
N.~Fomin, J.~Arrington, D.~Day, D.~Gaskell, A.~Daniel, {\em et~al.}, ``{Scaling
  of the $F_2$ structure function in nuclei and quark distributions at
  $x>1$},'' \href{http://dx.doi.org/10.1103/PhysRevLett.105.212502}{{\em
  Phys.Rev.Lett.} {\bfseries 105} (2010) 212502},
\href{http://arxiv.org/abs/1008.2713}{{\ttfamily arXiv:1008.2713 [nucl-ex]}}.
%%CITATION = ARXIV:1008.2713;%%.

\bibitem{deFlorian:2003qf}
D.~de~Florian and R.~Sassot, ``{Nuclear parton distributions at next-to-leading
  order},'' \href{http://dx.doi.org/10.1103/PhysRevD.69.074028}{{\em Phys.Rev.}
  {\bfseries D69} (2004) 074028},
\href{http://arxiv.org/abs/hep-ph/0311227}{{\ttfamily arXiv:hep-ph/0311227
  [hep-ph]}}.
%%CITATION = HEP-PH/0311227;%%.

\bibitem{pyminuit}
``{pyMinuit package}.'' \url{https://code.google.com/p/pyminuit/}.

\bibitem{Cminuit}
``{SEAL-Minuit}.''
  \url{http://seal.web.cern.ch/seal/work-packages/mathlibs/minuit/release/download.html}.

\bibitem{minuit}
F.~James. \url{https://wwwasdoc.web.cern.ch/wwwasdoc/minuit/minmain.html}.
\newblock ``MINUIT Reference Manual'', CERN Program Library Writeup D506.
  \url{https://wwwasdoc.web.cern.ch/wwwasdoc/minuit/minmain.html}.

\bibitem{Pumplin:2000vx}
J.~Pumplin, D.~Stump, and W.~Tung, ``{Multivariate fitting and the error matrix
  in global analysis of data},''
  \href{http://dx.doi.org/10.1103/PhysRevD.65.014011}{{\em Phys.Rev.}
  {\bfseries D65} (2001) 014011},
\href{http://arxiv.org/abs/hep-ph/0008191}{{\ttfamily arXiv:hep-ph/0008191
  [hep-ph]}}.
%%CITATION = HEP-PH/0008191;%%.

\bibitem{Pumplin:2001ct}
J.~Pumplin, D.~Stump, R.~Brock, D.~Casey, J.~Huston, {\em et~al.},
  ``{Uncertainties of predictions from parton distribution functions. 2. The
  Hessian method},'' \href{http://dx.doi.org/10.1103/PhysRevD.65.014013}{{\em
  Phys.Rev.} {\bfseries D65} (2001) 014013},
\href{http://arxiv.org/abs/hep-ph/0101032}{{\ttfamily arXiv:hep-ph/0101032
  [hep-ph]}}.
%%CITATION = HEP-PH/0101032;%%.

\bibitem{lanczosdif}
``Low-noise lanczos differentiators.''
  \url{http://www.holoborodko.com/pavel/numerical-methods/numerical-derivative/lanczos-low-noise-differentiators/}.

\bibitem{Stump:2001gu}
D.~Stump, J.~Pumplin, R.~Brock, D.~Casey, J.~Huston, {\em et~al.},
  ``{Uncertainties of predictions from parton distribution functions. 1. The
  Lagrange multiplier method},''
  \href{http://dx.doi.org/10.1103/PhysRevD.65.014012}{{\em Phys.Rev.}
  {\bfseries D65} (2001) 014012},
\href{http://arxiv.org/abs/hep-ph/0101051}{{\ttfamily arXiv:hep-ph/0101051
  [hep-ph]}}.
%%CITATION = HEP-PH/0101051;%%.

\bibitem{Martin:2009iq}
A.~Martin, W.~Stirling, R.~Thorne, and G.~Watt, ``{Parton distributions for the
  LHC},'' \href{http://dx.doi.org/10.1140/epjc/s10052-009-1072-5}{{\em
  Eur.Phys.J.} {\bfseries C63} (2009) 189--285},
\href{http://arxiv.org/abs/0901.0002}{{\ttfamily arXiv:0901.0002 [hep-ph]}}.
%%CITATION = ARXIV:0901.0002;%%.

\bibitem{Arneodo:1996qe}
{\bfseries New Muon} Collaboration, M.~Arneodo {\em et~al.}, ``{Measurement of
  the proton and deuteron structure functions, F2(p) and F2(d), and of the
  ratio sigma(L)/sigma(T)},''
  \href{http://dx.doi.org/10.1016/S0550-3213(96)00538-X}{{\em Nucl. Phys.}
  {\bfseries B483} (1997) 3--43},
\href{http://arxiv.org/abs/hep-ph/9610231}{{\ttfamily arXiv:hep-ph/9610231}}.
%%CITATION = HEP-PH/9610231;%%.

\bibitem{Airapetian:2002fx}
{\bfseries HERMES} Collaboration, A.~Airapetian {\em et~al.}, ``Measurement of
  ${R = \sigma_L/\sigma_T}$ in deep-inelastic scattering on nuclei,''
\href{http://arxiv.org/abs/hep-ex/0210068}{{\ttfamily hep-ex/0210068}}.
%%CITATION = HEP-EX 0210068;%%.

\bibitem{Amaudruz:1995tq}
{\bfseries New Muon} Collaboration, P.~Amaudruz {\em et~al.}, ``{A reevaluation
  of the nuclear structure function ratios for D, He, Li-6, C and Ca},'' {\em
  Nucl. Phys.} {\bfseries B441} (1995) 3--11,
\href{http://arxiv.org/abs/hep-ph/9503291}{{\ttfamily hep-ph/9503291}}.
%%CITATION = HEP-PH 9503291;%%.

\bibitem{Gomez:1993ri}
J.~Gomez {\em et~al.}, ``{Measurement of the $A$-dependence of deep inelastic
  electron scattering},''
{\em Phys. Rev.} {\bfseries D49} (1994) 4348--4372.
%%CITATION = PHRVA,D49,4348;%%.

\bibitem{Arneodo:1995cs}
{\bfseries New Muon.} Collaboration, M.~Arneodo {\em et~al.}, ``{The structure
  function ratios $F_2^{Li} / F_2^{D}$ and $F_2^{C} / F_2^{D}$ at small $x$},''
  {\em Nucl. Phys.} {\bfseries B441} (1995) 12--30,
\href{http://arxiv.org/abs/hep-ex/9504002}{{\ttfamily hep-ex/9504002}}.
%%CITATION = HEP-EX 9504002;%%.

\bibitem{Adams:1995is}
{\bfseries E665} Collaboration, M.~R. Adams {\em et~al.}, ``Shadowing in
  inelastic scattering of muons on carbon, calcium and lead at low $x_{Bj}$,''
  {\em Z. Phys.} {\bfseries C67} (1995) 403--410,
\href{http://arxiv.org/abs/hep-ex/9505006}{{\ttfamily hep-ex/9505006}}.
%%CITATION = HEP-EX 9505006;%%.

\bibitem{Ashman:1988bf}
{\bfseries European Muon} Collaboration, J.~Ashman {\em et~al.}, ``Measurement
  of the ratios of deep inelastic muon - nucleus cross-sections on various
  nuclei compared to deuterium,''
{\em Phys. Lett.} {\bfseries B202} (1988) 603.
%%CITATION = PHLTA,B202,603;%%.

\bibitem{Arneodo:1989sy}
{\bfseries European Muon} Collaboration, M.~Arneodo {\em et~al.},
  ``Measurements of the nucleon structure function in the range $0.002 < x <
  0.17$ and {$0.2\ {\rm GeV}^2 < Q^2 < 8\ {\rm GeV}^2$} in deuterium, carbon
  and calcium,''
{\em Nucl. Phys.} {\bfseries B333} (1990) 1.
%%CITATION = NUPHA,B333,1;%%.

\bibitem{Bari:1985ga}
{\bfseries BCDMS} Collaboration, G.~Bari {\em et~al.}, ``A measurement of
  nuclear effects in deep inelastic muon scattering on deuterium, nitrogen and
  iron targets,''
{\em Phys. Lett.} {\bfseries B163} (1985) 282.
%%CITATION = PHLTA,B163,282;%%.

\bibitem{Bodek:1983ec}
A.~Bodek {\em et~al.}, ``A comparison of the deep inelastic structure functions
  of deuterium and aluminum nuclei,''
{\em Phys. Rev. Lett.} {\bfseries 51} (1983) 534.
%%CITATION = PRLTA,51,534;%%.

\bibitem{Bodek:1983qn}
A.~Bodek {\em et~al.}, ``Electron scattering from nuclear targets and quark
  distributions in nuclei,''
{\em Phys. Rev. Lett.} {\bfseries 50} (1983) 1431.
%%CITATION = PRLTA,50,1431;%%.

\bibitem{Dasu:1993vk}
S.~Dasu {\em et~al.}, ``{Measurement of kinematic and nuclear dependence of $R
  = \sigma_L/ \sigma_T$ in deep inelastic electron scattering},''
{\em Phys. Rev.} {\bfseries D49} (1994) 5641--5670.
%%CITATION = PHRVA,D49,5641;%%.

\bibitem{Benvenuti:1987az}
{\bfseries BCDMS} Collaboration, A.~C. Benvenuti {\em et~al.}, ``Nuclear
  effects in deep inelastic muon scattering on deuterium and iron targets,''
{\em Phys. Lett.} {\bfseries B189} (1987) 483.
%%CITATION = PHLTA,B189,483;%%.

\bibitem{Ashman:1992kv}
{\bfseries European Muon} Collaboration, J.~Ashman {\em et~al.}, ``A
  measurement of the ratio of the nucleon structure function in copper and
  deuterium,''
{\em Z. Phys.} {\bfseries C57} (1993) 211--218.
%%CITATION = ZEPYA,C57,211;%%.

\bibitem{Adams:1992nf}
{\bfseries E665} Collaboration, M.~R. Adams {\em et~al.}, ``Saturation of
  shadowing at very low $x_{Bj}$,''
{\em Phys. Rev. Lett.} {\bfseries 68} (1992) 3266--3269.
%%CITATION = PRLTA,68,3266;%%.

\bibitem{Arneodo:1996rv}
{\bfseries New Muon} Collaboration, M.~Arneodo {\em et~al.}, ``{The $A$
  dependence of the nuclear structure function ratios},''
{\em Nucl. Phys.} {\bfseries B481} (1996) 3--22.
%%CITATION = NUPHA,B481,3;%%.

\bibitem{Arneodo:1996ru}
{\bfseries New Muon} Collaboration, M.~Arneodo {\em et~al.}, ``{The $Q^2$
  dependence of the structure function ratio $F_2^{Sn} / F_2^{C}$ and the
  difference $R^{Sn} - R^{C}$ in deep inelastic muon scattering},''
{\em Nucl. Phys.} {\bfseries B481} (1996) 23--39.
%%CITATION = NUPHA,B481,23;%%.

\bibitem{Alde:1990im}
D.~M. Alde {\em et~al.}, ``{Nuclear dependence of dimuon production at 800 GeV.
  FNAL-772 experiment},''
{\em Phys. Rev. Lett.} {\bfseries 64} (1990) 2479--2482.
%%CITATION = PRLTA,64,2479;%%.

\bibitem{Vasilev:1999fa}
{\bfseries FNAL E866} Collaboration, M.~A. Vasilev {\em et~al.}, ``{Parton
  energy loss limits and shadowing in Drell-Yan dimuon production},'' {\em
  Phys. Rev. Lett.} {\bfseries 83} (1999) 2304--2307,
\href{http://arxiv.org/abs/hep-ex/9906010}{{\ttfamily hep-ex/9906010}}.
%%CITATION = HEP-EX 9906010;%%.

\bibitem{Adler:2006wg}
{\bfseries PHENIX} Collaboration, S.~Adler {\em et~al.}, ``{Centrality
  dependence of pi0 and eta production at large transverse momentum in
  s(NN)**(1/2) = 200-GeV d+Au collisions},''
  \href{http://dx.doi.org/10.1103/PhysRevLett.98.172302}{{\em Phys.Rev.Lett.}
  {\bfseries 98} (2007) 172302},
\href{http://arxiv.org/abs/nucl-ex/0610036}{{\ttfamily arXiv:nucl-ex/0610036
  [nucl-ex]}}.
%%CITATION = NUCL-EX/0610036;%%.

\bibitem{Abelev:2009hx}
{\bfseries STAR} Collaboration, B.~Abelev {\em et~al.}, ``{Inclusive $\pi^0$,
  $\eta$, and direct photon production at high transverse momentum in $p+p$ and
  $d+$Au collisions at $\sqrt{s_{NN}}=200$ GeV},''
  \href{http://dx.doi.org/10.1103/PhysRevC.81.064904}{{\em Phys.Rev.}
  {\bfseries C81} (2010) 064904},
\href{http://arxiv.org/abs/0912.3838}{{\ttfamily arXiv:0912.3838 [hep-ex]}}.
%%CITATION = ARXIV:0912.3838;%%.

\bibitem{Arrington:2003nt}
J.~Arrington, R.~Ent, C.~E. Keppel, J.~Mammei, and I.~Niculescu, ``{Low Q
  scaling, duality, and the EMC effect},''
  \href{http://dx.doi.org/10.1103/PhysRevC.73.035205}{{\em Phys. Rev.}
  {\bfseries C73} (2006) 035205},
  \href{http://arxiv.org/abs/nucl-ex/0307012}{{\ttfamily arXiv:nucl-ex/0307012
  [nucl-ex]}}.

\bibitem{Malace:2014uea}
S.~Malace, D.~Gaskell, D.~W. Higinbotham, and I.~Cloet, ``{The Challenge of the
  EMC Effect: existing data and future directions},''
  \href{http://dx.doi.org/10.1142/S0218301314300136}{{\em Int. J. Mod. Phys.}
  {\bfseries E23} (2014) 1430013},
\href{http://arxiv.org/abs/1405.1270}{{\ttfamily arXiv:1405.1270 [nucl-ex]}}.
%%CITATION = ARXIV:1405.1270;%%.

\bibitem{Berge:1989hr}
J.~Berge, H.~Burkhardt, F.~Dydak, R.~Hagelberg, M.~Krasny, {\em et~al.}, ``{A
  Measurement of Differential Cross-Sections and Nucleon Structure Functions in
  Charged Current Neutrino Interactions on Iron},''
\href{http://dx.doi.org/10.1007/BF01555493}{{\em Z.Phys.} {\bfseries C49}
  (1991) 187--224}.
%%CITATION = ZEPYA,C49,187;%%.

\bibitem{Onengut:2005kv}
{\bfseries CHORUS} Collaboration, G.~Onengut {\em et~al.}, ``{Measurement of
  nucleon structure functions in neutrino scattering},''
\href{http://dx.doi.org/10.1016/j.physletb.2005.10.062}{{\em Phys.Lett.}
  {\bfseries B632} (2006) 65--75}.
%%CITATION = PHLTA,B632,65;%%.

\bibitem{Tzanov:2005kr}
{\bfseries NuTeV} Collaboration, M.~Tzanov {\em et~al.}, ``Precise measurement
  of neutrino and anti-neutrino differential cross sections,'' {\em Phys. Rev.}
  {\bfseries D74} (2006) 012008,
\href{http://arxiv.org/abs/hep-ex/0509010}{{\ttfamily hep-ex/0509010}}.
%%CITATION = HEP-EX 0509010;%%.

\bibitem{Paukkunen:2010hb}
H.~Paukkunen and C.~A. Salgado, ``{Compatibility of neutrino DIS data and
  global analyses of parton distribution functions},''
  \href{http://dx.doi.org/10.1007/JHEP07(2010)032}{{\em JHEP} {\bfseries 1007}
  (2010) 032},
\href{http://arxiv.org/abs/1004.3140}{{\ttfamily arXiv:1004.3140 [hep-ph]}}.
%%CITATION = ARXIV:1004.3140;%%.

\bibitem{Aad:2014bha}
{\bfseries ATLAS} Collaboration, G.~Aad {\em et~al.}, ``{Measurement of the
  production and lepton charge asymmetry of $W$ bosons in Pb+Pb collisions at
  $\mathbf {\sqrt{\mathbf {s}_{\mathrm {\mathbf {NN}}}}=2.76\;TeV}$ with the
  ATLAS detector},''
  \href{http://dx.doi.org/10.1140/epjc/s10052-014-3231-6}{{\em Eur.Phys.J.}
  {\bfseries C75} no.~1, (2015) 23},
\href{http://arxiv.org/abs/1408.4674}{{\ttfamily arXiv:1408.4674 [hep-ex]}}.
%%CITATION = ARXIV:1408.4674;%%.

\bibitem{ATLAS-CONF-2014-020}
``{$Z$ Boson Production in p+Pb Collisions at $\sqrt{s_{\mathrm{NN}}}=5.02$ TeV
  with the ATLAS Detector},'' Tech. Rep. ATLAS-CONF-2014-020, CERN, Geneva,
  May, 2014.

\bibitem{Chatrchyan:2014csa}
{\bfseries CMS} Collaboration, S.~Chatrchyan {\em et~al.}, ``{Study of Z
  production in PbPb and pp collisions at $ \sqrt{s_{\mathrm{NN}}}=2.76 $ TeV
  in the dimuon and dielectron decay channels},''
  \href{http://dx.doi.org/10.1007/JHEP03(2015)022}{{\em JHEP} {\bfseries 1503}
  (2015) 022},
\href{http://arxiv.org/abs/1410.4825}{{\ttfamily arXiv:1410.4825 [nucl-ex]}}.
%%CITATION = ARXIV:1410.4825;%%.

\bibitem{Khachatryan:2015hha}
{\bfseries CMS Collaboration} Collaboration, V.~Khachatryan {\em et~al.},
  ``{Study of W boson production in pPb collisions at sqrt(s[NN]) = 5.02
  TeV},''
\href{http://arxiv.org/abs/1503.05825}{{\ttfamily arXiv:1503.05825 [nucl-ex]}}.
%%CITATION = ARXIV:1503.05825;%%.

\bibitem{STARdata}
{\bfseries STAR} Collaboration, ``{Data from: Phys.Rev. C 81 (2010) 064904;
  arXiv:0912.3838}.''
  \url{https://drupal.star.bnl.gov/STAR/files/starpublications/151/data.html}.

\bibitem{Cloet:2009qs}
I.~C. Cloet, W.~Bentz, and A.~W. Thomas, ``{Isovector EMC effect explains the
  NuTeV anomaly},''
  \href{http://dx.doi.org/10.1103/PhysRevLett.102.252301}{{\em Phys. Rev.
  Lett.} {\bfseries 102} (2009) 252301},
\href{http://arxiv.org/abs/0901.3559}{{\ttfamily arXiv:0901.3559 [nucl-th]}}.
%%CITATION = ARXIV:0901.3559;%%.

\bibitem{Dutta:2010pg}
D.~Dutta, J.~C. Peng, I.~C. Cloet, and D.~Gaskell, ``{Pion-induced Drell-Yan
  processes and the flavor-dependent EMC effect},''
  \href{http://dx.doi.org/10.1103/PhysRevC.83.042201}{{\em Phys. Rev.}
  {\bfseries C83} (2011) 042201},
\href{http://arxiv.org/abs/1007.3916}{{\ttfamily arXiv:1007.3916 [nucl-ex]}}.
%%CITATION = ARXIV:1007.3916;%%.

\bibitem{Accardi:2009br}
A.~Accardi, M.~Christy, C.~Keppel, P.~Monaghan, W.~Melnitchouk, {\em et~al.},
  ``{New parton distributions from large-$x$ and low-$Q^2$ data},''
  \href{http://dx.doi.org/10.1103/PhysRevD.81.034016}{{\em Phys.Rev.}
  {\bfseries D81} (2010) 034016},
\href{http://arxiv.org/abs/0911.2254}{{\ttfamily arXiv:0911.2254 [hep-ph]}}.
%%CITATION = ARXIV:0911.2254;%%.

\bibitem{Accardi:2011fa}
A.~Accardi, W.~Melnitchouk, J.~Owens, M.~Christy, C.~Keppel, {\em et~al.},
  ``{Uncertainties in determining parton distributions at large x},''
  \href{http://dx.doi.org/10.1103/PhysRevD.84.014008}{{\em Phys.Rev.}
  {\bfseries D84} (2011) 014008},
\href{http://arxiv.org/abs/1102.3686}{{\ttfamily arXiv:1102.3686 [hep-ph]}}.
%%CITATION = ARXIV:1102.3686;%%.

\bibitem{Binnewies:1994ju}
J.~Binnewies, B.~A. Kniehl, and G.~Kramer, ``{Next-to-leading order
  fragmentation functions for pions and kaons},''
  \href{http://dx.doi.org/10.1007/BF01556135}{{\em Z.Phys.} {\bfseries C65}
  (1995) 471--480},
\href{http://arxiv.org/abs/hep-ph/9407347}{{\ttfamily arXiv:hep-ph/9407347
  [hep-ph]}}.
%%CITATION = HEP-PH/9407347;%%.

\bibitem{Kniehl:2000fe}
B.~A. Kniehl, G.~Kramer, and B.~Potter, ``{Fragmentation functions for pions,
  kaons, and protons at next-to-leading order},''
  \href{http://dx.doi.org/10.1016/S0550-3213(00)00303-5}{{\em Nucl.Phys.}
  {\bfseries B582} (2000) 514--536},
\href{http://arxiv.org/abs/hep-ph/0010289}{{\ttfamily arXiv:hep-ph/0010289
  [hep-ph]}}.
%%CITATION = HEP-PH/0010289;%%.

\bibitem{Enterria:2013vba}
D.~d'Enterria, K.~J. Eskola, I.~Helenius, and H.~Paukkunen, ``{Confronting
  current NLO parton fragmentation functions with inclusive charged-particle
  spectra at hadron colliders},''
  \href{http://dx.doi.org/10.1016/j.nuclphysb.2014.04.006}{{\em Nucl.Phys.}
  {\bfseries B883} (2014) 615--628},
\href{http://arxiv.org/abs/1311.1415}{{\ttfamily arXiv:1311.1415 [hep-ph]}}.
%%CITATION = ARXIV:1311.1415;%%.

\bibitem{Nadolsky:2008zw}
P.~M. Nadolsky, H.-L. Lai, Q.-H. Cao, J.~Huston, J.~Pumplin, {\em et~al.},
  ``{Implications of CTEQ global analysis for collider observables},''
  \href{http://dx.doi.org/10.1103/PhysRevD.78.013004}{{\em Phys.Rev.}
  {\bfseries D78} (2008) 013004},
\href{http://arxiv.org/abs/0802.0007}{{\ttfamily arXiv:0802.0007 [hep-ph]}}.
%%CITATION = ARXIV:0802.0007;%%.

\bibitem{Khanpour:2016pph}
H.~Khanpour and S.~A. Tehrani, ``{Global analysis of nuclear parton
  distribution functions and their uncertainties at next-to-next-to-leading
  order},'' \href{http://dx.doi.org/10.1103/PhysRevD.93.014026}{{\em Phys.
  Rev.} {\bfseries D93} no.~1, (2016) 014026},
\href{http://arxiv.org/abs/1601.00939}{{\ttfamily arXiv:1601.00939 [hep-ph]}}.
%%CITATION = ARXIV:1601.00939;%%.

\bibitem{nCTEQwebpage}
``{\tt nCTEQ} website.'' \url{http://ncteq.hepforge.org/}.

\bibitem{Buckley:2014ana}
A.~Buckley, J.~Ferrando, S.~Lloyd, K.~Nordstr{\"o}m, B.~Page, {\em et~al.},
  ``{LHAPDF6: parton density access in the LHC precision era},''
  \href{http://dx.doi.org/10.1140/epjc/s10052-015-3318-8}{{\em Eur.Phys.J.}
  {\bfseries C75} no.~3, (2015) 132},
\href{http://arxiv.org/abs/1412.7420}{{\ttfamily arXiv:1412.7420 [hep-ph]}}.
%%CITATION = ARXIV:1412.7420;%%.

\bibitem{LHAPDFwebpage}
``{\tt LHAPDF} website.'' \url{https://lhapdf.hepforge.org/}.

\end{thebibliography}
%%%%%%%%%%%%%%%%%%%%%%%%%%%%%

%%%%%%%%%%%%%%%%%%%%%%%%%%%%%%%%%%%%%%%%%%%%%%%%%%%%%%%%%%
%%%%%%%%%%%%%%%%%%%%%%%%%%%%%%%%%%%%%%%%%%%%%%%%%%%%%%%%%%
\end{document}